\documentclass{jfm}

\usepackage{natbib}

\usepackage{amsmath}
\usepackage{graphicx}
\usepackage{caption}
\usepackage{subcaption}
\usepackage{float}
\usepackage{multirow}
\usepackage{bm}
\usepackage{color}
\usepackage{tikz}

\newcommand\solidrule[1][0.5cm]{\rule[0.5ex]{#1}{1pt}}
\newcommand\dashedrule{\mbox{%
  \solidrule[1mm]\hspace{0.7mm}\solidrule[1mm]}}

\begin{document}

\newtheorem{lemma}{Lemma}
\newtheorem{corollary}{Corollary}

\shorttitle{Global modes and nonlinear analysis of inverted-flag flapping} 
\shortauthor{A. Goza, T. Colonius, and J. Sader} 

\title{Global modes and nonlinear analysis of inverted-flag flapping}

\author
 {
Andres Goza\aff{1}
  \corresp{\email{ajgoza@gmail.com}},
  Tim Colonius\aff{1},
  \and 
  John E. Sader\aff{2,3}
  }

\affiliation
{
\aff{1}
Department of Mechanical and Civil Engineering, California Institute of Technology, Pasadena, CA 91125, USA
\aff{2}
ARC Centre of Excellence in Exciton Science, School of Mathematics and Statistics, University of Melbourne, Victoria 3010, Australia
\aff{3}
Department of Physics, California Institute of Technology, Pasadena, CA 91125, USA
}

\maketitle

\begin{abstract}

An inverted flag has its trailing edge clamped and exhibits dynamics distinct from that of a conventional flag, whose leading edge is restrained. We perform nonlinear simulations and a global stability analysis of the inverted-flag system for a range of Reynolds numbers, flag masses and stiffnesses. Our global stability analysis is based on a linearisation of the fully-coupled fluid-structure system of equations. The calculated equilibria are steady-state solutions of the fully-coupled nonlinear equations. By implementing this approach, we (i) explore the mechanisms that initiate flapping, (ii) study the role of vortex shedding and vortex-induced vibration (VIV) in large-amplitude flapping, and (iii) characterise the chaotic flapping regime. For point (i), we identify a deformed-equilibrium state and show through a global stability analysis that the onset of flapping is due to a supercritical Hopf bifurcation. For large-amplitude flapping, point (ii), we confirm the arguments of \citet{Sader2016a} that for a range of parameters this regime is a VIV. We also show that there are other flow regimes for which large-amplitude flapping persists and is not a VIV. Specifically, flapping can occur at low Reynolds numbers ($<50$), albeit via a previously unexplored mechanism. Finally, with respect to point (iii), chaotic flapping has been observed experimentally for Reynolds numbers of $O(10^4)$, and here we show that chaos also persists at a moderate Reynolds number of 200. We characterise this chaotic regime and calculate its strange attractor, whose structure is controlled by the above-mentioned deformed equilibria and is similar to a Lorenz attractor. These results are contextualised with bifurcation diagrams that depict the different equilibria and various flapping regimes. 

\end{abstract}

\section{Introduction}

\begin{figure}
    \begin{subfigure}[b]{0.19\textwidth}
        \includegraphics[width=\textwidth]{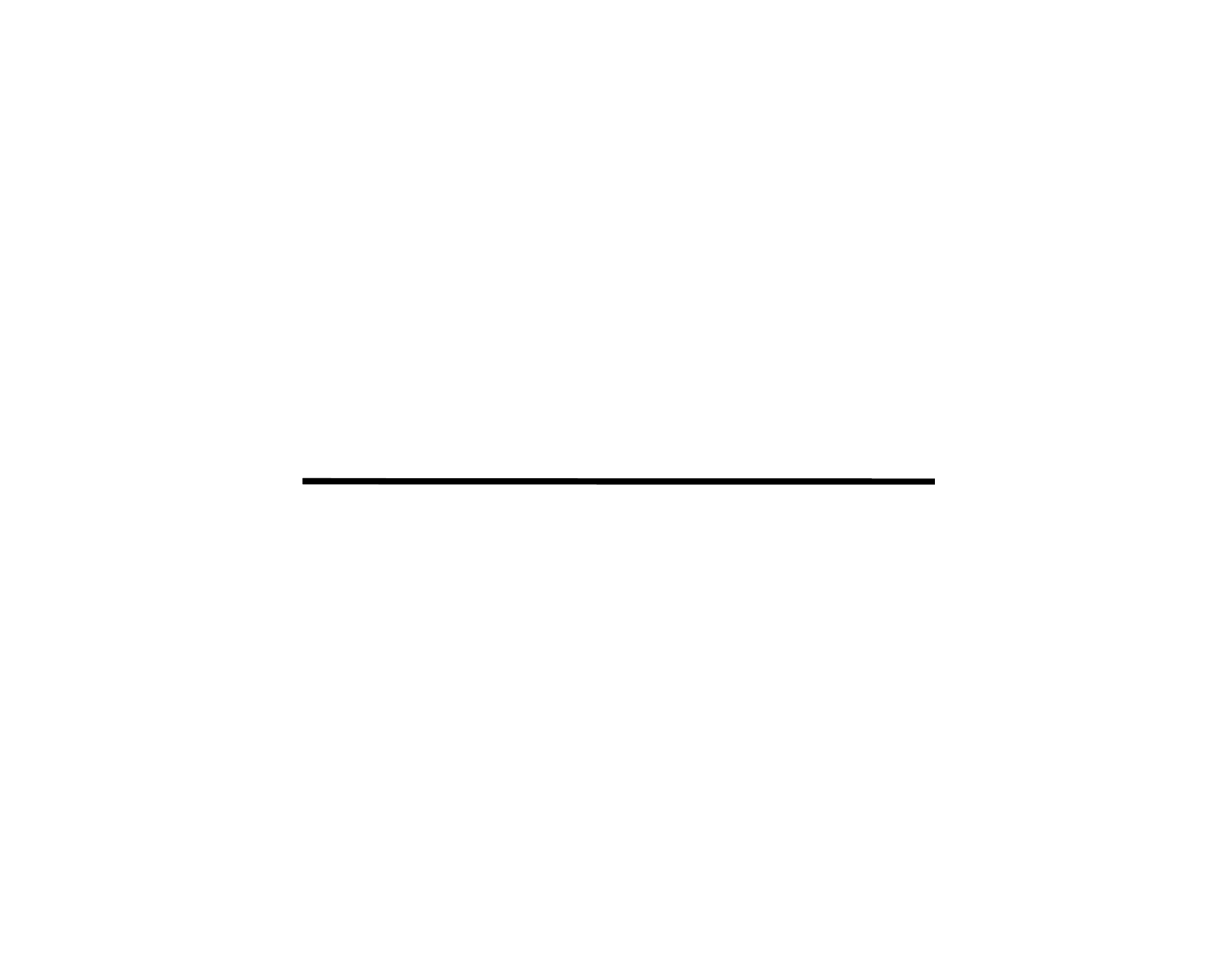}
        \caption{}
        \label{fig:hor_EP}
    \end{subfigure}
    \begin{subfigure}[b]{0.19\textwidth}
        \includegraphics[width=\textwidth]{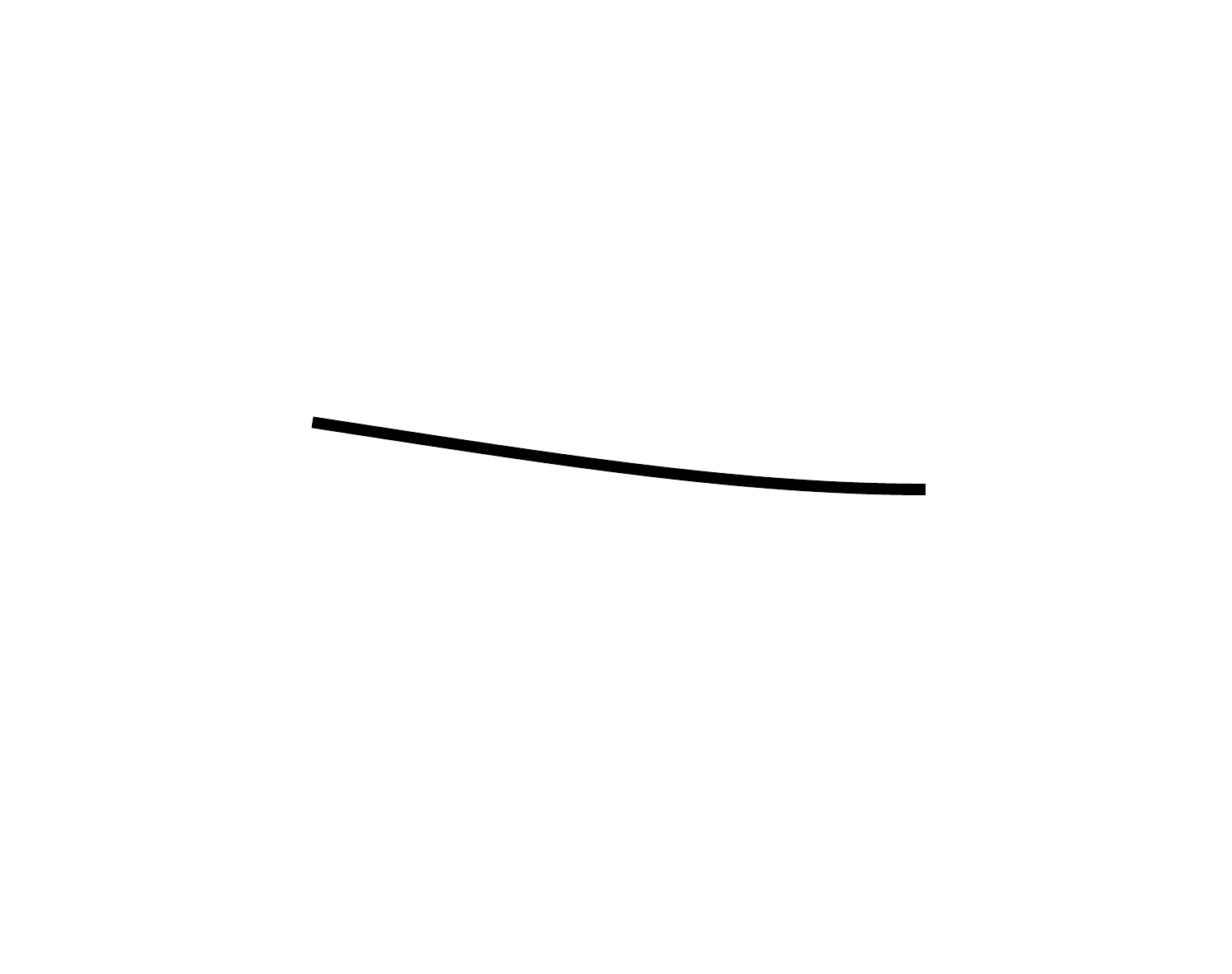}
        \caption{}
        \label{fig:def_EP}
    \end{subfigure}
    \begin{subfigure}[b]{0.19\textwidth}
        \includegraphics[width=\textwidth]{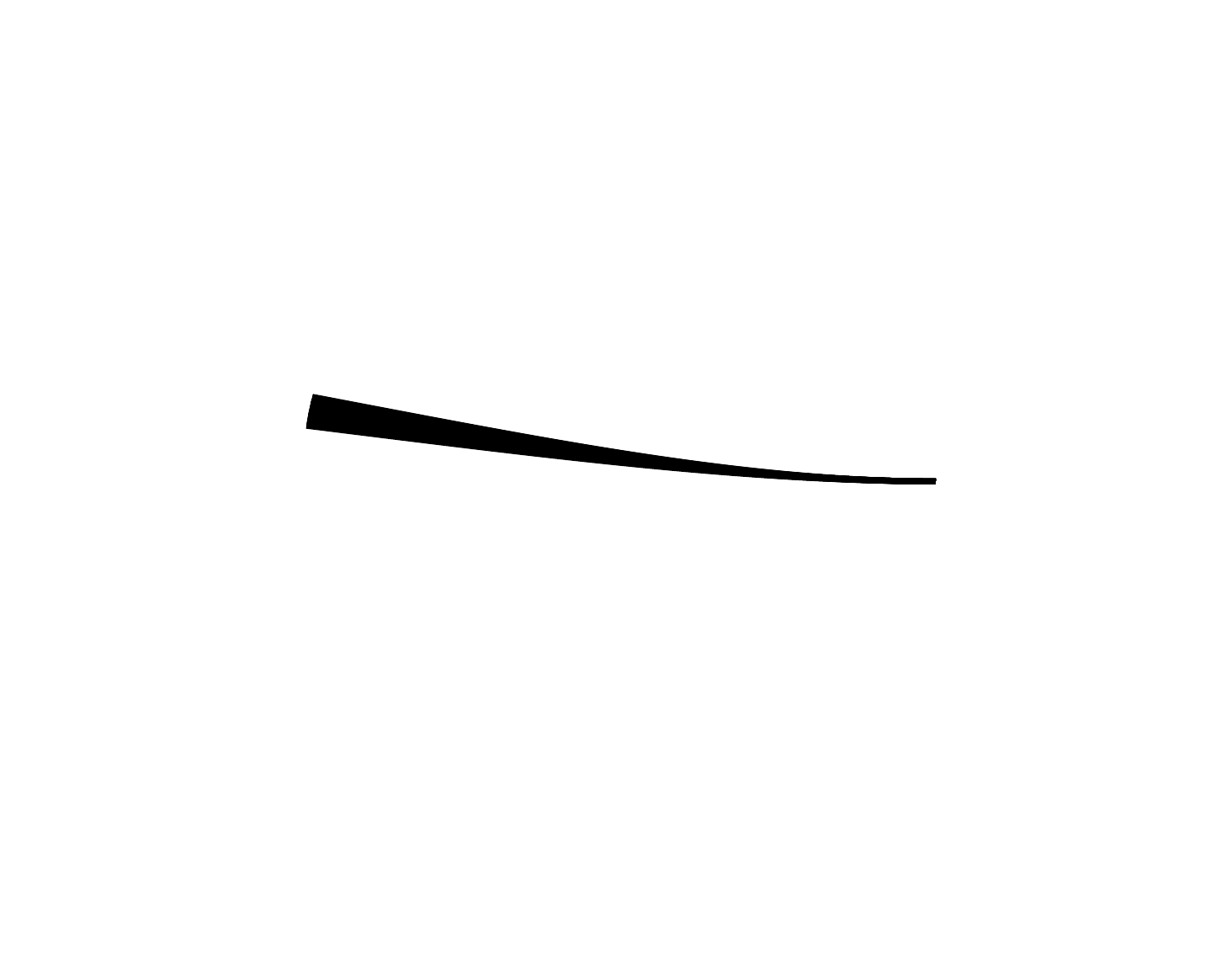}
        \caption{}
        \label{fig:def_LC_s}
    \end{subfigure}
    \begin{subfigure}[b]{0.19\textwidth}
        \includegraphics[width=\textwidth]{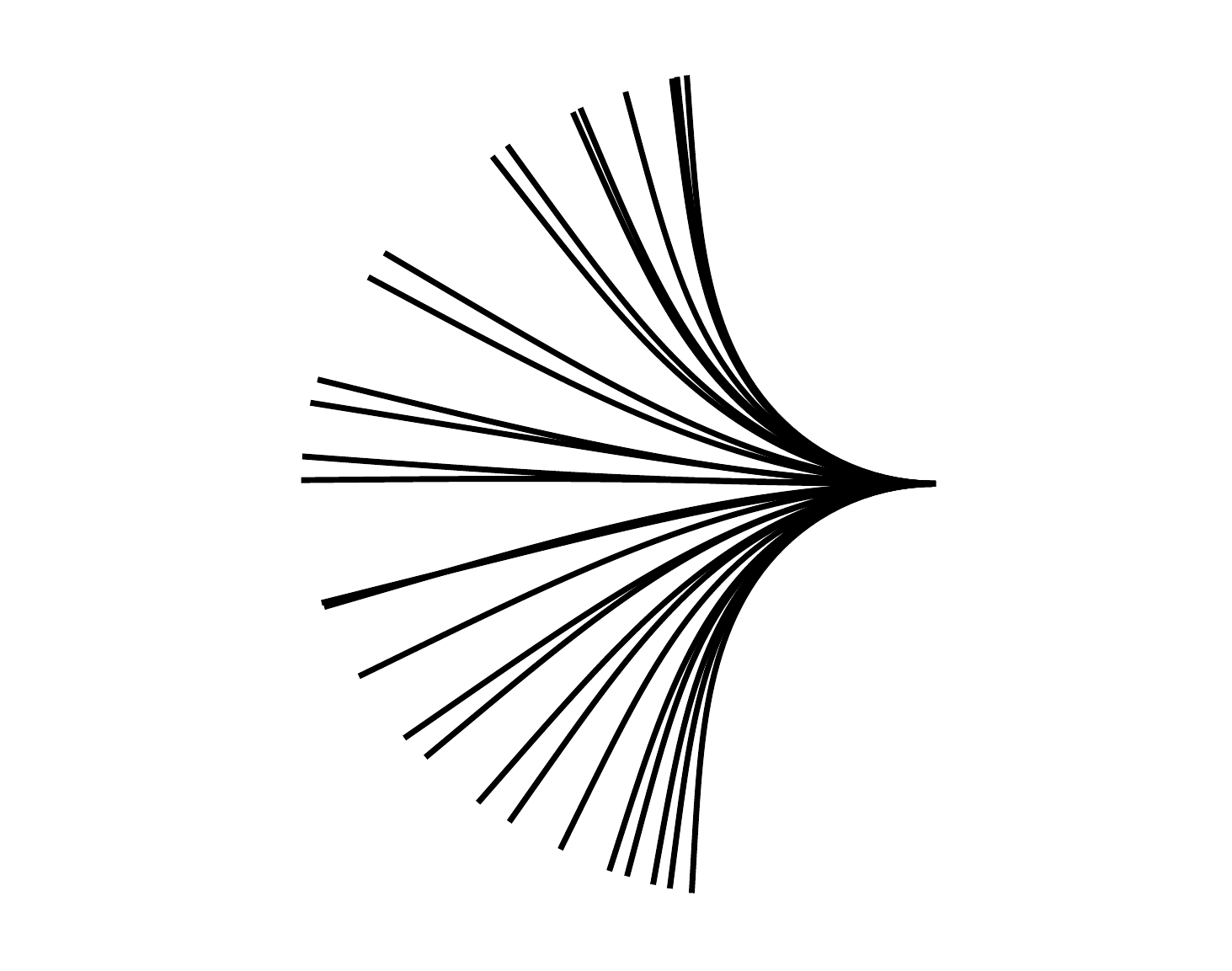}
        \caption{}
        \label{fig:large_LC}
    \end{subfigure}
    \begin{subfigure}[b]{0.19\textwidth}
        \includegraphics[width=\textwidth]{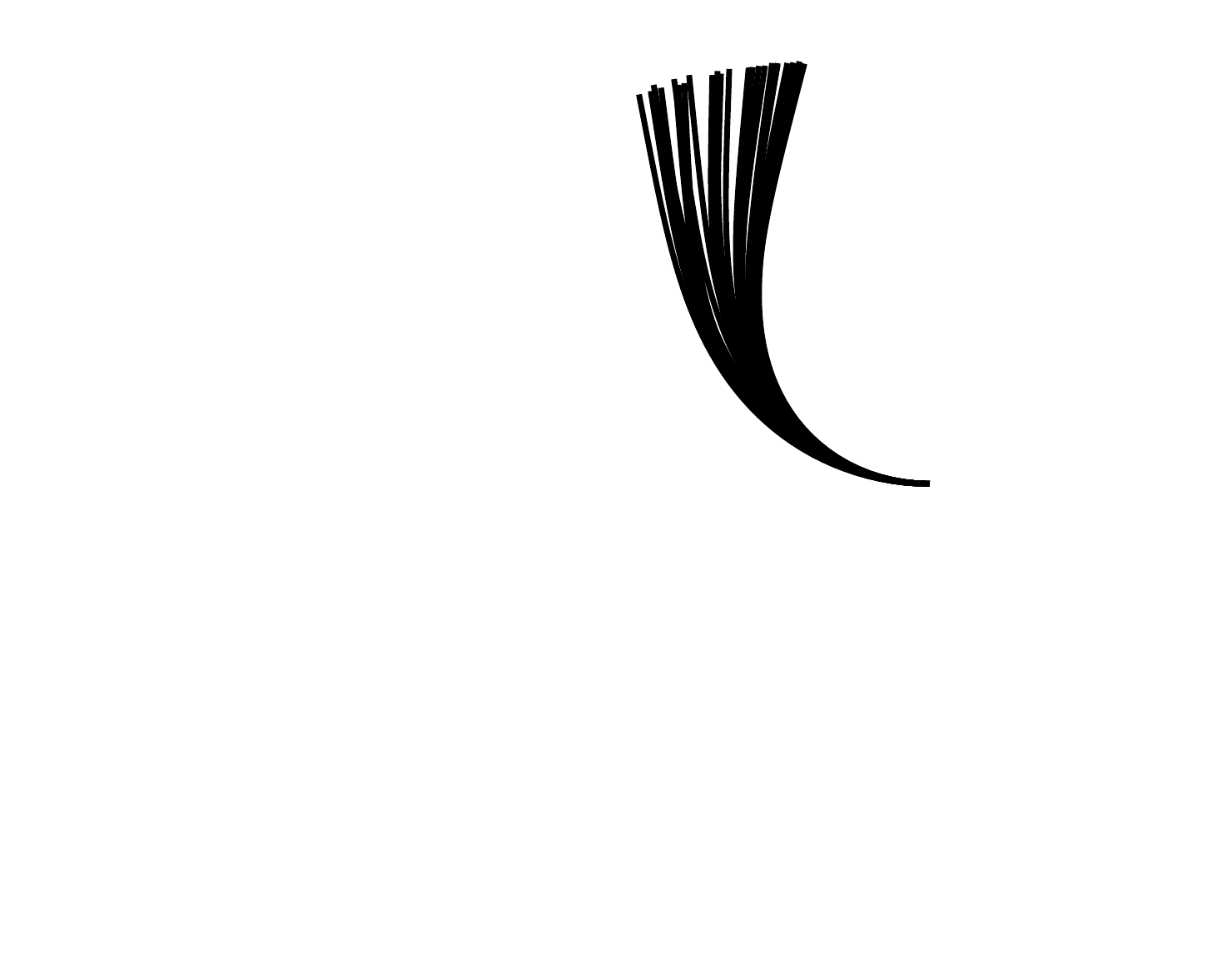}
        \caption{}
        \label{fig:def_LC_L}
    \end{subfigure}
    \caption{Time lapses of flag position for (a) the undeformed equilibrium, (b) small-deflection stable, (c) small-deflection deformed flapping, (d) large-amplitude flapping, and (e) deflected-mode regimes. In all figures, the flag is clamped at its right edge and the flow direction is from left to right.}
    \label{fig:flag_schem}
\end{figure}

Uniform flow past a conventional flag---where the flag is pinned or clamped at its leading edge with respect to the oncoming flow---has been studied widely beginning with the early work of \citet{Taneda1968} (see \citet{Shelley2011} for a recent review). By contrast, studies of flow past an inverted flag, in which the flag is clamped at its trailing edge, have only been reported recently. The inverted-flag system displays a wide range of dynamical regimes \citep{Kim2013, Gurugubelli2015, Ryu2015}, many of which are depicted in figure \ref{fig:flag_schem}. This figure is produced from the numerical simulations described in section \ref{sec:math}.

One of the dynamical regimes depicted is large-amplitude flapping (figure \ref{fig:large_LC}), which is associated with a larger strain energy than that of conventional flag flapping. These large bending strains make the inverted-flag system a promising candidate for energy harvesting technologies that convert strain energy to electricity, \emph{e.g.}, by using piezoelectric materials. \cite{Shoele2016} studied this energy harvesting potential in detail by performing numerical simulations of a fully-coupled fluid-structure-piezoelectric model.

Transitions between the various regimes in figure \ref{fig:flag_schem} depend on the Reynolds number ($Re$), dimensionless mass ratio ($M_\rho$), and dimensionless bending stiffness ($K_B$), defined as
\begin{equation}
Re = \frac{\rho_f U L}{\mu}, \; M_\rho = \frac{\rho_s h}{\rho_f L}, \; K_B = \frac{EI}{\rho_f U^2 L^3}
\label{eqn:params}
\end{equation}
where $\rho_f$ ($\rho_s$) is the fluid (structure) density, $U$ is the freestream velocity, $L$ is the flag length, $\mu$ is the shear viscosity of the fluid, $h$ is the flag thickness, and $EI$ is the flexural rigidity of the flag. In experiments, regime transitions are  triggered by increasing the flow rate \citep{Kim2013}. This coincides with a decrease in $K_B$ and an increase in $Re$ for fixed $M_\rho$, by virtue of (\ref{eqn:params}). In contrast, numerical simulations often decrease the flag's stiffness at fixed $Re$ and $M_\rho$, which isolates the effect of various parameters and facilitates comparison to previous numerical studies of flow-induced vibration.

Simulations show that for moderate Reynolds numbers ($\lesssim$ 1000), a systematic decrease in $K_B$ causes a change from a stable undeformed equilibrium state (figure \ref{fig:hor_EP}) to a small-deflection stable state (figure \ref{fig:def_EP}). This is followed by a transition to small-deflection deformed flapping (figure \ref{fig:def_LC_s}), then to large-amplitude flapping (figure \ref{fig:large_LC}), and finally to a deflected-mode regime (figure \ref{fig:def_LC_L}) \citep{Gurugubelli2015,Ryu2015}. These simulations have been performed primarily for $M_\rho \le O(1)$ (heavy fluid loading), though \citet{Shoele2016} considered large values of $M_\rho$.
 
The same regime transitions persist at higher Reynolds numbers, $Re \sim O(10^4)$, except that the small-deflection stable and small-deflection deformed flapping regimes discussed above are no longer present. That is, the undeformed equilibrium directly gives way to large-amplitude flapping \citep{Kim2013}. Moreover,  \citet{Sader2016a} experimentally identified a chaotic flapping regime (not shown in figure \ref{fig:flag_schem}) at these higher Reynolds numbers that has yet to be reported using numerical simulations with $Re \le O(1000)$.

At low Reynolds numbers ($Re < 50$), numerical simulations have shown that the inverted flag's dynamics can change significantly: no flapping occurs, with the only observed regimes being the undeformed equilibrium and stable deflected states \citep{Ryu2015}. These simulations were performed over a wide range of $K_B$ for only one value of $M_\rho$, and the system's dependence on these two parameters remains an open question at these lower Reynolds numbers.

Several driving mechanisms of the various regimes illustrated in figures \ref{fig:flag_schem}(a)--(e) have been identified. The bifurcation from the undeformed equilibrium is caused by a divergence instability (\emph{i.e.}, the instability is independent of $M_\rho$). This mechanism was originally suggested by \citet{Kim2013}, and subsequently found computationally \citep{Gurugubelli2015} and mathematically via a linear stability analysis \citep{Sader2016a}. For large-amplitude flapping, \cite{Sader2016a} used experiments and a scaling analysis to argue that this regime is a vortex-induced vibration (VIV) for a distinct range of parameters. The primary role of vortex shedding in large-amplitude flapping is further evidenced by the above-mentioned observation of \citet{Ryu2015} that flapping does not occur below $Re \approx 50$ (for certain values of $M_\rho$). Based on a scaling analysis, \cite{Sader2016a} also predicted that VIV should cease as the mass ratio, $M_\rho$, increases---a prediction that is yet to be verified.  With respect to the deflected-mode regime, small-amplitude flapping about a large mean-deflected position occurs, and \citet{Shoele2016} showed that the flapping frequency is identical to that of the vortex shedding caused by the flag's bluffness.

In this article, we use high-fidelity nonlinear simulations and a global linear stability analysis to further characterise the regimes in figure \ref{fig:flag_schem} and explore their driving physical mechanisms. We emphasise that our global stability analysis is based on a linearisation of the fully-coupled fluid-structure system of equations. Moreover, the computed equilibria are steady-state solutions of the fully-coupled nonlinear equations described in section \ref{sec:math}. Our results are presented for Reynolds numbers of 20 and 200, various values of $K_B$, and values of $M_\rho$ spanning four orders of magnitude.

Using this approach, we (i) study the mechanisms responsible for the onset of small-deflection deformed flapping, (ii) probe the role of vortex shedding and VIV in large-amplitude flapping, and (iii) investigate whether chaotic flapping occurs at low-to-moderate Reynolds numbers ($Re = $ 20 and 200). To explore (i), we first demonstrate that the small-deflection stable state is an equilibrium of the fully-coupled fluid-structure system. Through a global stability analysis, we show that the subsequent transition to small-deflection deformed flapping (figure \ref{fig:def_LC_s}) as the bending stiffness decreases is a supercritical Hopf bifurcation of this deformed equilibrium. For point (ii), we confirm the arguments of \citet{Sader2016a} that large-amplitude flapping is a VIV for the higher Reynolds number of $Re = 200$ and lower values of the mass ratio, $M_\rho < O(1)$. VIV is also shown to cease for sufficiently large $M_\rho$, consistent with the scaling analysis of \citet{Sader2016a}. Moreover, we show that large-amplitude flapping persists at these large values of $M_\rho$  despite the absence of VIV, albeit by a previously unidentified mechanism. This non-VIV large-amplitude flapping regime also occurs for large $M_\rho$ at the lower Reynolds number of $Re = 20$. Consistent with the simulation results of \citet{Ryu2015}, we find no flapping at this low Reynolds number for $M_\rho < O(1)$. Finally, with respect to (iii), we confirm that chaotic flapping persists at moderate Reynolds numbers ($Re = 200$) for light flags with $M_\rho < 1$, and demonstrate that the structure of the associated strange attractor is controlled by a combination of the large-amplitude and deflected-mode regimes. Chaos does not occur for heavy flags at $Re = 200$ or for any mass ratio considered at $Re=20$. Thus, chaos is associated with parameters for which VIV flapping occurs.



\begin{figure}
	\centering
        \includegraphics[scale=0.5,trim={0cm 0cm 0cm 0cm},clip]{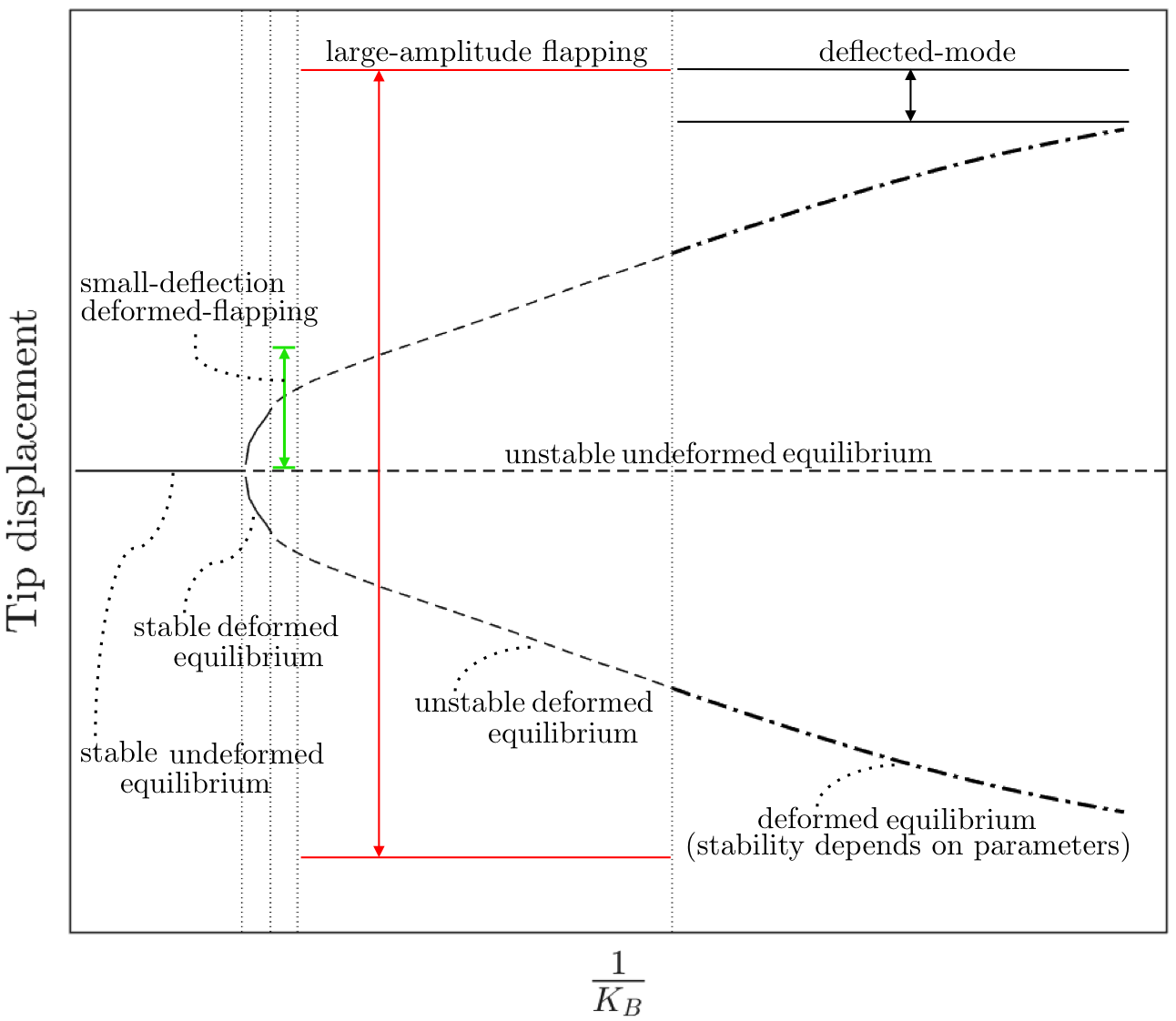}
       \caption{A schematic bifurcation diagram that summarises the results obtained for various parameters considered in the present work. Equilibria are presented by lines ({\color{black} \rule[0.6 pt]{10 pt}{1.2 pt}}, stable equilibria; {\color{black} \rule[0.6 pt]{5 pt}{1.2 pt}}\;{\color{black} \rule[0.6 pt]{5 pt}{1.2 pt}}, unstable equilibria; \protect\tikz \protect\fill[black] (0.5ex,0.5ex) circle (0.3ex); {\color{black} \rule[0.6 pt]{10 pt}{1.2 pt}}, stability depends on parameters). The lines with the double arrows indicate regimes where flapping occurs, with the top and bottom lines representing the peak-to-peak flapping amplitude. The diagram shows that with decreasing $K_B$ (moving left to right), the system transitions from the undeformed equilibrium to a stable deformed equilibrium. Following this, the system bifurcates to small-deflection deformed flapping, then large-amplitude flapping (chaotic flapping can also occur in this regime depending on parameters), and finally to a deflected-mode regime whose dynamics depend on Reynolds number: for $Re = 20$, no flapping occurs and the large-deflection state is an equilibrium of the fully-coupled system; for $Re = 200$, the large-deflection state is characterised by small-amplitude flapping. The above diagram only corresponds to cases when flapping occurs. For $Re = 20$ and $M_\rho < O(1)$, no flapping occurs and the only two regimes are the undeformed and deformed equilibria.}
    \label{fig:bif_schem}
\end{figure}

We contextualise the simulation results over this wide range of parameters using bifurcation diagrams. These provide an overview of the  equilibria, their stability, and the flapping dynamics. Figure \ref{fig:bif_schem} shows an \emph{illustrative}  bifurcation diagram that summarises what will be shown in later sections. In these bifurcation diagrams, the leading edge transverse displacement (tip displacement) is plotted versus the flag flexibility ($1/K_B$) for a particular choice of $Re$ and $M_\rho$ (see the caption for details). Note that even though the undeformed and deformed equilibria become unstable with a decrease in $K_B$, they nonetheless remain as equilibria of the system. We demonstrate below through a global stability analysis that these unstable deformed equilibria are key to understanding a variety of flapping behaviour of the inverted-flag system.

Two-dimensional (2D) simulations are presented throughout. As mentioned above, many similarities exist between the 3D experiments of \cite{Kim2013} and 2D simulations of \cite{Gurugubelli2015, Ryu2015, Shoele2016}. This suggests  that features of the 2D dynamics persist in 3D (though \citet{Sader2016b} demonstrated that substantial differences occur for low-aspect ratio flags). Exploring these similarities and differences between 2D and 3D geometries is a subject of future work and is not considered here. Quantities presented below are dimensionless, with length scales, velocity scales, and time scales nondimensionalised by $L$, $U$, and $L/U$, respectively.

\section{Numerical methods: nonlinear solver and global stability analysis}
\label{sec:math}

Our nonlinear simulations use the immersed boundary method of \cite{Goza2017}. The method treats the fluid with the 2D Navier-Stokes equations, and the flag with the geometrically nonlinear Euler-Bernoulli beam equation. The method is strongly-coupled (\emph{i.e.}, it accounts for the nonlinear coupling between the flag and the fluid), and therefore allows for arbitrarily large flag displacements and rotations. We have validated our method against a variety of test problems involving conventional and inverted flags \citep{Goza2017}. The global stability analysis is based on a linearisation of the nonlinear, fully-coupled flow-structure interaction system, and therefore reveals instability-driving mechanisms in both the flag and the fluid.

In what follows, we review the nonlinear solver (see \cite{Goza2017} for more details) and derive the linearised equations. We then describe the global mode solution approach, the procedure used to compute equilibria of the flow-flag system, and the grid spacing and domain size used for our simulations.

\subsection{Nonlinear solver}

We define the fluid domain as $\Omega$ and the flag surface as $\Gamma$. We let $\textbf{x}$ denote the Eulerian coordinate representing a position in space, and $\bm{\chi}(\theta,t)$ be the Lagrangian coordinate attached to the body $\Gamma$ ($\theta$ is a variable that parametrizes the surface). The dimensionless governing equations are written as
\begin{gather}
 \frac{\partial \textbf{u}}{\partial t} = - \textbf{u}\cdot \nabla \textbf{u} - \nabla p + \frac{1}{\text{Re}} \nabla^2 \textbf{u} + \int_{    \Gamma} \textbf{f}(\bm{\chi}(\theta,t)) \delta(\bm{\chi}(\theta,t) - \textbf{x}) d\theta  \label{eqn:NS} \\
 \nabla \cdot \textbf{u} = 0 \label{eqn:incomp} \\
 \frac{\rho_s}{\rho_f} \frac{\partial^2 \bm{\chi}}{\partial t^2} = \frac{1}{\rho_f U^2}\nabla \cdot \boldsymbol\sigma + \textbf{g}(\bm{\chi}) - \textbf{f}(\bm{\chi}) \label{eqn:sol_eqn} \\
 \int_\Omega \textbf{u}(\textbf{x})\delta(\textbf{x} - \bm{\chi}(\theta,t)) d\textbf{x} = \frac{\partial \bm{\chi}(\theta,t)}{\partial t} \label{eqn:BC}
\end{gather} 

In the above, (\ref{eqn:NS}) expresses the Navier-Stokes equations in an immersed boundary formulation, (\ref{eqn:incomp}) is the continuity equation for the fluid, (\ref{eqn:sol_eqn}) represents the structural equations governing the motion of the flag ($\textbf{g}$ is a body force term), and (\ref{eqn:BC}) is the no-slip boundary condition enforcing that the fluid velocity matches the flag velocity on the flag surface. Note that $\textbf{f}$ represents the effect from the flag surface stresses on the fluid, and is present in both (\ref{eqn:NS}) and (\ref{eqn:sol_eqn}) since by Newton's third law its negative imparts the fluid stresses on the flag surface \citep{Goza2017}. In (\ref{eqn:sol_eqn}), the time derivative is a Lagrangian derivative and the stress tensor is the Cauchy tensor in terms of the deformed flag configuration.


The fluid equations are spatially discretised with the immersed boundary discrete-streamfunction formulation of \citet{Colonius2008}, which removes the pressure and eliminates the continuity equation. The flag equations are treated with a finite element corotational formulation \citep{Crisfield1991}. The spatially discrete, temporally continuous equations written as a first order system of differential-algebraic equations are
\begin{gather}
C^TC\dot{s} = - C^T\mathcal{N}(s) + \frac{1}{Re}C^TLCs - C^TE^T(\chi)f \label{eqn:eq_s} \\
M \dot{\zeta} = -R(\chi) + Q(g + W(\chi)f) \label{eqn:eq_u} \\
\dot{\chi} = \zeta \label{eqn:eq_x} \\
0 = E(\chi)Cs - \zeta \label{eqn:eq_c} 
\end{gather}
where $\chi$ and $f$ are discrete analogues to their continuous counterparts, $s$ is the discrete streamfunction, $\zeta$ is the flag velocity, and all other variables are defined below.  

Equation (\ref{eqn:eq_s}) represents the Navier-Stokes equations written in a discrete-streamfunction formulation, (\ref{eqn:eq_u}) is the geometrically nonlinear Euler-Bernoulli beam equation, (\ref{eqn:eq_x}) matches the time derivative of the flag position to the flag velocity, and (\ref{eqn:eq_c}) is the interface constraint that the fluid and flag must satisfy the no-slip boundary condition on the flag surface.

In (\ref{eqn:eq_s})--(\ref{eqn:eq_c}), $C$ and $C^T$ are discrete curl operators that mimic $\nabla \times (\cdot)$; $\mathcal{N}(s)$ is a discretization of the advection operator $\textbf{u}\cdot\nabla\textbf{u}$ written in terms of the discrete streamfunction \citep{Colonius2008}; $L$ is a discrete Laplacian associated with the viscous diffusion term; $E^Tf$ is a ``smearing'' operator (arising from the immersed boundary treatment) that applies the surface stresses from the flag onto the fluid; $M$ is a mass matrix associated with the flag's inertia; $R(\chi)$ is the internal stress within the flag; $Qg$ is a body force term (\emph{e.g.}, gravity); and $QWf$ is the stress imposed on the flag from the fluid. 

Equation (\ref{eqn:eq_s}) is discretised in time using an Adams Bashforth AB2 scheme for the convective term and a second order Crank-Nicholson scheme for the diffusive term. The flag equations (\ref{eqn:eq_u})--(\ref{eqn:eq_x}) are discretized using an implicit Newmark scheme. The method is strongly coupled, so the constraint equation (\ref{eqn:eq_c}) is enforced at each time step including the present one.

A novel feature of our method is the efficient iterative procedure used to treat the nonlinear coupling between the flag and fluid. Many methods use a block-Gauss Seidel iterative procedure, which converges slowly (or not at all) for light structures \citep{Tian2014}. Other methods use a Newton-Raphson scheme, which exhibits fast convergence behaviour but requires the solution of linear systems involving large Jacobian matrices \citep{Degroote2009}. Our method employs the latter approach, but we use a block-LU factorization of the Jacobian matrix to restrict all iterations to subsystems whose dimensions scale with the number of discretisation points on the flag, rather than on the entire flow domain. Thus, our algorithm inherits the fast convergence behaviour of Newton-Raphson methods while substantially reducing the cost of performing an iteration. 

\subsection{Linearised equations and global modes}

For ease of notation, we define the state vector $y = [s, \zeta, \chi, f]^T$ and let $r(y)$ be the right hand side of (\ref{eqn:eq_s})--(\ref{eqn:eq_c}). We write the state as $y = y_b + y_p$, where $y_b=[s_b, \zeta_b, \chi_b, f_b]^T$ is a base state and $y_p=[s_p, \zeta_p, \chi_p, f_p]^T$ is a perturbation. Plugging this expression for $y$ into (\ref{eqn:eq_s})--(\ref{eqn:eq_c}), Taylor expanding about $y_b$, and retaining only first order terms in the perturbation variables gives the linearised equations:
\begin{equation}
B\dot{y}_p = A({y_b})y_p
\label{eqn:lin}
\end{equation}
where 
\begin{gather}
B = \begin{bmatrix} C^TC & & & \\ & M & & \\ & & I & \\ & & & & 0 \end{bmatrix}, \;
A({y_b}) = \begin{bmatrix} J^{ss} & 0 & -J^{\chi s} & -C^TE^T \\ 0 & 0 & -K + J^{\chi \chi} & QW \\ 0 & I & 0 & 0 \\ EC & -I & J^{\chi c} & 0\end{bmatrix}_{y = y_b}
\label{eqn:A}
\end{gather}
and the remaining sub-blocks of the Jacobian matrix $A$ are given in index notation as
\begin{gather}
(J^{ss})_{ik} = -(C^TC)_{ik}^2 - C^T_{ij}\frac{\partial \mathcal{N}_{j} }{\partial s_k}  \\
(J^{\chi s})_{ik} = C^T_{ij} \frac{\partial E^T_{jl}}{\partial \chi_k} (f_b)_l  \\
(J^{\chi \chi})_{ik} = Q_{ij}\frac{\partial W_{jl}}{\partial \chi_k}(f_b)_l \\
(J^{\chi c})_{ik} = \frac{\partial E_{ij}}{\partial \chi_k}C_{jl}(s_b)_l
\end{gather}
Note that we used $B\dot{y}_b = r(y_b)$ in arriving at the linearised equations (\ref{eqn:lin}).

Global modes are eigenvectors $v$ of the generalised eigenvalue problem $Av = \lambda Bv$, where $\lambda$ is the corresponding eigenvalue. We build and store $A$ and $B$ sparsely and solve the generalised eigenvalue problem using an implicitly restarted Arnoldi algorithm (see \cite{Lehoucq1998} for more details). 

In the results below, $1\times10^{-10}$ was used as the tolerance for convergence of the computed eigenvalues and eigenvectors. Global eigenfunctions are unique to a scalar multiple, and were scaled to unit norm, $|| y ||_2 = 1$.

\subsection{Equilibrium computations}

Undeformed and deformed equilibria are steady state solutions to the fully-coupled equations (\ref{eqn:eq_s})-(\ref{eqn:eq_c}) with all time derivate terms set to zero; \emph{i.e.}, these equilibria satisfy $0 = r(y)$, where $y = [s, \zeta, \chi, f]^T$ is the state vector and $r(y)$ is the right hand side of (\ref{eqn:eq_s})-(\ref{eqn:eq_c}). This is a nonlinear algebraic system of equations that we solve using a Newton-Raphson method. With this method, the $k^{th}$ guess for the base state, $y^{(k)}$, is updated as $y^{(k+1)} = y^{(k)} + \Delta y$, where 
\begin{equation}
\Delta y = -( A({y^{(k)}}) )^{-1} r( y^{(k)}) 
\label{eqn:dy}
\end{equation}
Note that the Jacobian matrix $A$ in (\ref{eqn:dy}) is the same matrix as in (\ref{eqn:A}) evaluated at $y=y^{(k)}$.  

The guess for the state $y$ is updated until the residual at the current guess is less than a desired threshold (\emph{i.e.}, until $||r(y^{(k))}||_2 / ||y^{(k)}||_2  < \epsilon$). In the results shown below we used $\epsilon = 1\times10^{-6}$.

\subsection{Domain size and grid resolution}

The flow equations are treated using a multidomain approach: the finest grid surrounds the body and grids of increasing coarseness are used at progressively larger distances \citep{Colonius2008}. In all computations below, the domain size of the finest sub-domain is $[-0.2, 1.8] \times [-1.1, 1.1]$ and the total domain size is $[-15.04, 16.64] \times [-17.44, 17.44]$. The grid spacing on the finest domain is $h = 0.01$ and the grid spacing for the flag is $\Delta s = 0.02$. For computations involving time marching, the time step is $\Delta t = 0.001$, which gives a maximum Courant-Friedrichs-Levy number of $\approx 0.15$. 

To determine the suitability of these parameters, we performed a grid convergence study of the nonlinear solver using $Re = 200, M_\rho = 0.5, K_B = 0.35$. For these parameters the flag enters limit cycle flapping of fixed amplitude and frequency. Using the grid described above, the amplitude and frequency of these oscillations were $a = \pm 0.81, f = 0.180$, respectively.  Refining the grid spacing to $h = 0.0075$ on the finest domain and increasing the domain such that the finest sub-domain size was $[-0.2, 2.8]\times[-1.5,1.5]$ and the total domain size was $[-22.58, 25.18]\times[-23.88,23.88]$ changed these values to $a = \pm 0.80, f = 0.183$, respectively.  

\section{Dynamics for $Re = 200$}
\label{sec:Re_200}

We now consider the inverted-flag system for $Re = 200$. We demonstrate the existence of a deformed equilibrium that is stable over a small range of stiffnesses and becomes unstable as $K_B$ is decreased. The transition to small-deflection deformed flapping associated with this decrease in $K_B$ is shown through a global stability analysis to be a supercritical Hopf bifurcation of the deformed equilibrium. We then consider the large-amplitude flapping regime, and confirm the arguments of \citet{Sader2016a} that this regime is a VIV for small values of $M_\rho$. VIV is shown to cease for larger mass ratios, consistent with the scaling analysis of \citet{Sader2016a}, and we demonstrate that large-amplitude flapping persists despite the absence of VIV. The potential mechanisms associated with this non-VIV large-amplitude flapping regime are discussed. We then use a global stability analysis to confirm the argument of \citet{Shoele2016} that small-amplitude flapping in the deflected-mode regime is driven by the bluff-body vortex-shedding instability. Finally, we show that for a range of $K_B$, light flags with $M_\rho \le O(1)$ exhibit chaotic flapping characterised by switching between large-amplitude flapping and the deflected-mode state. No chaotic flapping is observed for heavy flags, \emph{i.e.}, $M_\rho > O(1)$. 

\subsection{Bifurcation diagrams and general observations}
\label{sec:bif_Re200}

\begin{figure}
    \begin{subfigure}[b]{0.5\textwidth}
        \includegraphics[scale=0.5,trim={0cm 1.6cm 0cm 0cm},clip]{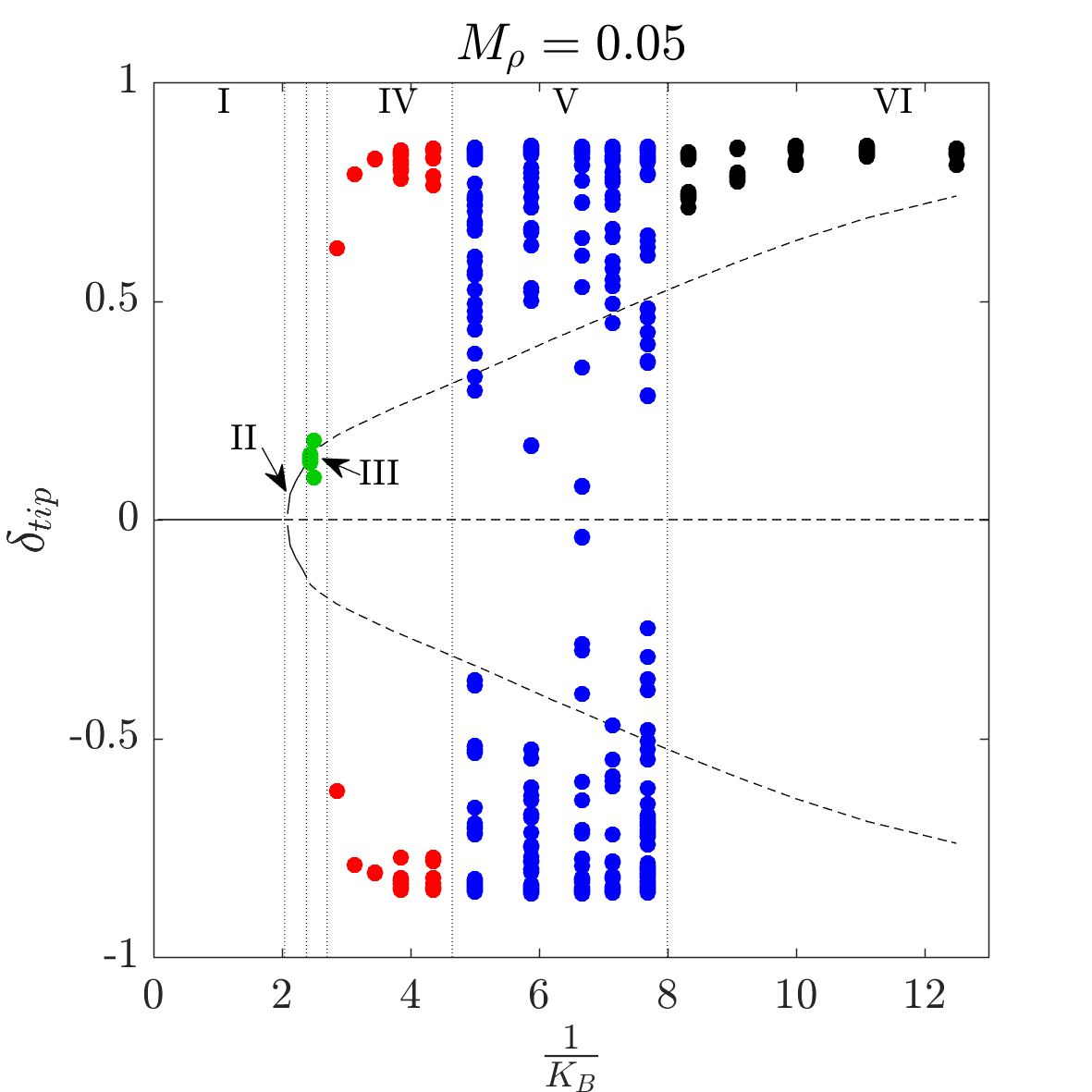}
    \end{subfigure}
    \hspace*{3mm}
    \begin{subfigure}[b]{0.5\textwidth}
        \includegraphics[scale=0.5,trim={1.95cm 1.6cm 0cm 0cm},clip]{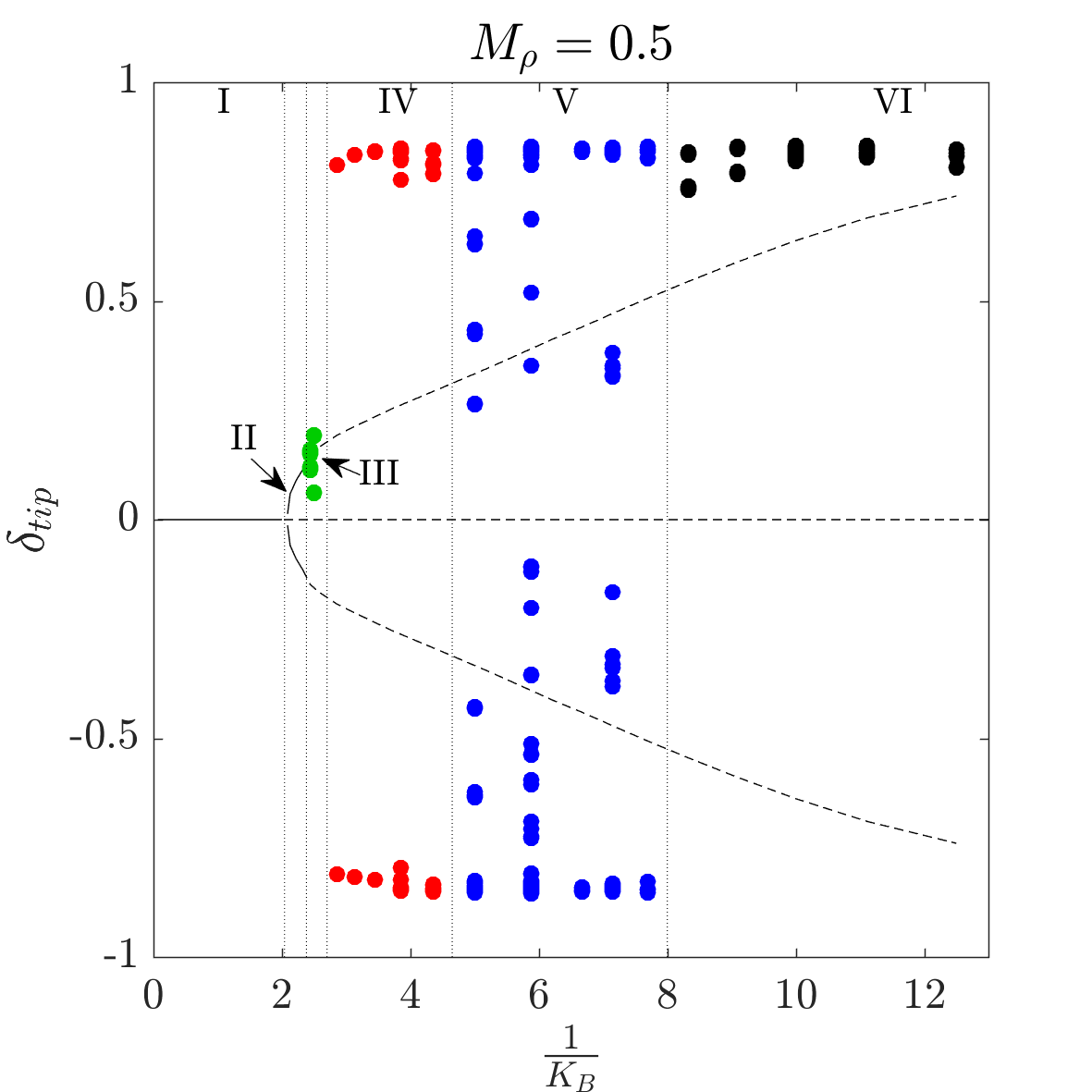}
    \end{subfigure}

	\begin{subfigure}[b]{0.5\textwidth}
	\vspace*{2mm}
        \includegraphics[scale=0.5,trim={0cm 0cm 0cm 0cm},clip]{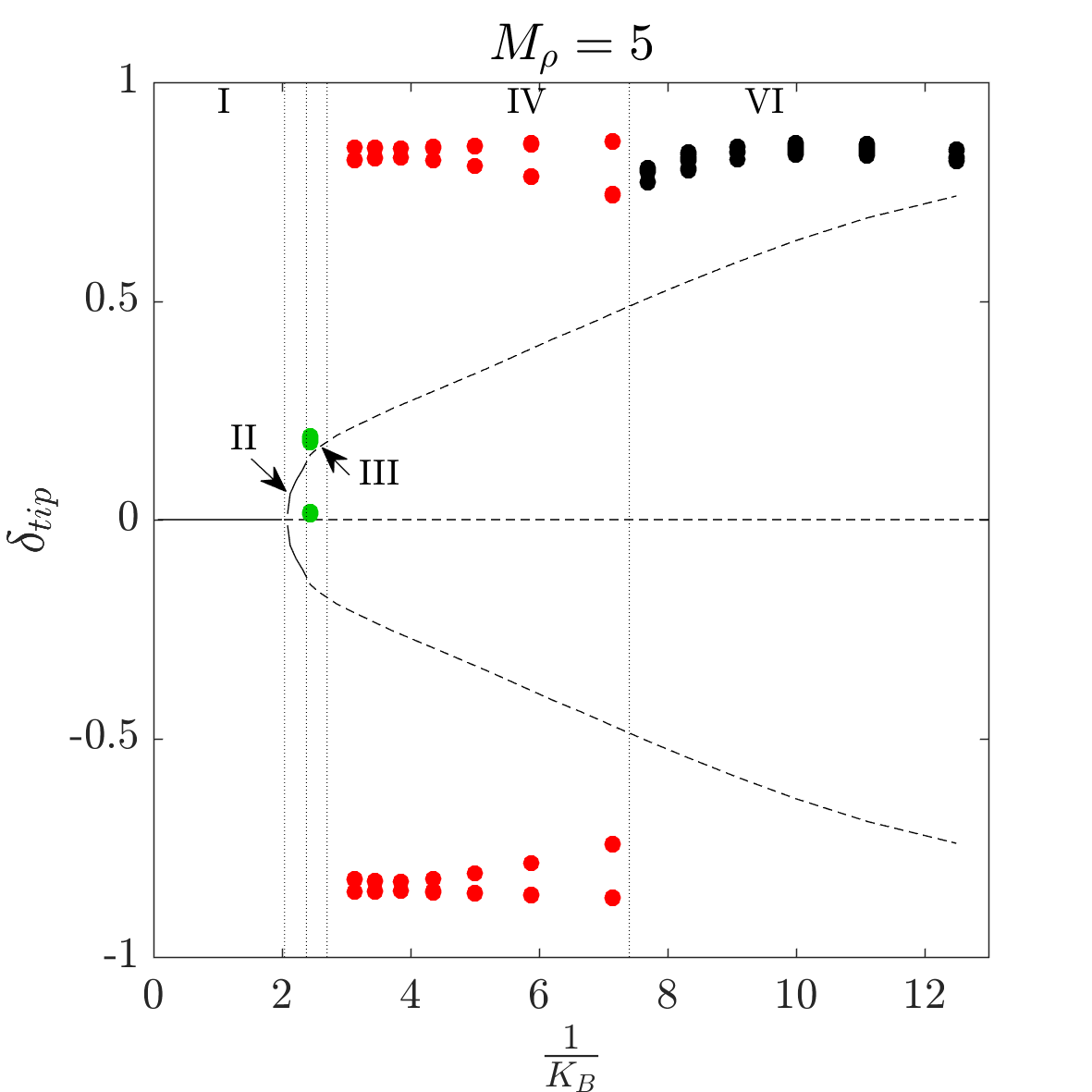}
    \end{subfigure}
        \hspace*{3mm}
	\begin{subfigure}[b]{0.5\textwidth}
       \vspace*{2mm}
	 \includegraphics[scale=0.5,trim={1.95cm 0cm 0cm 0cm},clip]{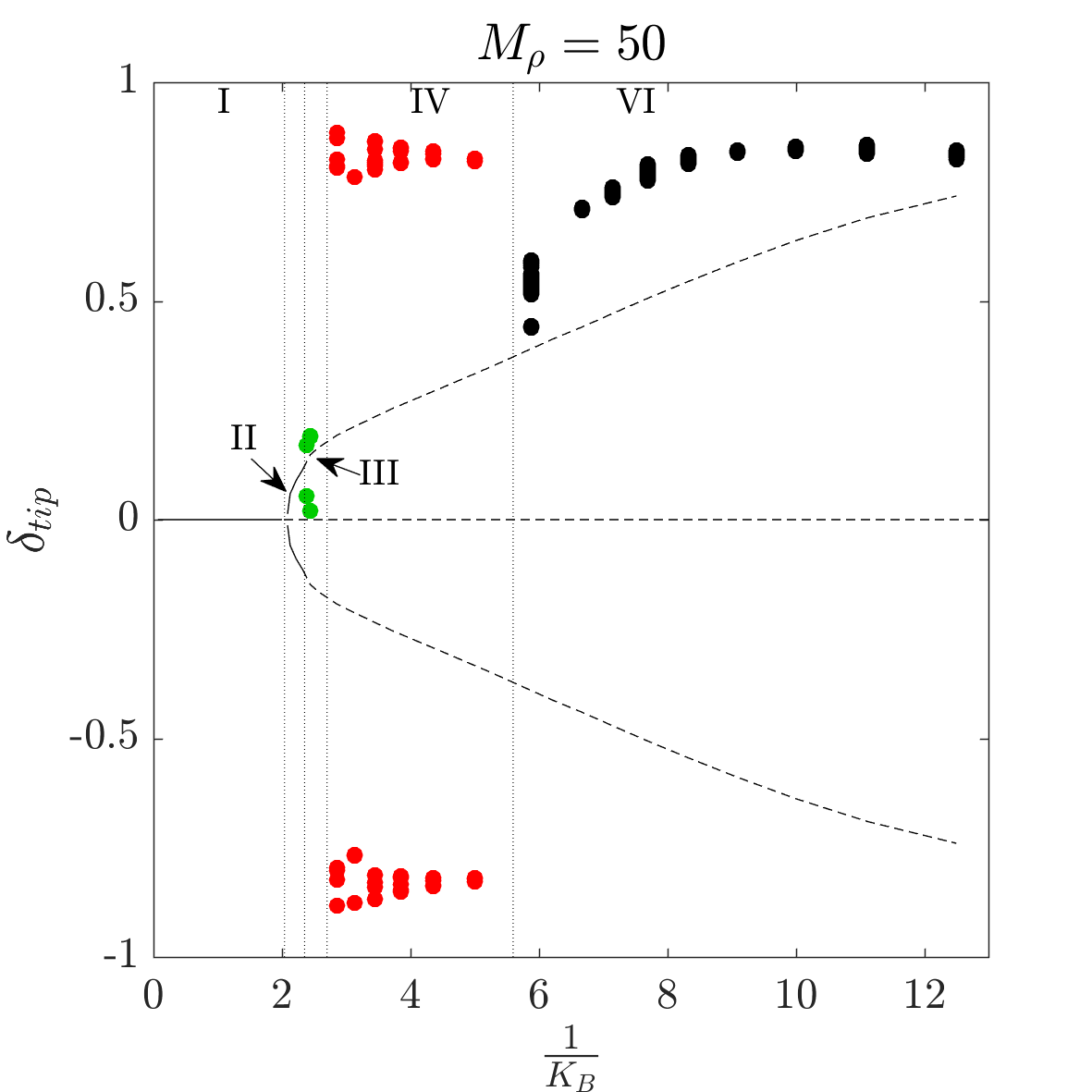}
    \end{subfigure}
       \caption{Bifurcation diagrams of inverted-flag dynamics at $Re = 200$ that show leading edge transverse displacement (tip deflection, $\delta_{tip}$) versus inverse stiffness ($1/K_B$). I: undeformed equilibrium, II: deformed equilibrium, III: small-deflection deformed flapping, IV: large-amplitude flapping, V: chaotic flapping, VI: deflected mode. See the main text for a description of the various lines and markers and details on how the diagrams were constructed.}
    \label{fig:bif_Re200}
\end{figure}

Figure \ref{fig:bif_Re200} shows bifurcation diagrams at four different masses for $Re = 200$. Each plot gives the transverse leading edge displacement (tip deflection, $\delta_{tip}$, nondimensionalised by the flag length $L$) as a function of the reciprocal stiffness ($1/K_B$). Solid lines represent stable equilibria, and dashed lines correspond to unstable equilibria. Information for unsteady regimes is conveyed through the markers. A set of markers at a given stiffness corresponds to tip deflection values from a single nonlinear simulation at moments when the tip velocity is zero (\emph{i.e.}, when the flag changes direction at the tip). From a dynamical systems perspective, the markers correspond to zero tip velocity Poincar\'{e} sections of a tip velocity-tip displacement phase portrait. All nonlinear simulations were started with the flag in its undeflected position and the flow impulsively started to its freestream value. A small body force was introduced at an early time to trigger any instabilities in the system. All simulations contain a minimum of 15 flapping cycles except for the chaotic flapping regime, where a minimum of 55 cycles were used. To avoid representing transient behaviour in the figures, we omit the first several flapping cycles in the diagrams. The bifurcation diagrams were insensitive to starting conditions---the results were unchanged by running a corresponding set of simulations with the flag initialised in its deformed equilibrium state. 

To illustrate the meaning of the markers in figure \ref{fig:bif_Re200} further, consider $1/K_B \approx4$ for $M_\rho = 0.5$. The system enters into large-amplitude limit cycle flapping with a fixed amplitude of $\approx \pm 0.8$, and the bifurcation diagram reflects this with a marker at these peak tip displacements, which are the only tip displacement values where the tip velocity is zero. Note that there are actually several markers superposed onto one another at this stiffness since multiple flapping periods were used to plot these diagrams, though only one marker is visible because of the limit cycle behaviour exhibited. As another example, the bifurcation diagram at $1/K_B \approx 6$ for $M_\rho = 0.05$ depicts chaotic flapping. Many markers are visible at this stiffness because the flag changes direction at several different values of $\delta_{tip}$. The value of using zero tip-velocity Poincar\'{e} sections for the bifurcation diagrams is seen through chaotic flapping: these Poincar\'{e} sections demonstrate the variety of transverse locations where the flag changes direction--- a fact not captured through, for example, plotting the peak-to-peak flapping amplitudes at a given stiffness.

The bifurcation diagrams in figure \ref{fig:bif_Re200} depict the undeformed equilibrium (I), deformed equilibrium (II), small-deflection deformed flapping (III), large-amplitude flapping (IV), deflected mode (VI), and chaotic flapping (V) regimes. In small-deflection deformed flapping, flapping is seen about the upward deflected equilibrium. There is a corresponding deformed equilibrium with a negative flag deflection, and different initial conditions would result in flapping about this equilibrium. We refrain from plotting this behaviour to avoid confusion with large-amplitude flapping.

\begin{figure}
	\begin{subfigure}[b]{0.245\textwidth}
        		\includegraphics[scale=0.26,trim={0cm 0cm 0cm 0cm},clip]{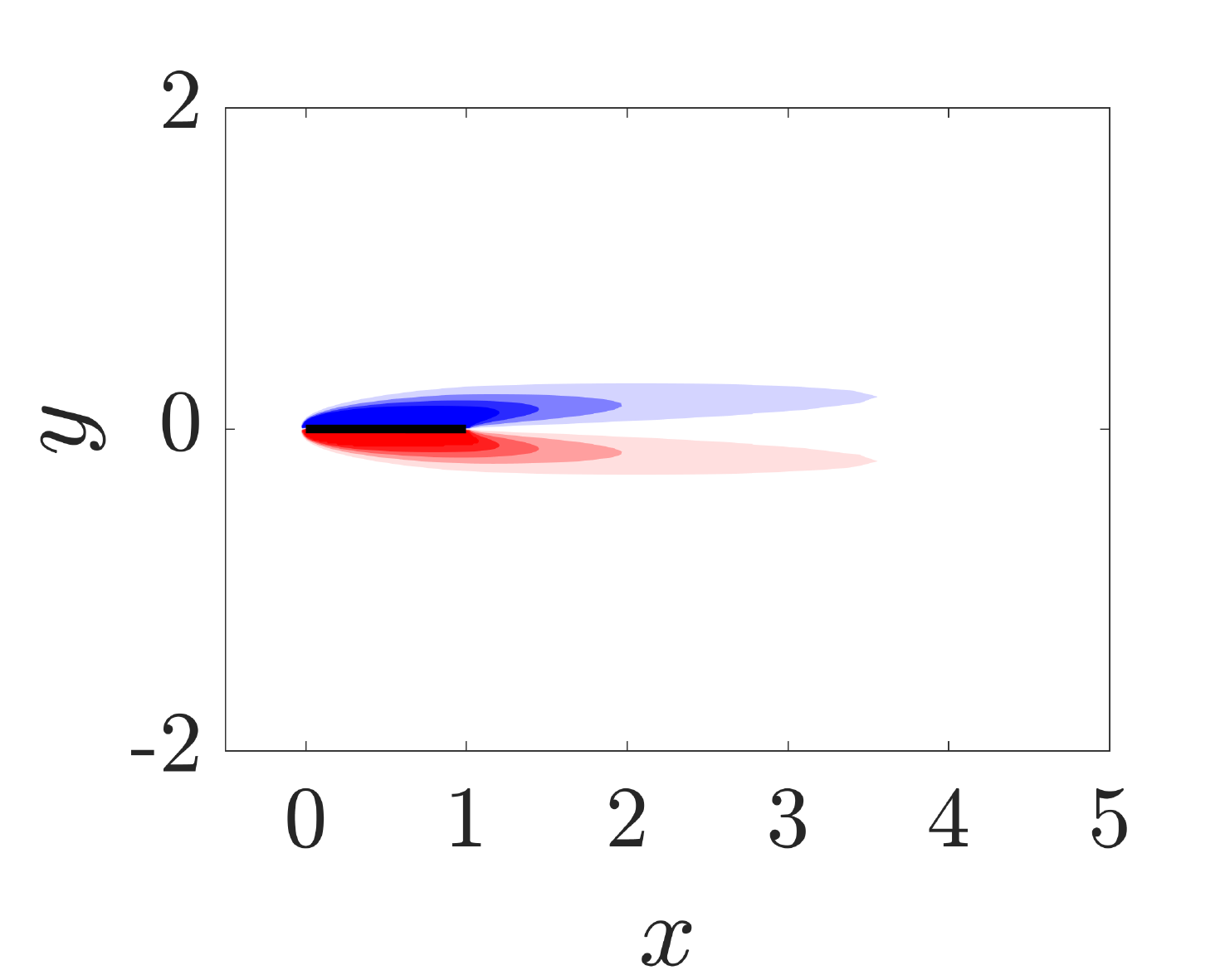}
	\end{subfigure}
	\begin{subfigure}[b]{0.245\textwidth}
		\hspace*{3.9mm}
        		\includegraphics[scale=0.26,trim={2.7cm 0cm 0cm 0cm},clip]{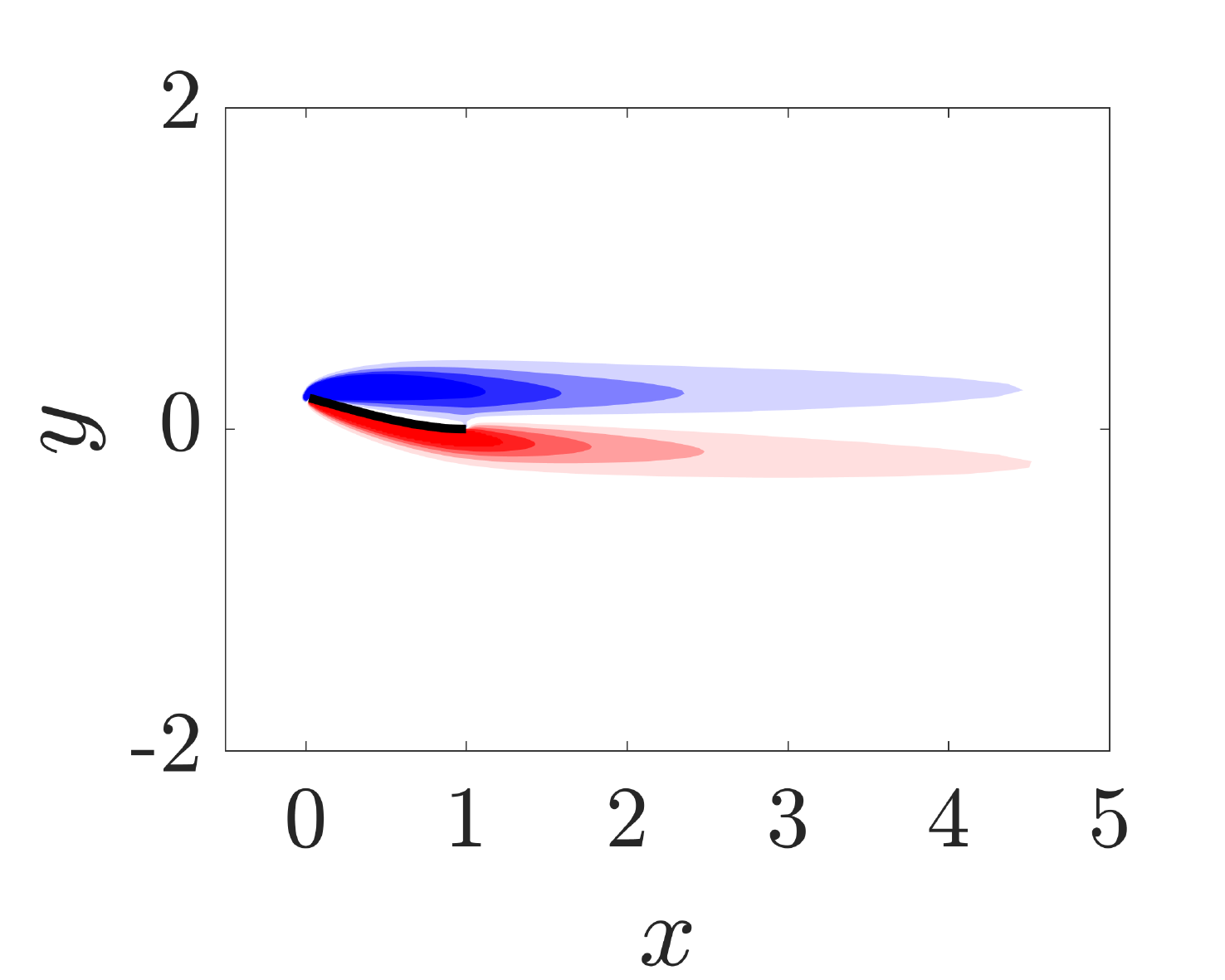}
	\end{subfigure}
    	\begin{subfigure}[b]{0.245\textwidth}
        		\hspace*{1.5mm}
        		\includegraphics[scale=0.26,trim={2.7cm 0cm 0cm 0cm},clip]{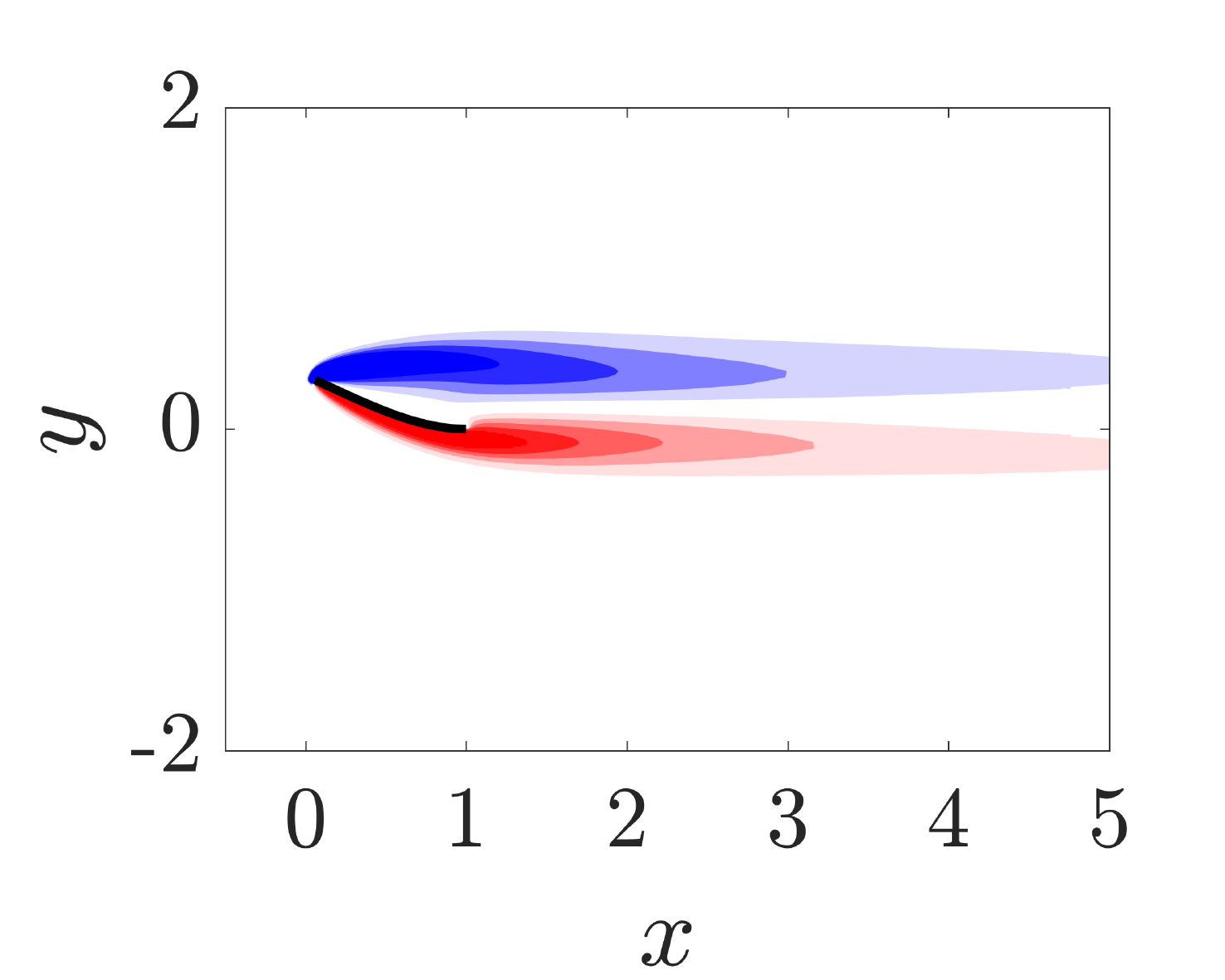}
	\end{subfigure}
	\begin{subfigure}[b]{0.245\textwidth}
        		\includegraphics[scale=0.26,trim={2.7cm 0cm 0cm 0cm},clip]{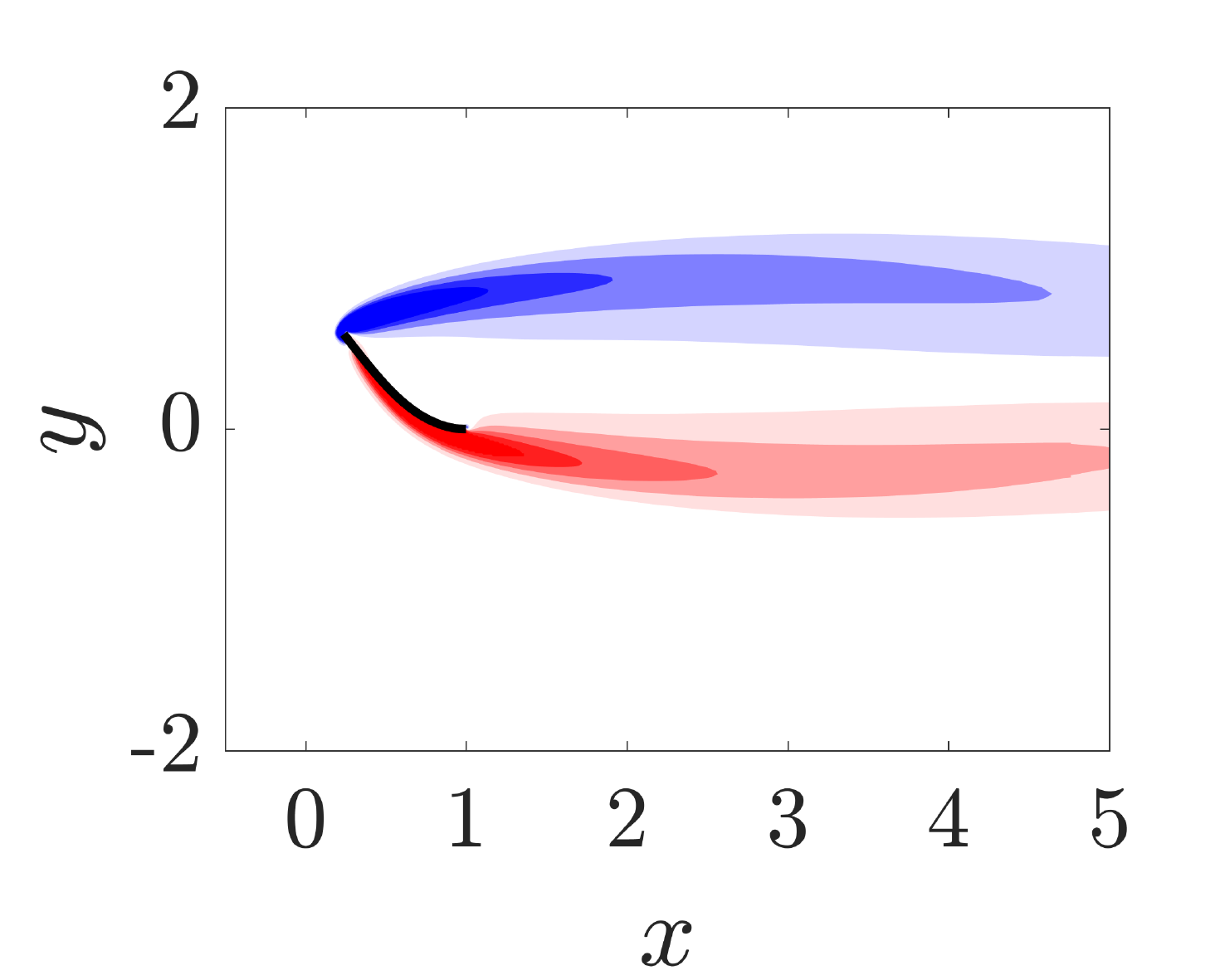}
	\end{subfigure}
    \caption{Vorticity contours for equilibrium states of the flow-inverted-flag system at $Re = 20$. From left to right: $K_B = 0.5, 0.35, 0.22, 0.11$.  The three rightmost equilibria are unstable for all masses considered. Contours are in 18 increments from -5 to 5.}
	\label{fig:DEP_Re200}
\end{figure}

The undeformed equilibrium becomes unstable with decreasing stiffness due to a divergence instability (the critical stiffness for instability is independent of the mass ratio) \citep{Kim2013,Gurugubelli2015,Sader2016a}. We see from figure \ref{fig:bif_Re200} that this instability causes a transition to a regime where the flag is in a steady deflected position. As stiffness is decreased, this steady deflected state is characterised by increasingly large tip deflections (see figure \ref{fig:DEP_Re200}). This regime was first observed by \cite{Ryu2015,Gurugubelli2015}, and we note that it represents a deformed equilibrium state (\emph{i.e.}, in the notation of section \ref{sec:math} it satisfies the steady state equations $r(y) = 0$). Moreover, even for masses where flapping occurs, the deformed equilibrium still exists as an unstable steady-state solution to the fully-coupled equations (\ref{eqn:eq_s})--(\ref{eqn:eq_c}). Note also that for a given stiffness the tip deflection of the deformed equilibrium is constant for all masses, since the equilibrium is a steady state solution of (\ref{eqn:eq_s})--(\ref{eqn:eq_c}) and therefore does not depend on flag inertia. Figure \ref{fig:DEP_Re200} provides illustrations of deformed equilibria for various stiffnesses (some of which are unstable). 

\begin{figure}
    	\begin{subfigure}[b]{0.5\textwidth}
	\vspace*{2mm}
        \includegraphics[scale=0.5,trim={0cm 1.7cm 0cm 0cm},clip]{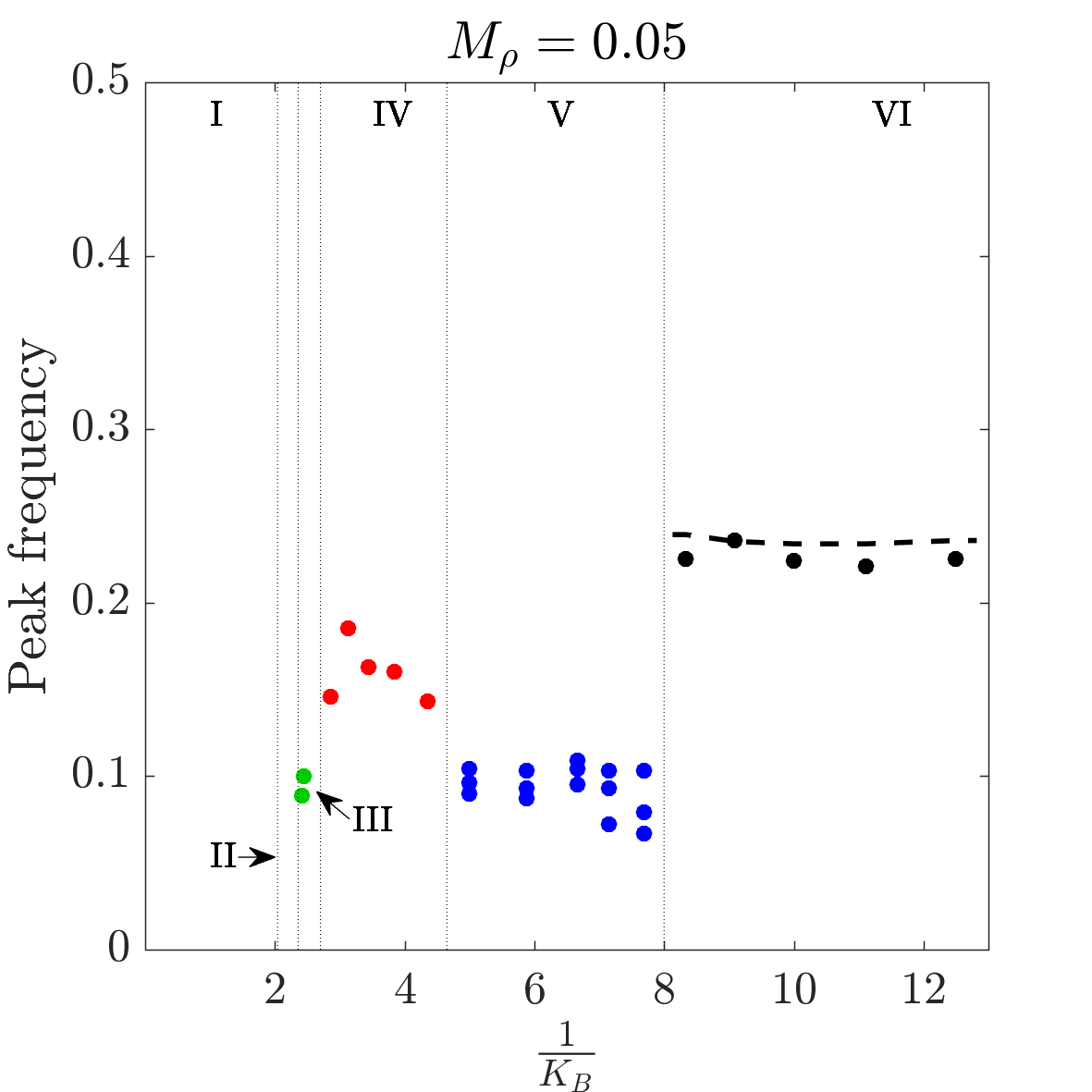}
    \end{subfigure}
        \hspace*{3mm}
	\begin{subfigure}[b]{0.5\textwidth}
       \vspace*{2mm}
	 \includegraphics[scale=0.5,trim={2.cm 1.7cm 0cm 0cm},clip]{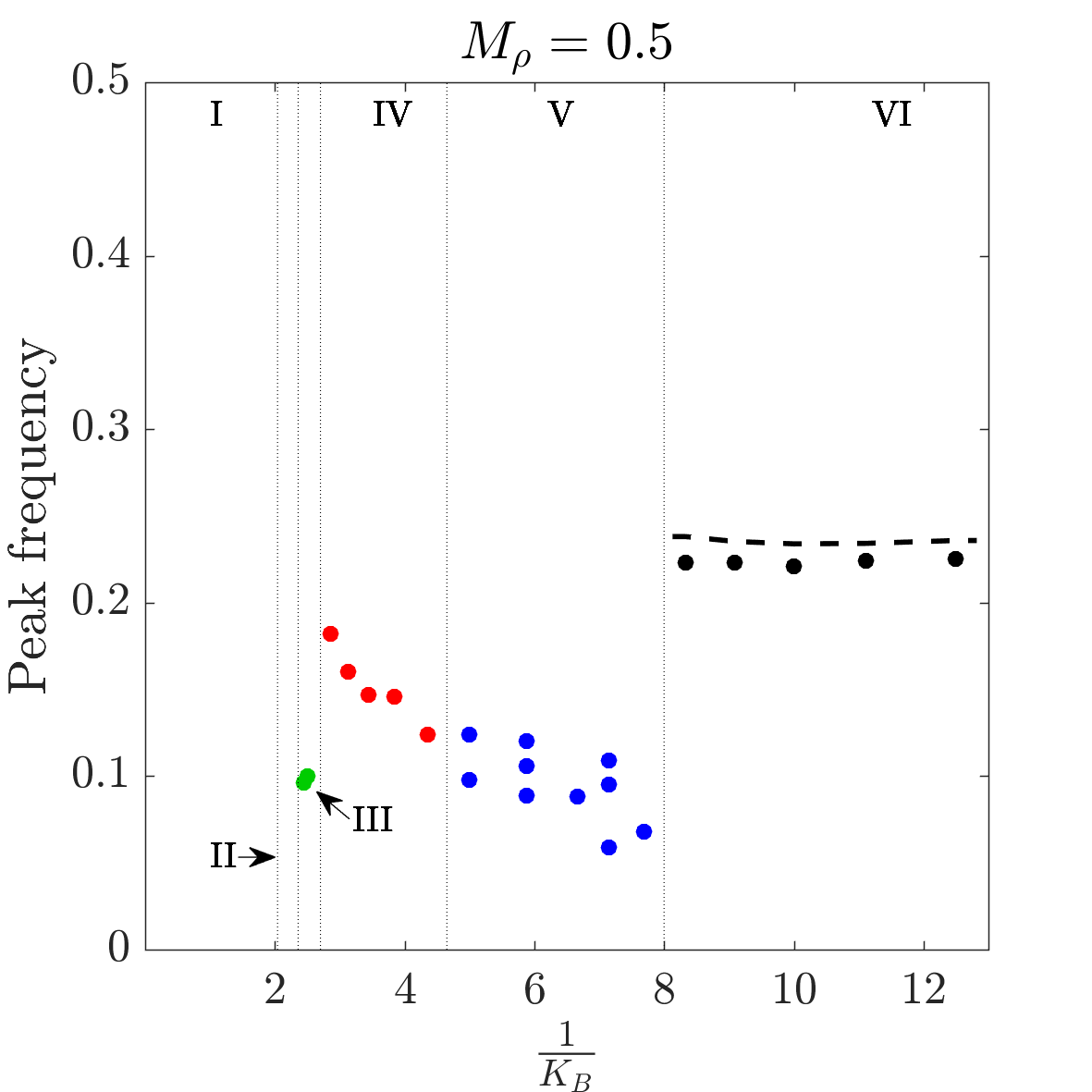}
    \end{subfigure}
       
       \begin{subfigure}[b]{0.5\textwidth}
	\vspace*{2mm}
        \includegraphics[scale=0.5,trim={0cm 0cm 0cm 0cm},clip]{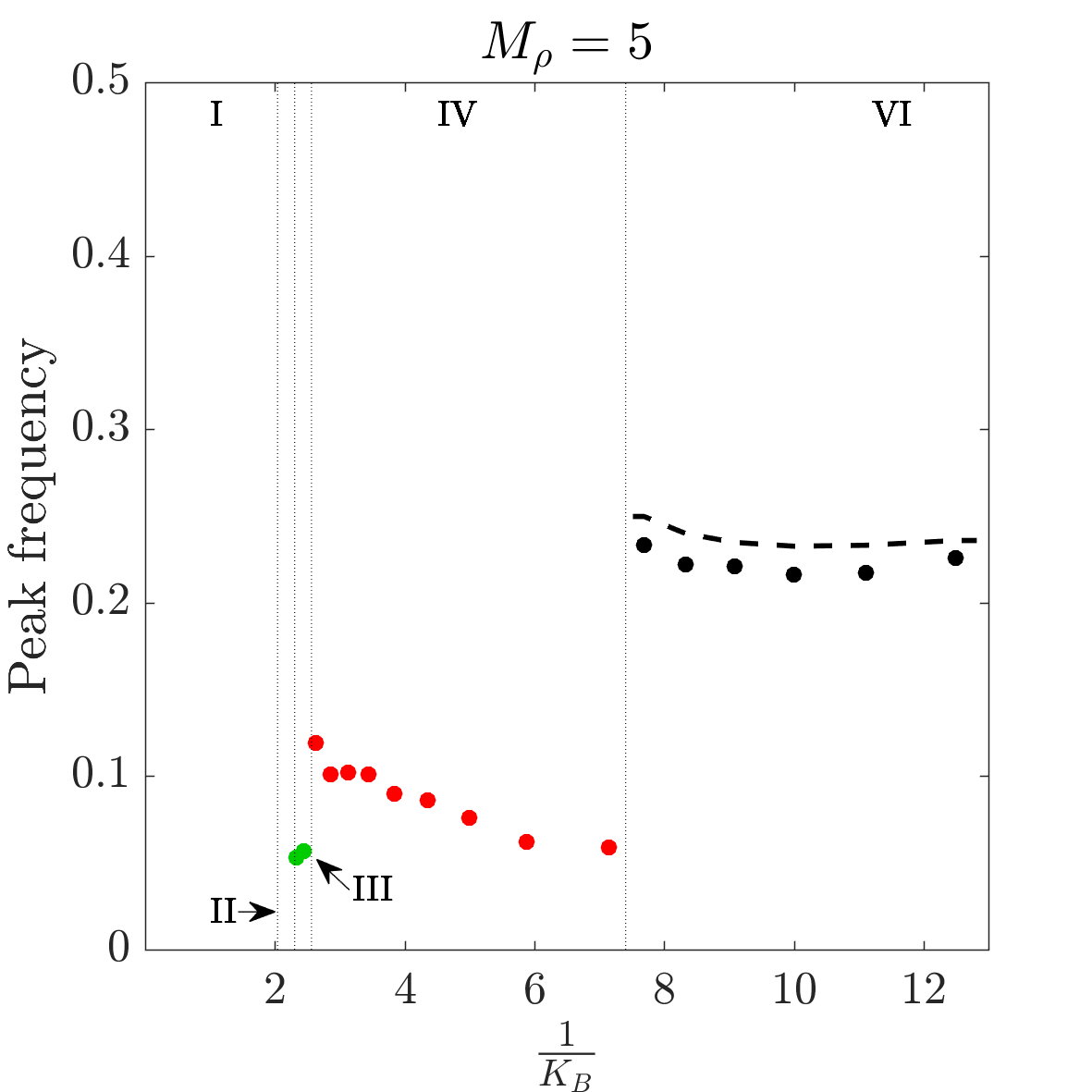}
    \end{subfigure}
        \hspace*{3mm}
	\begin{subfigure}[b]{0.5\textwidth}
       \vspace*{2mm}
	 \includegraphics[scale=0.5,trim={2.cm 0cm 0cm 0cm},clip]{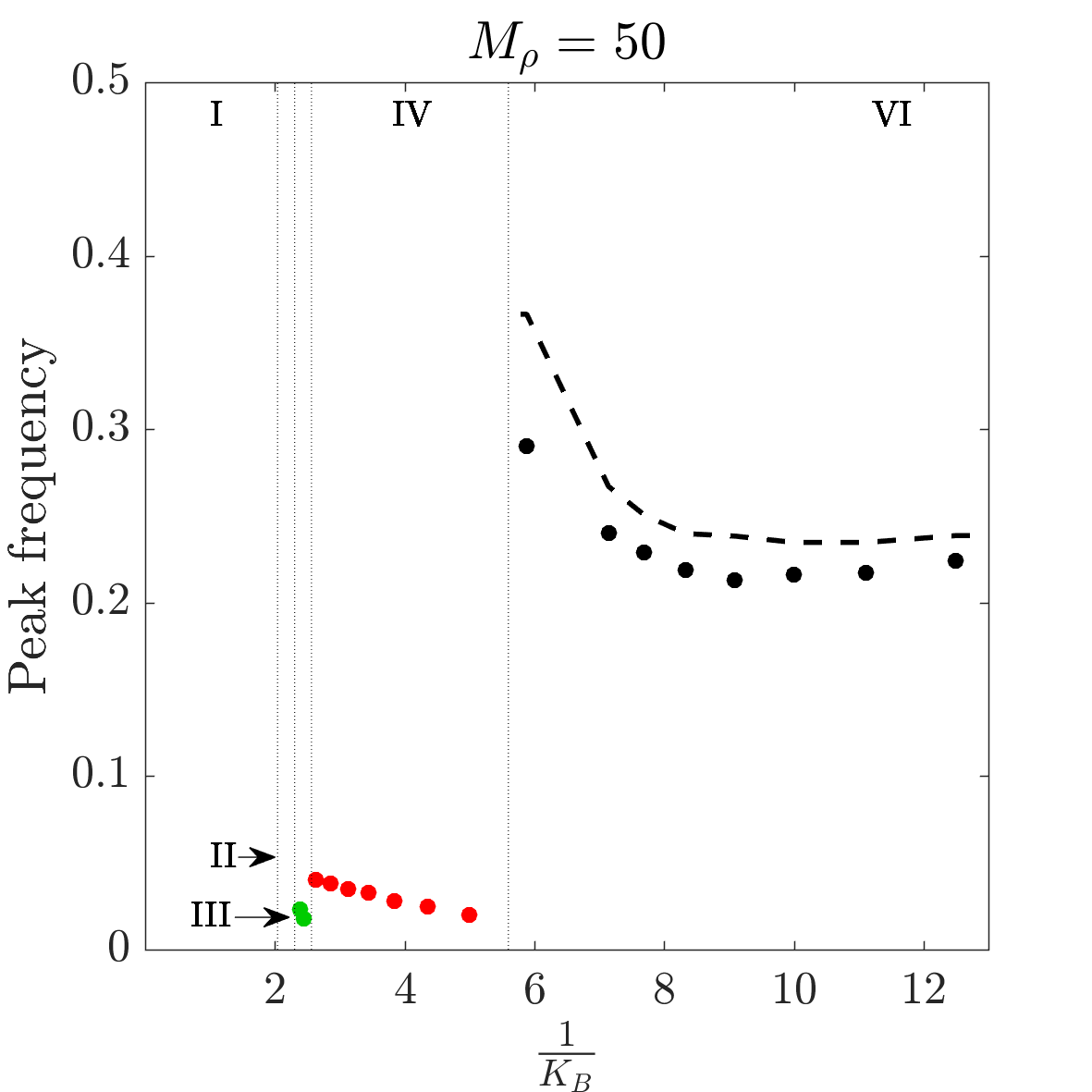}
    \end{subfigure}
       \caption{Markers: peak flapping frequency at $Re = 200$ for the parameters corresponding to the bifurcation diagrams shown in figure \ref{fig:bif_Re200}; (\dashedrule): bluff-body shedding frequency ($=0.2/L_p$, where $L_p$ is the projected length of the flag to the flow defined using the maximum tip deflection at a given stiffness).}
    \label{fig:freq_Re200}
\end{figure}

Figure \ref{fig:freq_Re200} gives the peak flapping frequency for the various regimes where flapping occurs. For all masses considered, small-deflection flapping is associated with a low frequency that is not indicative of VIV behaviour: using the maximal tip displacement as the length scale, the largest Strouhal number over all masses is 0.02. We show in the next section that this regime is caused by the transition to instability of the leading global mode of the deformed equilibrium. Note that the frequency is dependent on flag mass in this regime, illustrating the fully-coupled nature of the problem. 

Large-amplitude flapping is qualitatively different for light flags ($M_\rho = 0.05, 0.5$) and heavy flags ($M_\rho = 5, 50$). \textit{Light flags}: the flapping frequency is roughly constant across an order-of-magnitude change in $M_\rho$, which demonstrates the flow-driven nature of flapping in this regime. \cite{Sader2016a} found that for a range of parameters large-amplitude flapping exhibits several properties of a VIV, and we confirm below that for light flags the fluid forces on the flag synchronise with the flag's motion to form a VIV. \textit{Heavy flags}: the flapping frequency is decreased relative to light flags, and we show in a later section that there is a corresponding de-synchronisation between flapping and vortex shedding. Thus, for heavy flags large-amplitude flapping is not a VIV. This confirms the scaling analysis of \cite{Sader2016a} that VIV behaviour should cease for sufficiently heavy flags. We note from region IV of the bifurcation diagrams in figure \ref{fig:bif_Re200} that for $M_\rho = 5, 50$ large-amplitude flapping persist despite the absence of a VIV. We discuss the mechanism for flapping in this non-VIV regime in section \ref{sec:largeamp_Re200}.

For all masses, the deflected-mode regime (occuring at low stiffness/ high flow rate) has a peak frequency that matches the bluff-body shedding frequency (depicted by the dashed lines). This regime is therefore flow-driven and caused by the canonical bluff-body wake instability irrespective of flag mass \citep{Shoele2016}. We show in section \ref{sec:Deflectedmode_Re200} that the global stability analysis reflects this behaviour.

Finally we note that for light flags ($M_\rho = 0.05, 0.5$), large-amplitude flapping bifurcates (with decreasing stiffness) to chaotic flapping before entering into the deflected-mode regime. The frequency plot for $M_\rho = 0.05, 0.5$ in figure \ref{fig:freq_Re200} illustrates this transition further: in the large-amplitude flapping regime (region IV), decreasing stiffness leads to a corresponding decrease in flapping frequency. Eventually the decrease in frequency becomes significant enough that de-synchronisation between flag flapping and vortex shedding occurs. At this point, chaotic flapping (region V) characterised by multiple frequencies ensues. This regime is discussed in more detail in section \ref{sec:chaos_Re200}.

\subsection{Small-deflection deformed flapping}

Table \ref{tab:Hopf_Re200} demonstrates that the bifurcation to small-deflection deformed flapping is a supercritical Hopf bifurcation of the deformed equilibrium. For all four mass ratios ($M_\rho$), the onset of flapping is associated with the transition to instability of the leading mode of the deformed equilibrium. Table \ref{tab:Hopf_Re200} also shows that the leading mode accurately captures the flapping frequency observed in the nonlinear simulations near this stability boundary where flapping amplitudes remain small.

\begin{table}
\centering
\begin{tabular}{ c c c c c }
\multirow{2}{*}{$M_\rho$} & \multirow{2}{*}{$K_B$} & \multicolumn{2}{ c }{Leading mode} & {Peak frequency of } \\ 
                        &                  & \scriptsize{Growth rate} & \scriptsize{Frequency}                                &    nonlinear simulation              \\ \hline
 0.05 & 0.415 & -0.0021 & 0.110 & N/A (stable equilibrium) \\
 0.05 & 0.41 & 0.0052 & 0.110 & 0.108 \\
 0.5 & 0.42 & -0.0031 & 0.104 & N/A (stable equilibrium) \\
 0.5 & 0.415 & 0.0073 & 0.103 & 0.101 \\
 5 & 0.425 & -0.0014 & 0.073 & N/A (stable equilibrium) \\
 5 & 0.42 & 0.0039 & 0.072 & 0.071 \\ 
 50 & 0.435 & -0.0022 & 0.028 & N/A (stable equilibrium) \\
 50 & 0.43 & 0.0045 & 0.027 & 0.028
\end{tabular}
 \caption{Growth rate and frequency of the leading global mode of the deformed equilibrium compared with nonlinear behaviour for parameters near the onset of small-deflection deformed flapping.}
\label{tab:Hopf_Re200}
\end{table}

\begin{figure}
	\centering
	\begin{subfigure}[b]{0.45\textwidth}
		\centering
        		\includegraphics[scale=0.4, trim={0cm 1.4cm 0cm 0cm},clip]{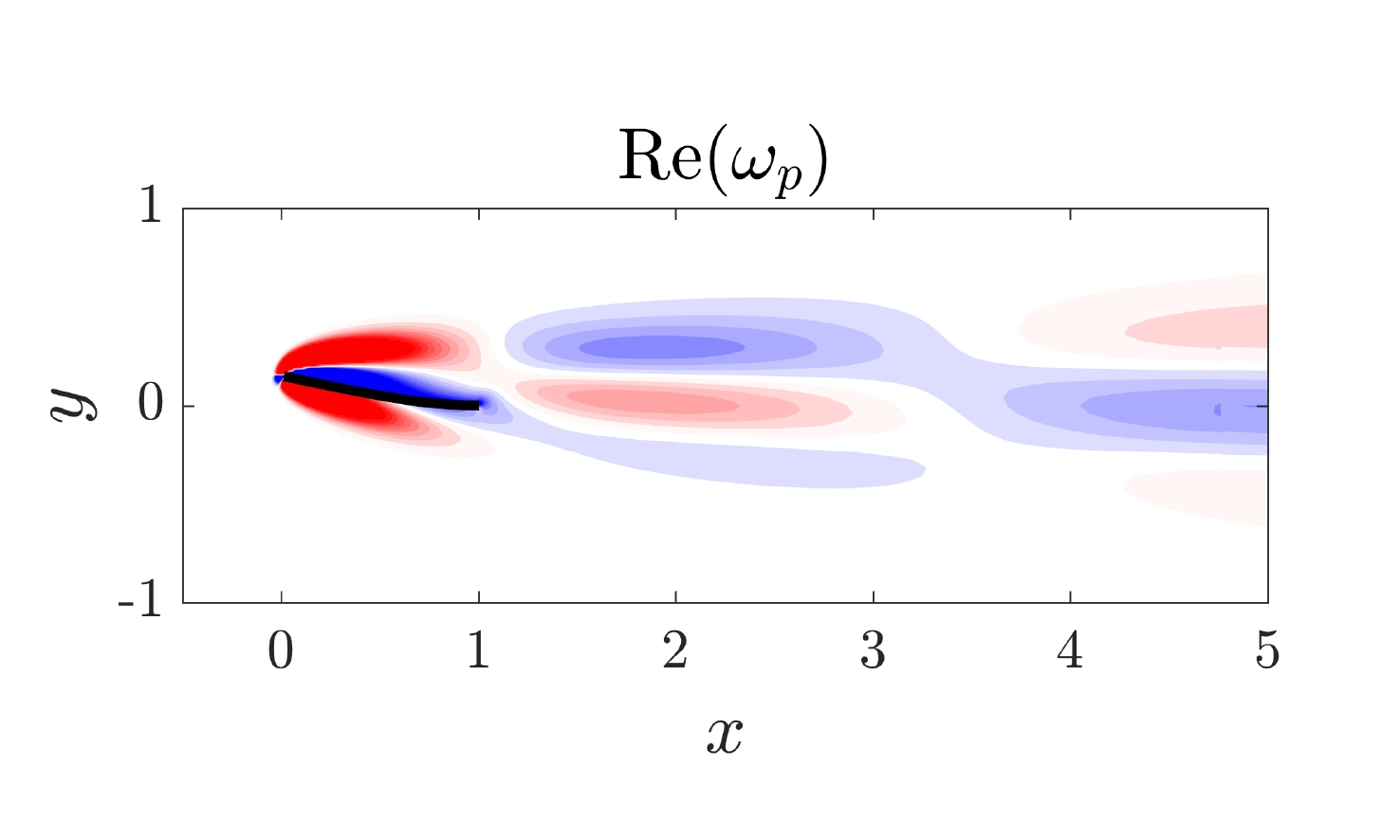}
	\end{subfigure}
	\begin{subfigure}[b]{0.45\textwidth}
		\centering
		\hspace*{3.9mm} 
        		\includegraphics[scale=0.32, trim={1cm 1.1cm 0cm 0cm},clip]{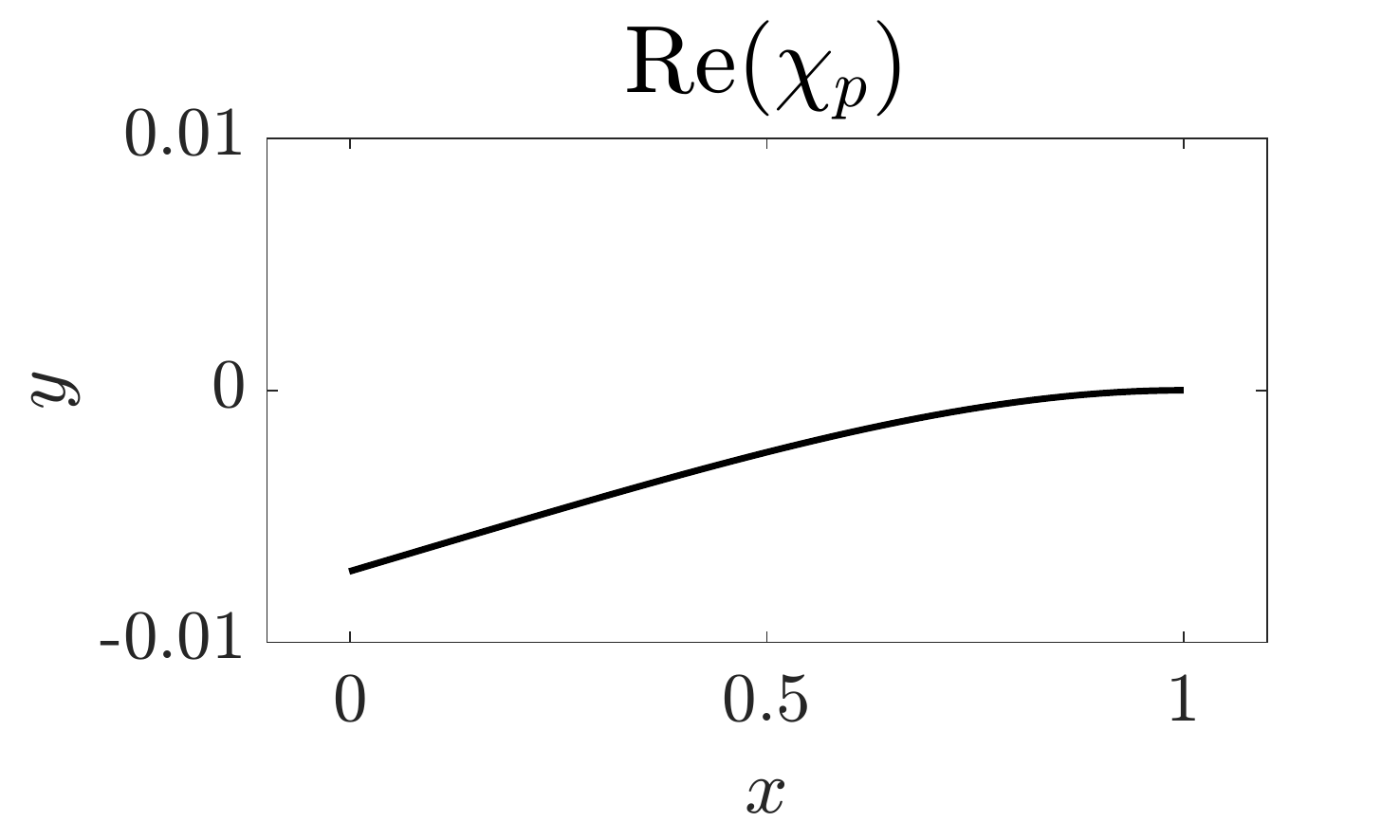}
		\vspace*{1.2mm}
	\end{subfigure}
	
    	\centering
	\begin{subfigure}[b]{0.45\textwidth}
		\centering
        		\includegraphics[scale=0.4, trim={0cm 0cm 0cm 0cm},clip]{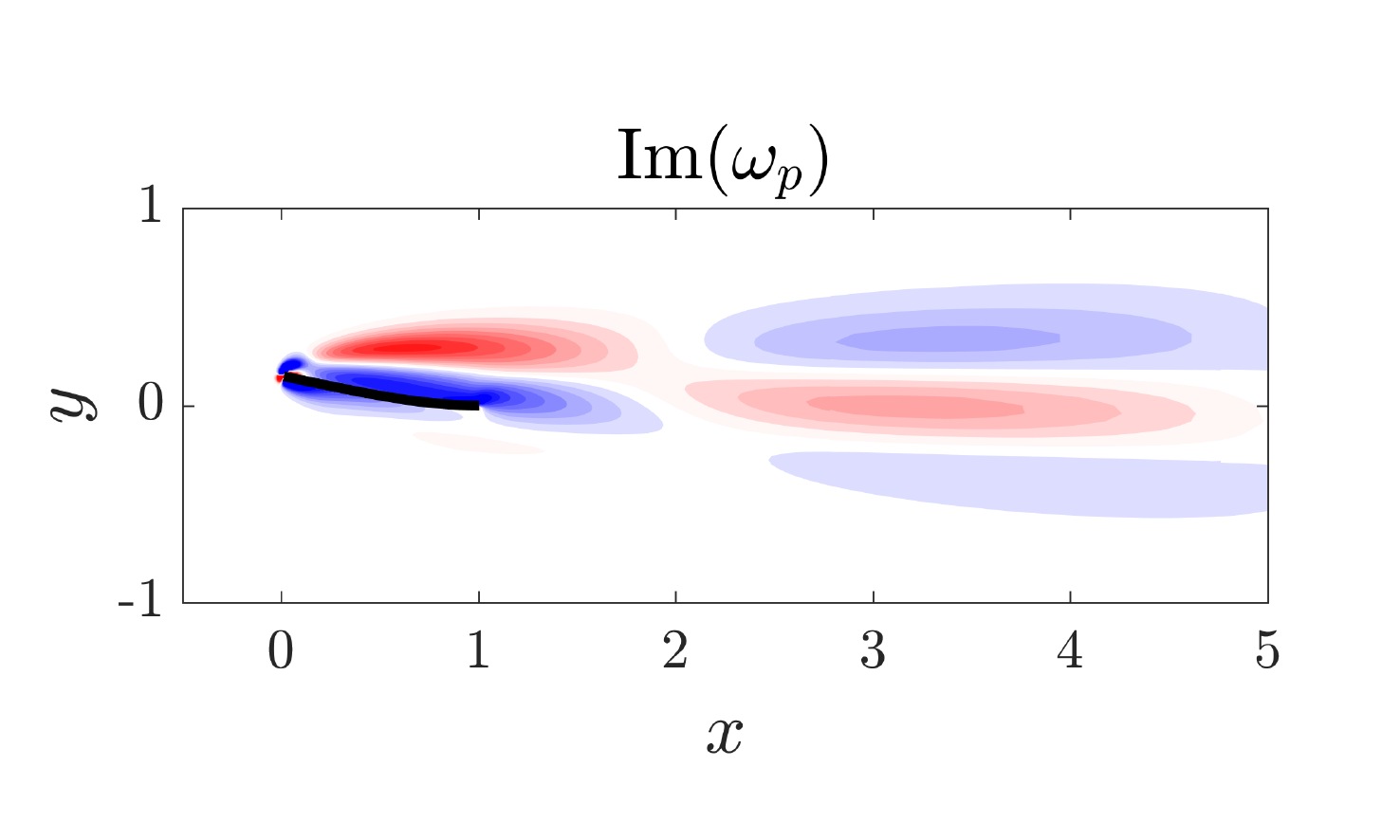}
	\end{subfigure}
	\begin{subfigure}[b]{0.45\textwidth}
		\centering
		\hspace*{3.9mm}
        		\includegraphics[scale=0.32, trim={1cm 0cm 0cm 0cm},clip]{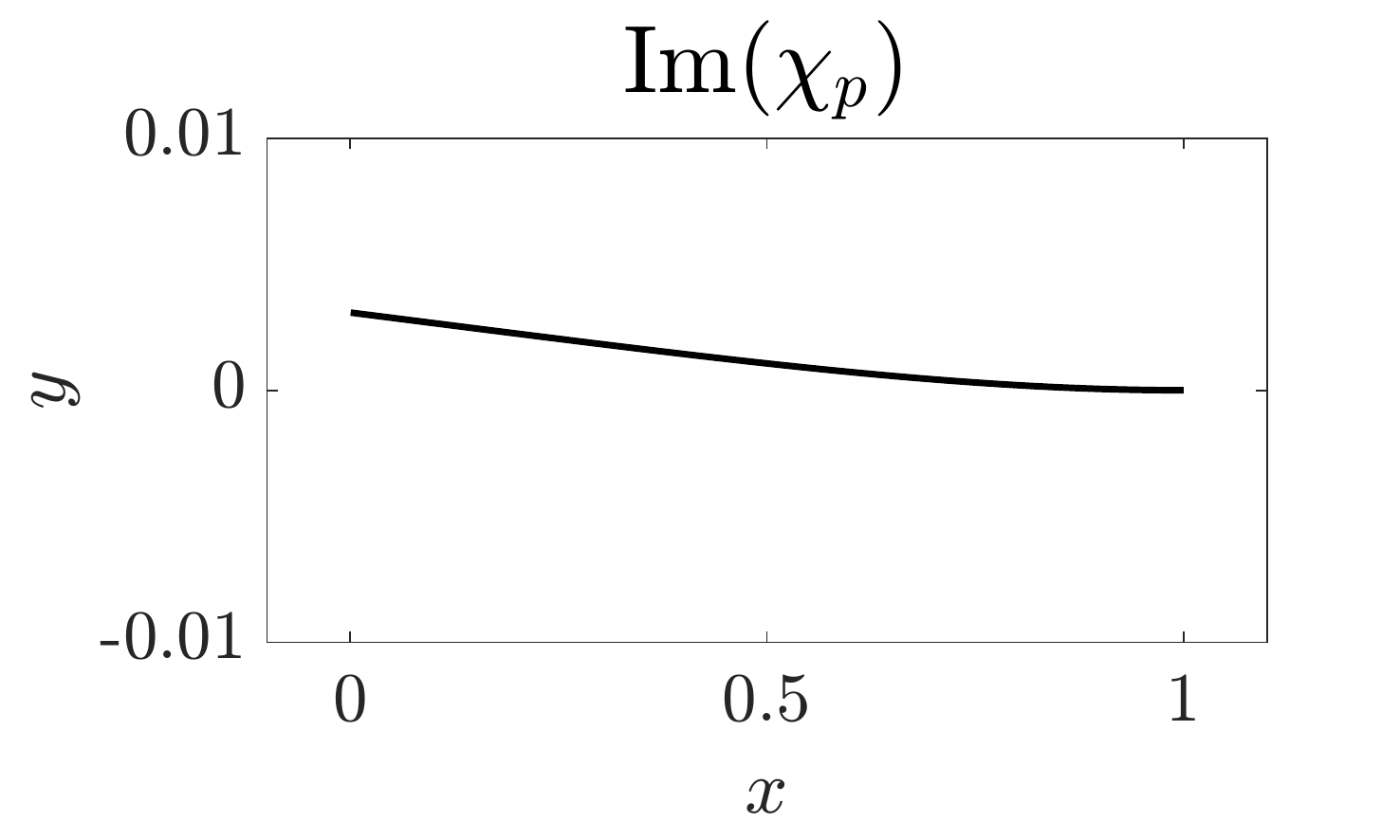}
		\vspace*{3.3mm}
	\end{subfigure}
   	\caption{Real (top) and imaginary (bottom) parts of vorticity (left) and flag displacement (right) of the leading global mode of the deformed equilibrium for $M_\rho = 0.5, K_B = 0.41$ and $Re = 200$ (corresponding to small-deflection deformed flapping). Vorticity contours are in 20 increments from -0.05 to 0.05.}
\label{fig:mode_Re200_def}
\end{figure}

\begin{figure}
	\centering
	\begin{subfigure}[b]{0.45\textwidth}
		\centering
        		\includegraphics[scale=0.4, trim={0cm 1.4cm 0cm 0cm},clip]{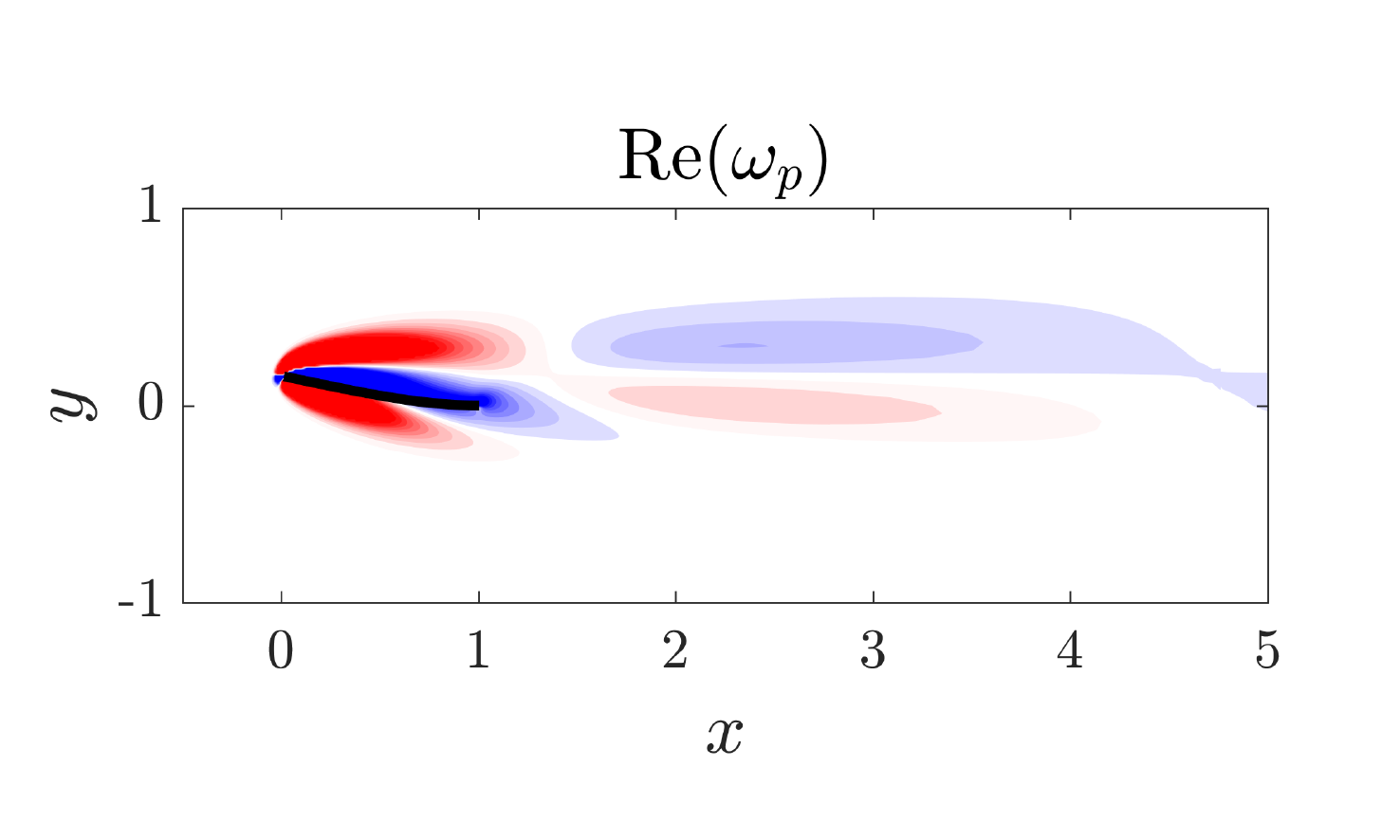}
	\end{subfigure}
	\begin{subfigure}[b]{0.45\textwidth}
		\centering
		\hspace*{3.9mm}
        		\includegraphics[scale=0.32, trim={1cm 1.1cm 0cm 0cm},clip]{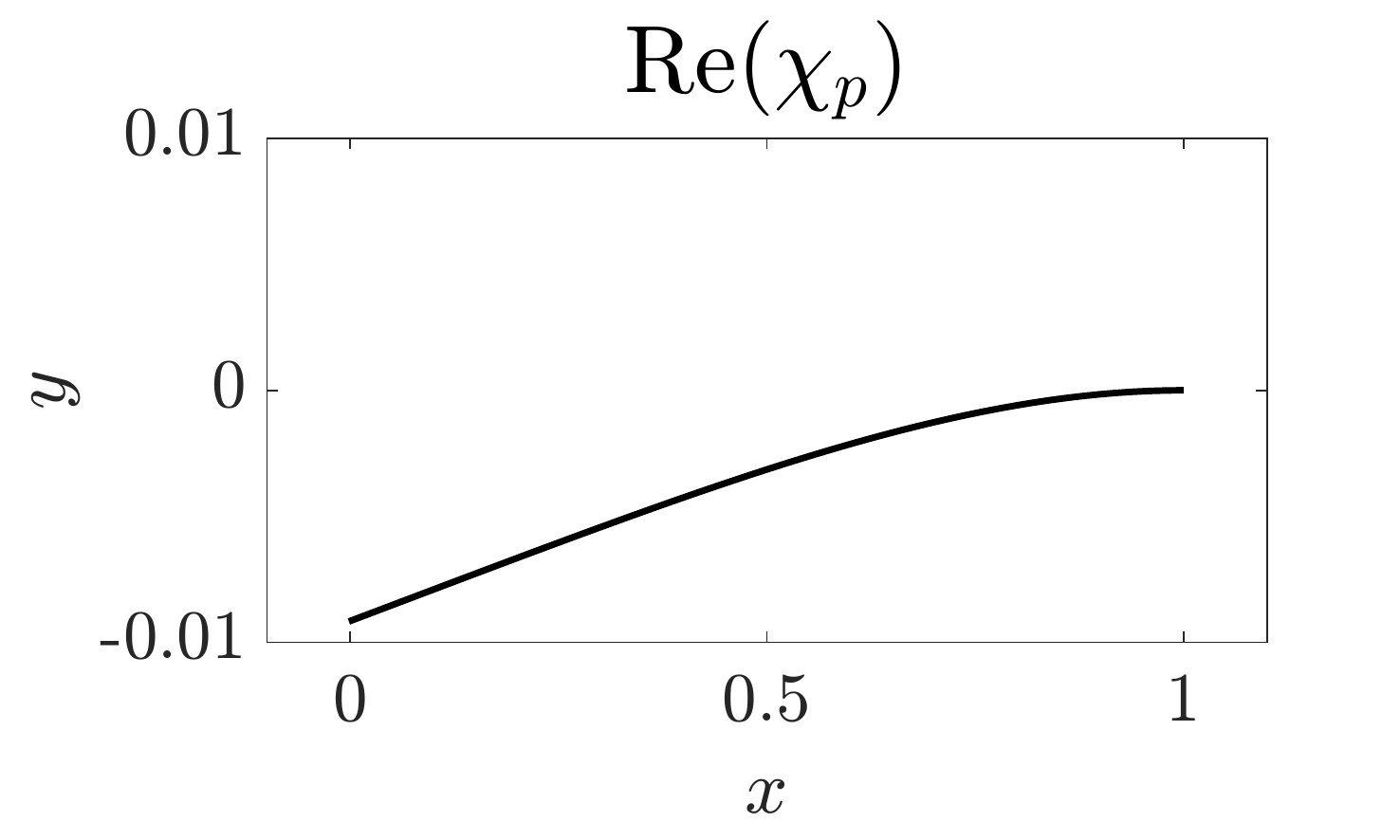}
		\vspace*{1.2mm}
	\end{subfigure}
	
    	\centering
	\begin{subfigure}[b]{0.45\textwidth}
		\centering
        		\includegraphics[scale=0.4, trim={0cm 0cm 0cm 0cm},clip]{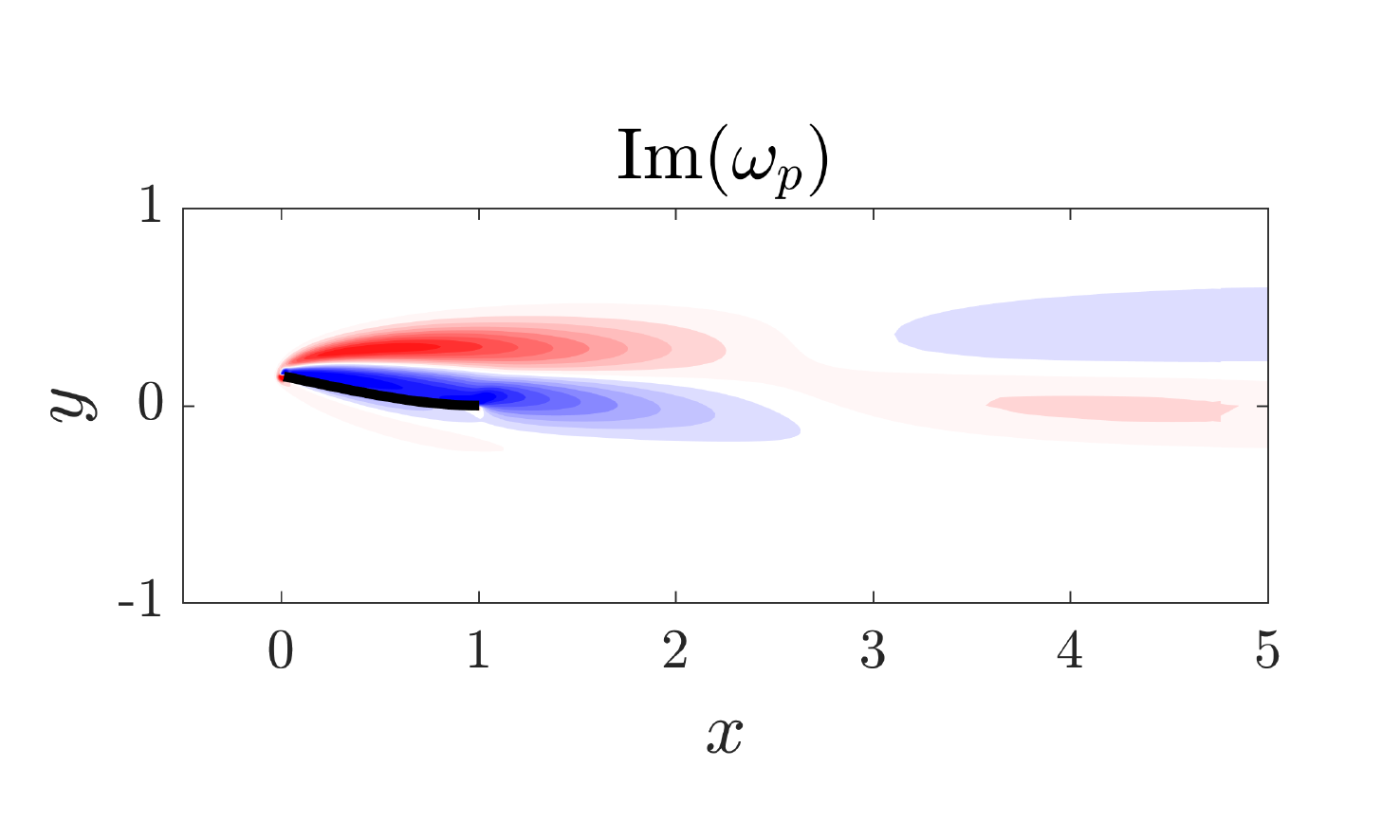}
	\end{subfigure}
	\begin{subfigure}[b]{0.45\textwidth}
		\centering
		\hspace*{3.9mm}
        		\includegraphics[scale=0.32, trim={1cm 0cm 0cm 0cm},clip]{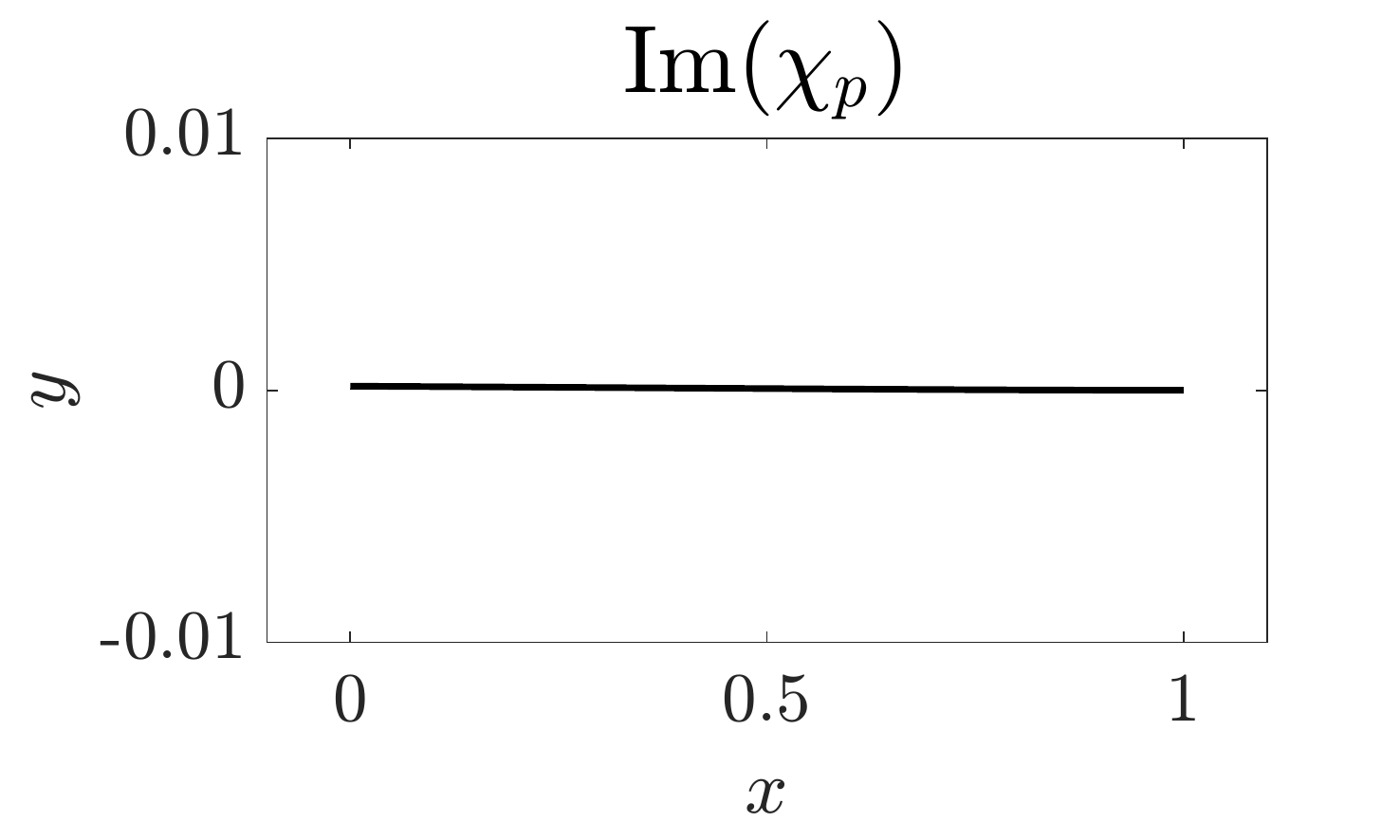}
		\vspace*{3.3mm}
	\end{subfigure}
\caption{Real (top) and imaginary (bottom) parts of vorticity (left) and flag displacement (right) of the leading global mode of the deformed equilibrium for $M_\rho = 5, K_B = 0.41$ and $Re = 200$ (corresponding to small-deflection deformed flapping). Vorticity contours are in 20 increments from -0.2 to 0.2.}
\label{fig:mode_Re200_def_M5}
\end{figure}

Figures \ref{fig:mode_Re200_def} and \ref{fig:mode_Re200_def_M5} show the leading mode of the deformed equilibrium near the critical stiffness values where bifurcation occurs for $M = 0.5$ and $M =5$, respectively. Flapping is associated with vortical structures isolated near the flag surface. Note that the vortical structures are longer for $M_\rho = 5$ (figure \ref{fig:mode_Re200_def_M5}) than for $M_\rho = 0.5$ (figure \ref{fig:mode_Re200_def}), which is commensurate with the lower flapping frequency seen for the more massive case. 

\begin{figure}
	\begin{subfigure}[b]{0.245\textwidth}
        		\includegraphics[scale=0.26,trim={0cm 0cm 0cm 0cm},clip]{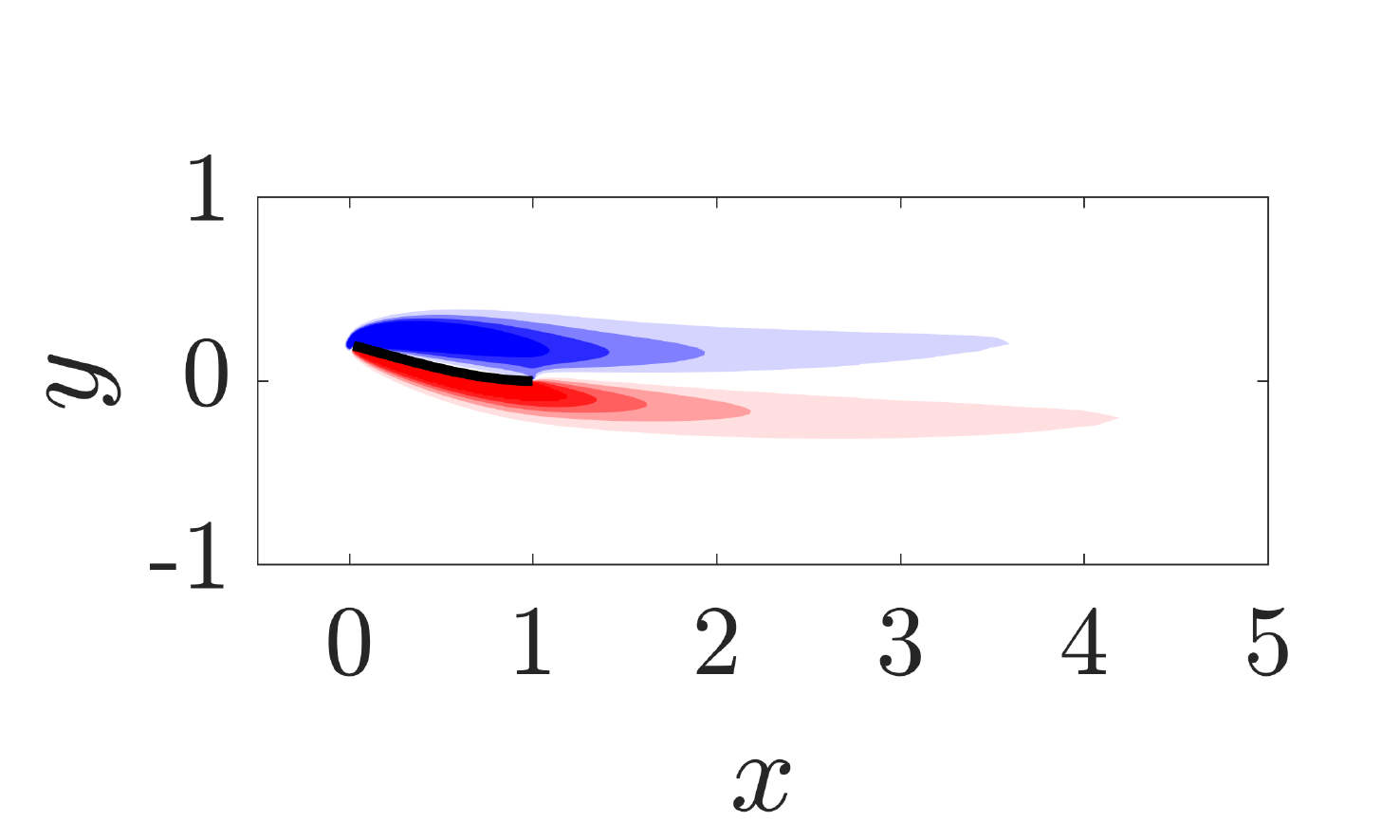}
	\end{subfigure}
	\begin{subfigure}[b]{0.245\textwidth}
		\hspace*{3.9mm}
        		\includegraphics[scale=0.26,trim={2.7cm 0cm 0cm 0cm},clip]{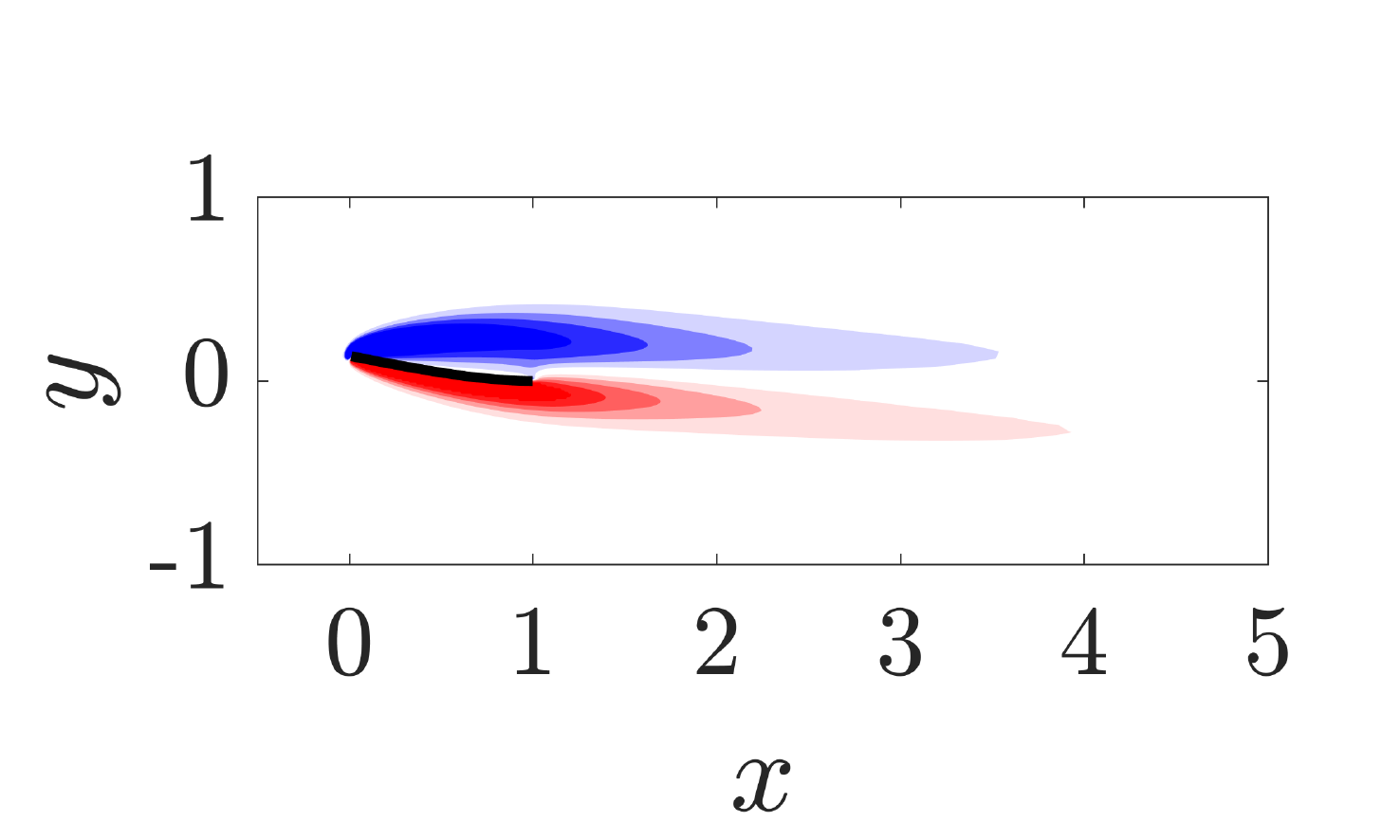}
	\end{subfigure}
    	\begin{subfigure}[b]{0.245\textwidth}
        		\hspace*{1.5mm}
        		\includegraphics[scale=0.26,trim={2.7cm 0cm 0cm 0cm},clip]{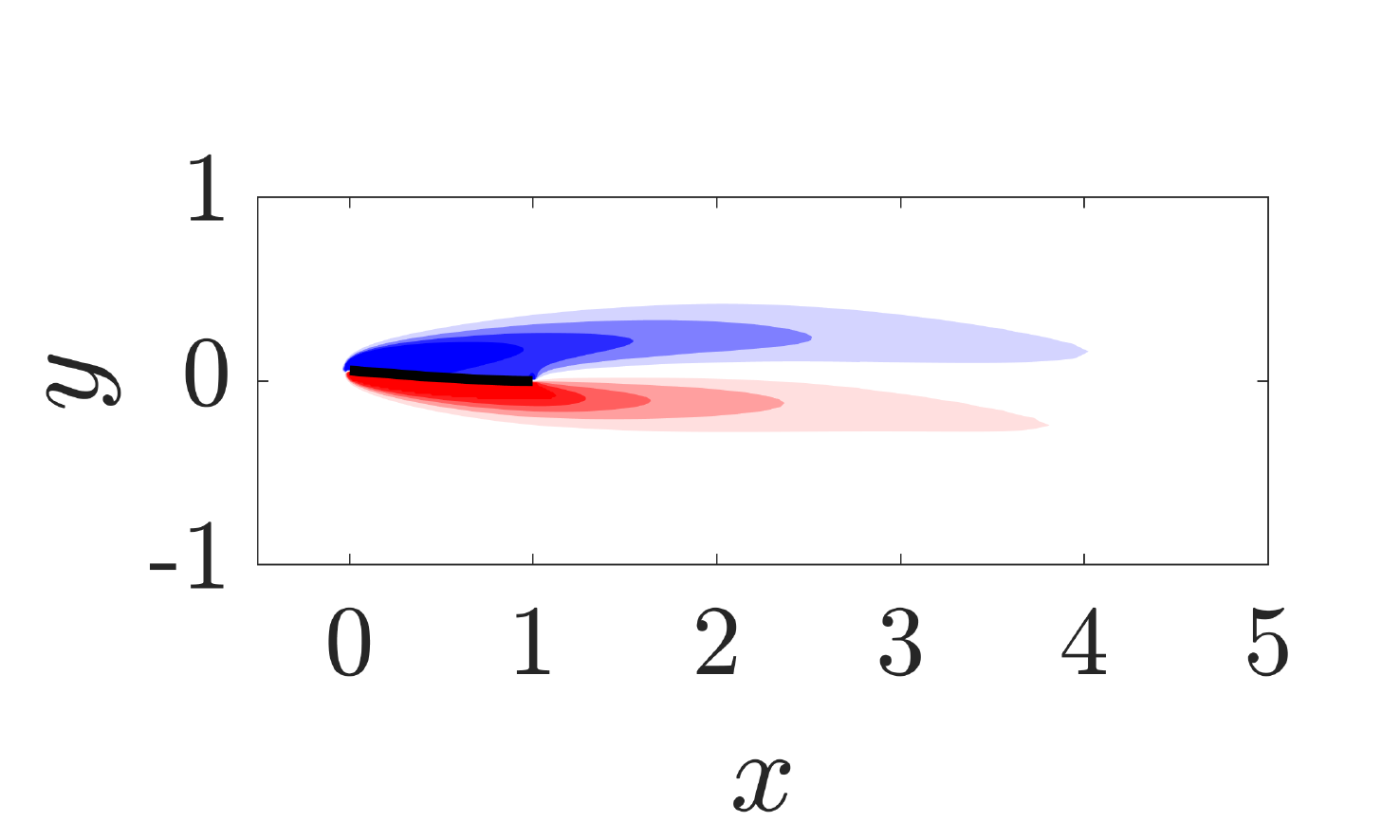}
	\end{subfigure}
	\begin{subfigure}[b]{0.245\textwidth}
        		\includegraphics[scale=0.26,trim={2.7cm 0cm 0cm 0cm},clip]{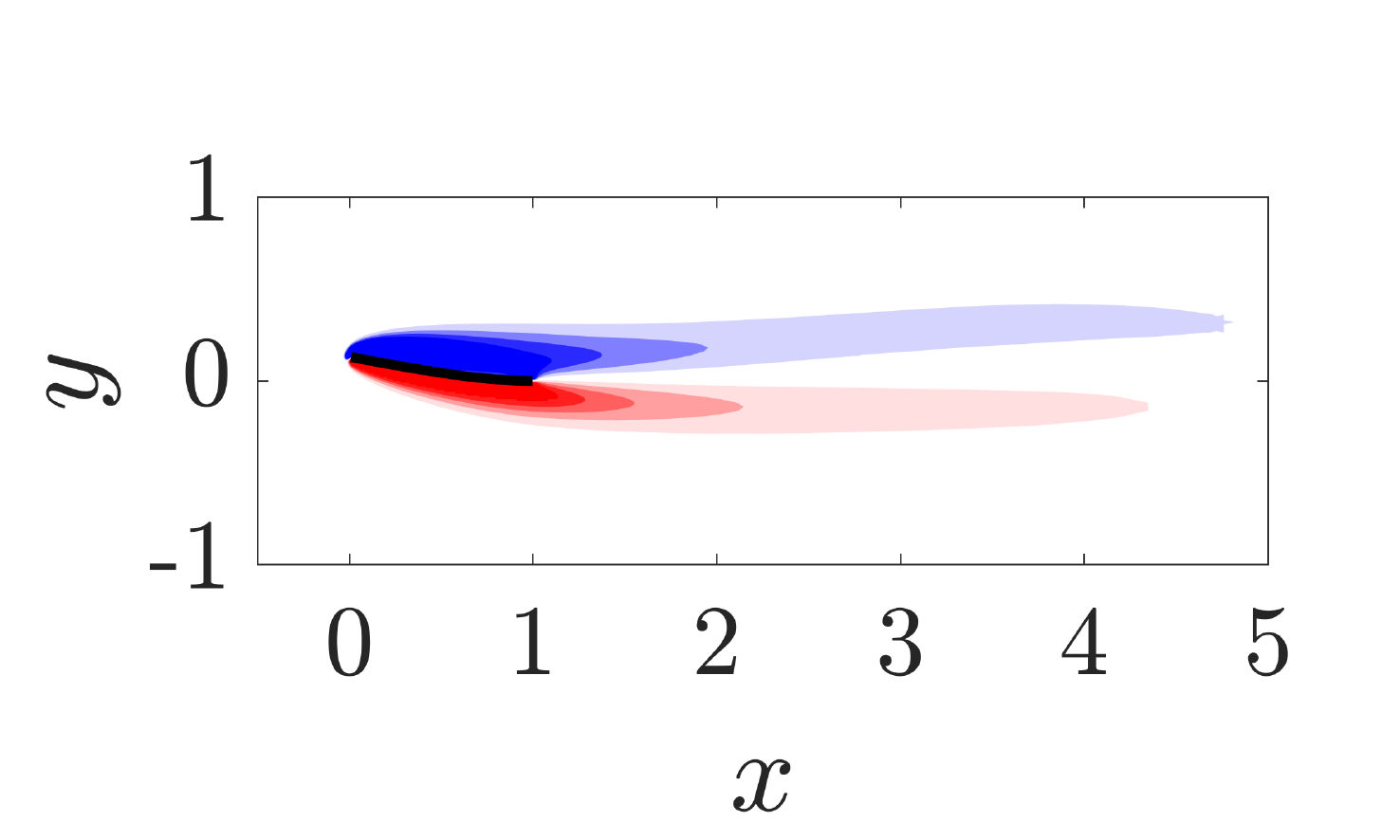}
	\end{subfigure}
    \caption{Vorticity contours at four snapshots of a flapping period of a flag in small-deflection deformed flapping. The figures were obtained from a nonlinear simulation with $Re = 200, M_\rho = 0.5, K_B = 0.41$. Contours are in 18 increments from -5 to 5.}
	\label{fig:smalldeflect_Re200}
\end{figure}

To demonstrate how the leading mode manifests itself in the nonlinear simulations, we show in figure \ref{fig:smalldeflect_Re200} snapshots during a flapping period of a flag with $M_\rho = 0.5, K_B = 0.41$ (\emph{i.e.}, in the small-deflection flapping regime). Note in particular the absence of vortex shedding---the entire flapping period is associated with long vortical structures that extend from the flag into the wake. This is distinct from the large-amplitude flapping behaviour discussed in the next section, and emphasises that even at the moderate Reynolds number of $Re = 200$, inverted flags have an intrinsic flapping mechanism devoid of vortex shedding.

\subsection{Large-amplitude flapping}
\label{sec:largeamp_Re200}
	
Decreasing stiffness in the small-deflection deformed flapping regime is associated with an increasingly unstable leading mode (see table \ref{tab:grow_Re200}) and a corresponding increase in flapping amplitude. Eventually, the amplitude is sufficiently large for the flag to reach past the centreline ($\delta_{tip} = 0$) position, and large-amplitude flapping ensues. 

\begin{table}
\centering
\begin{tabular}{ c c }
{$K_B$} & {Leading mode growth rate}  \\ \hline
 0.41 & 0.022 \\
 0.38 & 0.123  \\ 
 0.35 & 0.250 \\
 0.33 & 0.313
\end{tabular}
 \caption{Growth rate of the leading global mode of the deformed equilibrium for $M_\rho = 0.5$ for stiffnesses in the small-deflection deformed flapping and large-amplitude flapping regimes.}
\label{tab:grow_Re200}
\end{table}

At this Reynolds number, the large-amplitude behaviour is associated with sufficient bluffness to the flow that vortex shedding occurs. As discussed in section \ref{sec:bif_Re200}, the resulting dynamics are dependent on flag mass: light flags undergo the VIV behaviour identified by \cite{Sader2016a} and heavy flags do not, and we therefore consider them separately in what follows.

\subsubsection{Large-amplitude flapping of light flags}

\begin{figure}
	\begin{subfigure}[b]{0.245\textwidth}
        		\includegraphics[scale=0.345,trim={0cm 2.1cm 0cm 0cm},clip]{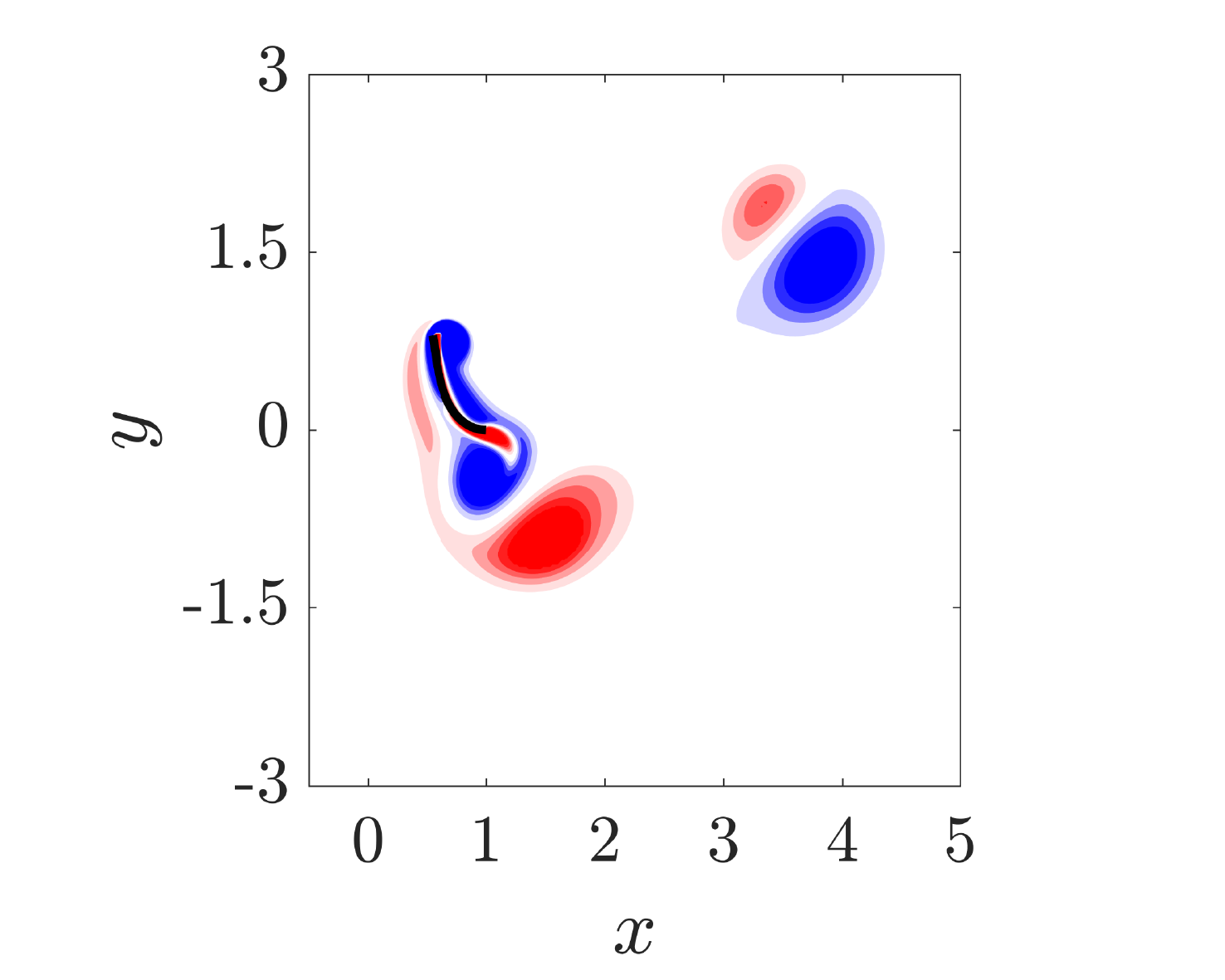}
	\end{subfigure}
	\begin{subfigure}[b]{0.245\textwidth}
		\hspace*{7.5mm}
        		\includegraphics[scale=0.345,trim={3.6cm 2.1cm 0cm 0cm},clip]{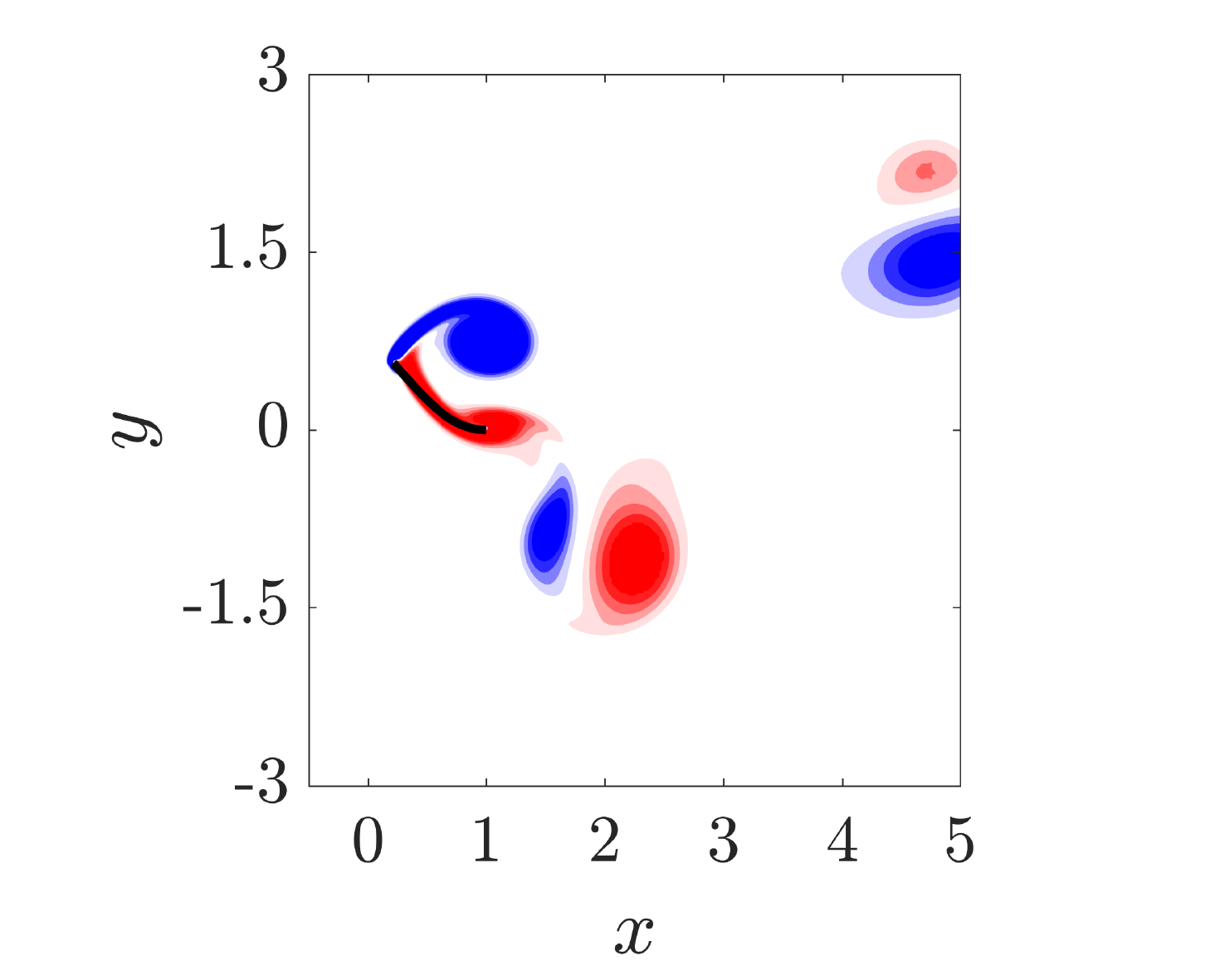}
	\end{subfigure}
    	\begin{subfigure}[b]{0.245\textwidth}
        		\hspace*{3.5mm}
        		\includegraphics[scale=0.345,trim={3.6cm 2.1cm 0cm 0cm},clip]{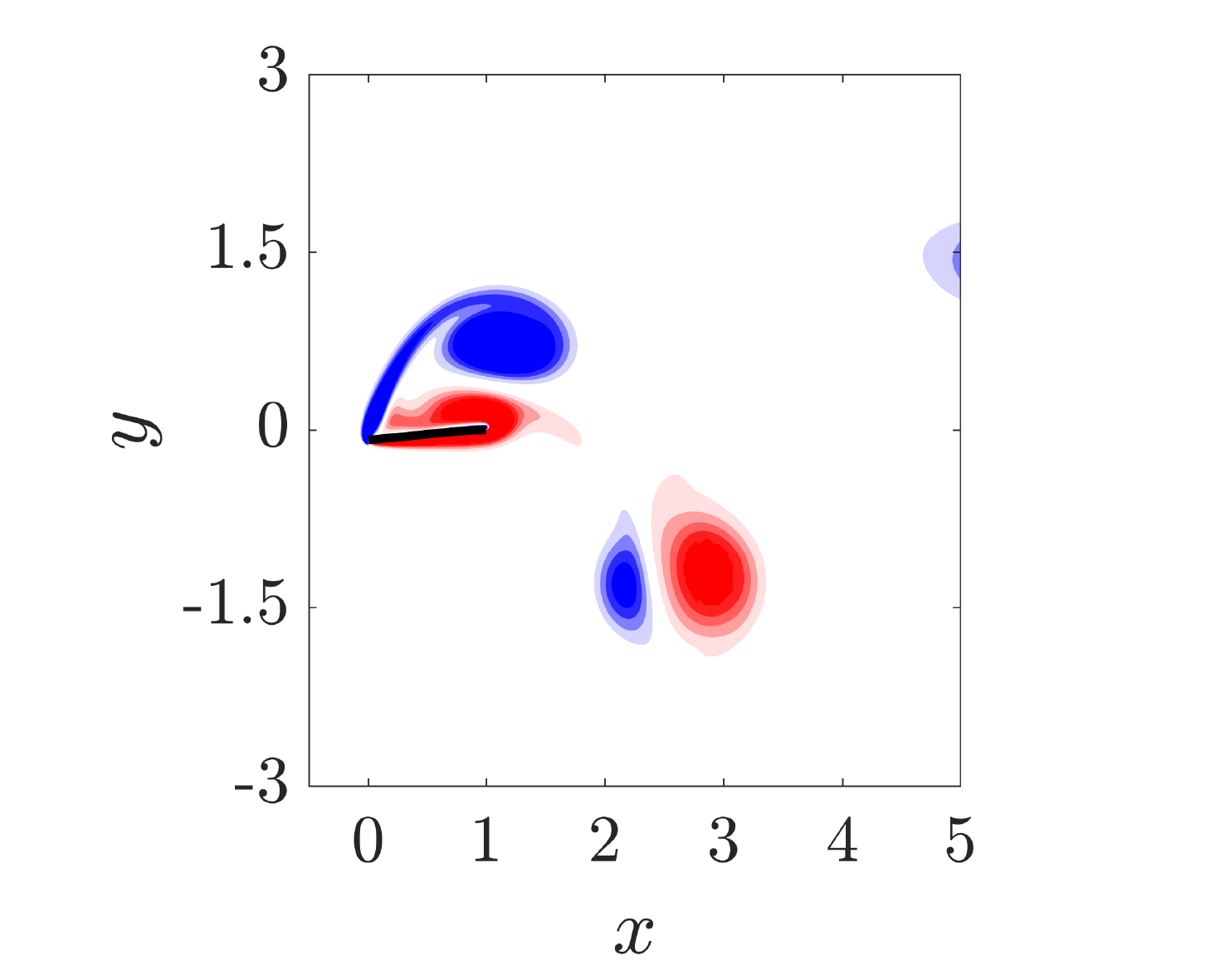}
	\end{subfigure}
	\begin{subfigure}[b]{0.245\textwidth}
        		\includegraphics[scale=0.345,trim={3.6cm 2.1cm 0cm 0cm},clip]{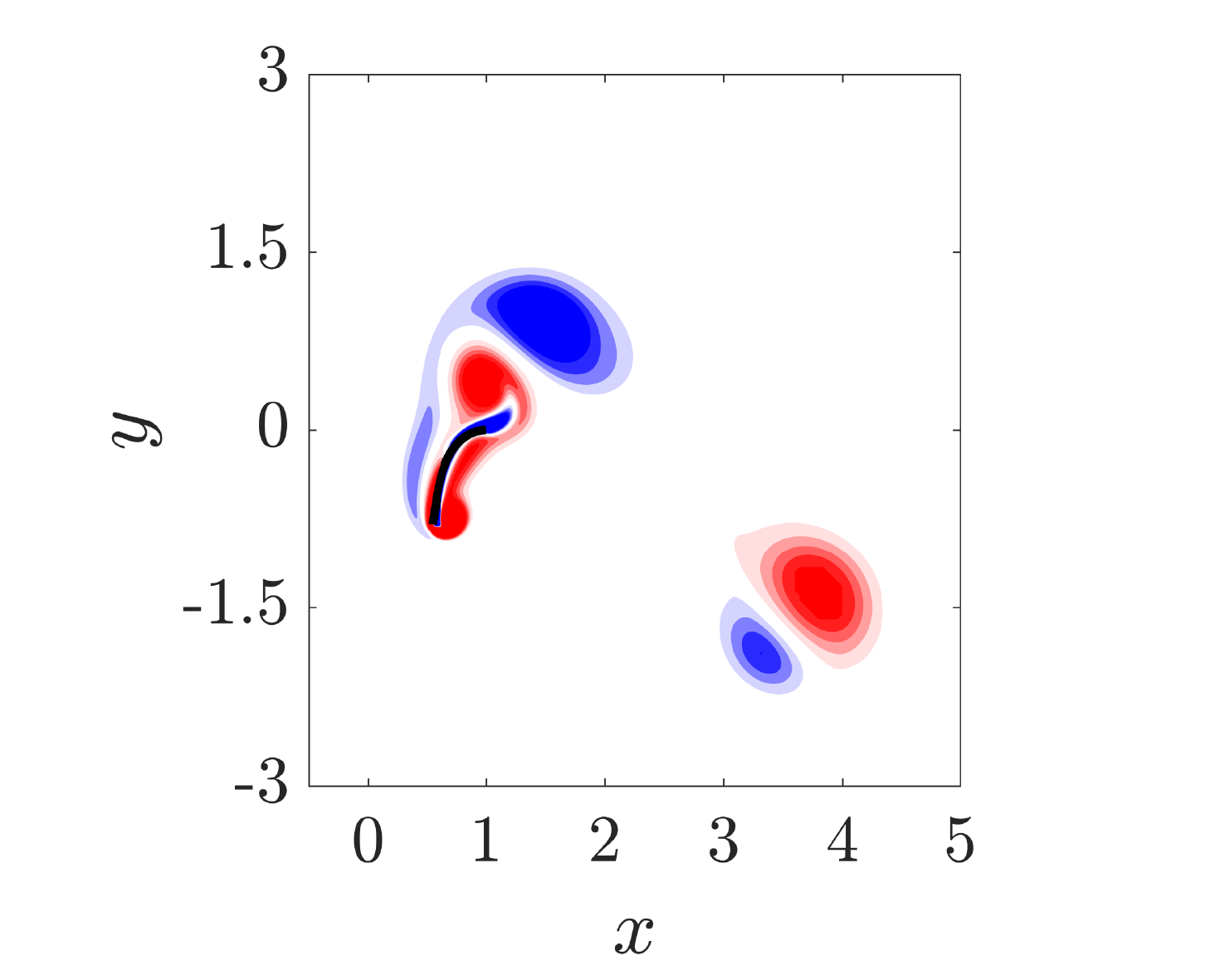}
	\end{subfigure}

	\begin{subfigure}[b]{0.245\textwidth}
        		\includegraphics[scale=0.345,trim={0cm 0cm 0cm 0cm},clip]{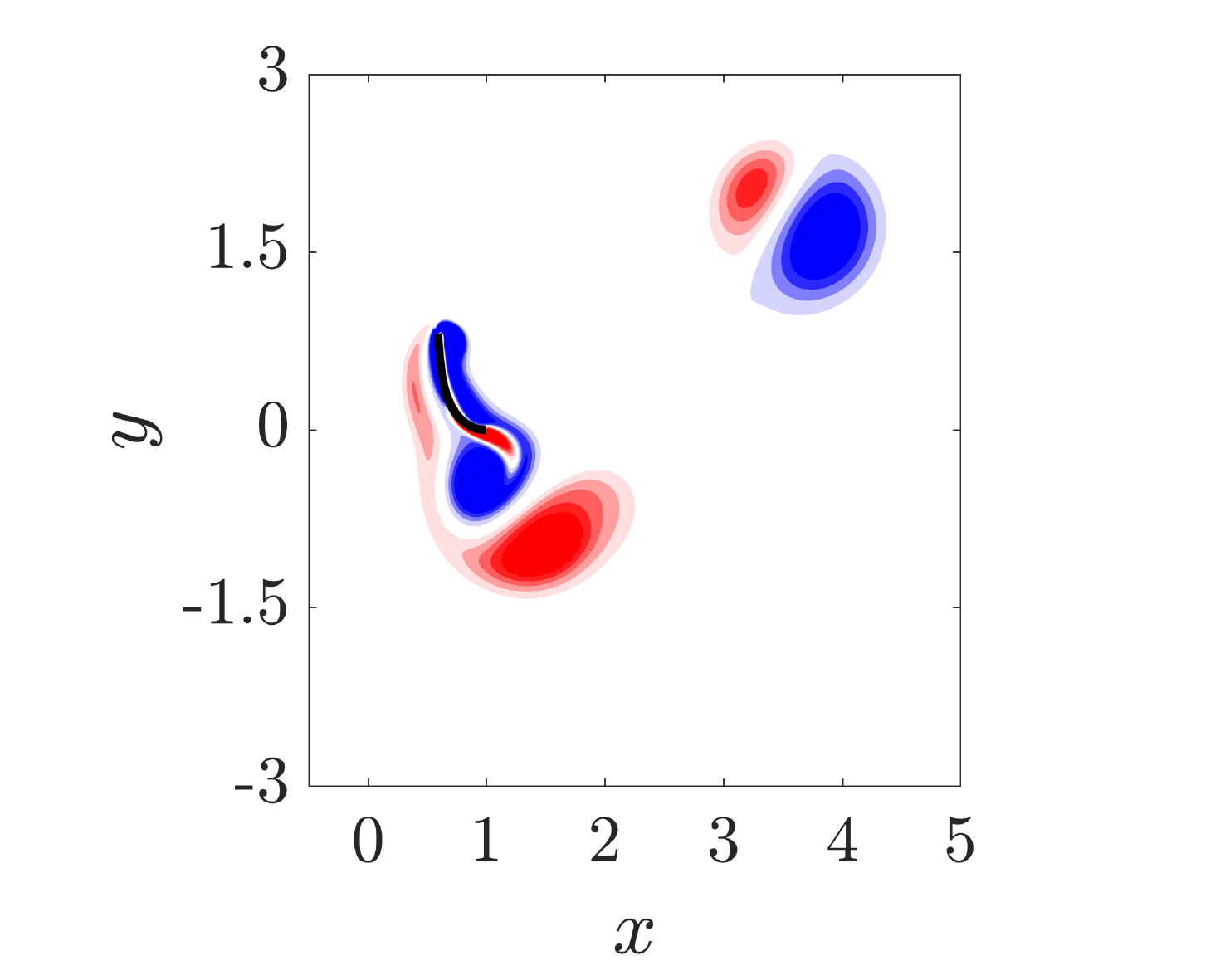}
	\end{subfigure}
	\begin{subfigure}[b]{0.245\textwidth}
		\hspace*{7.5mm}
        		\includegraphics[scale=0.345,trim={3.6cm 0cm 0cm 0cm},clip]{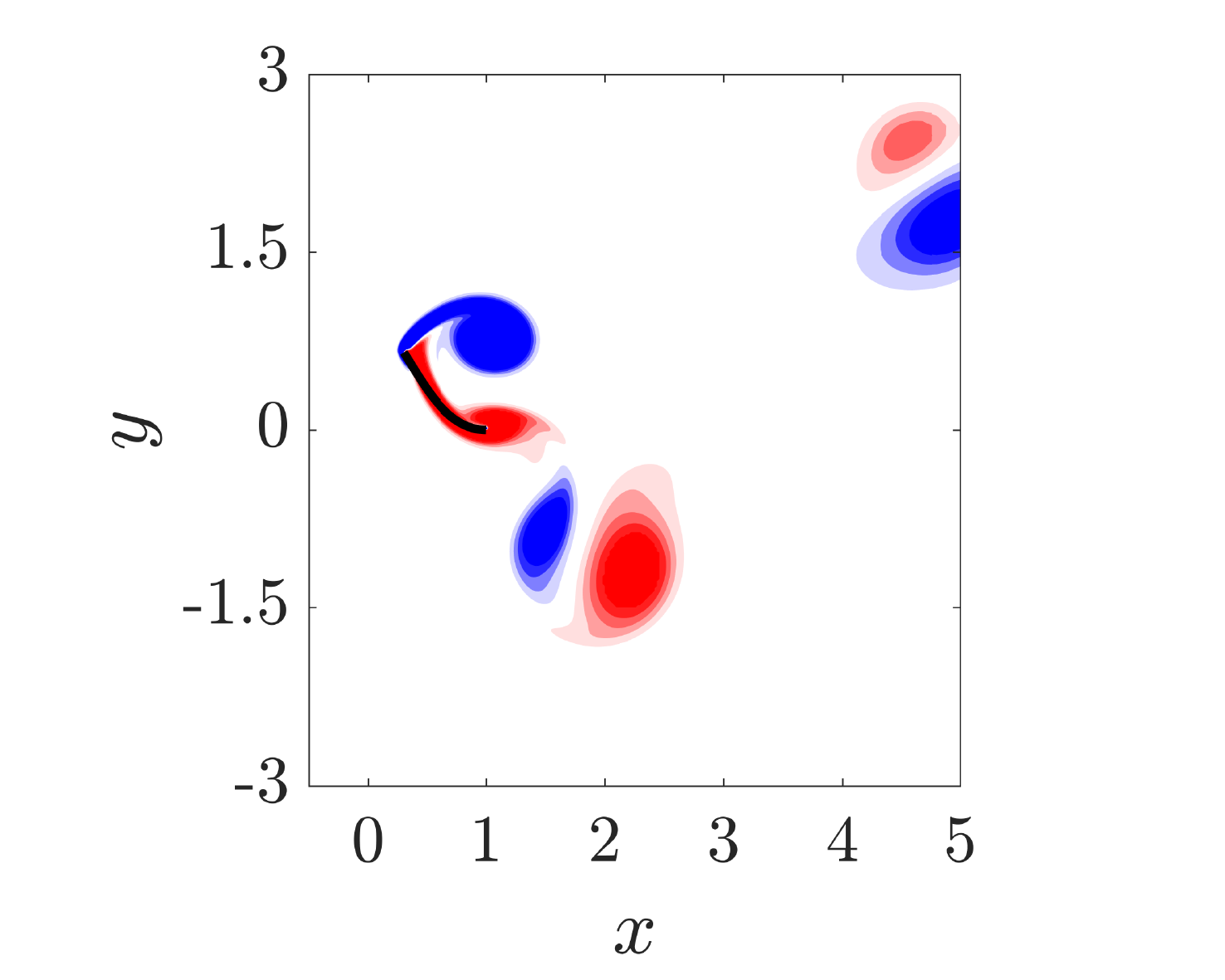}
	\end{subfigure}
    	\begin{subfigure}[b]{0.245\textwidth}
        		\hspace*{3.5mm}
        		\includegraphics[scale=0.345,trim={3.6cm 0cm 0cm 0cm},clip]{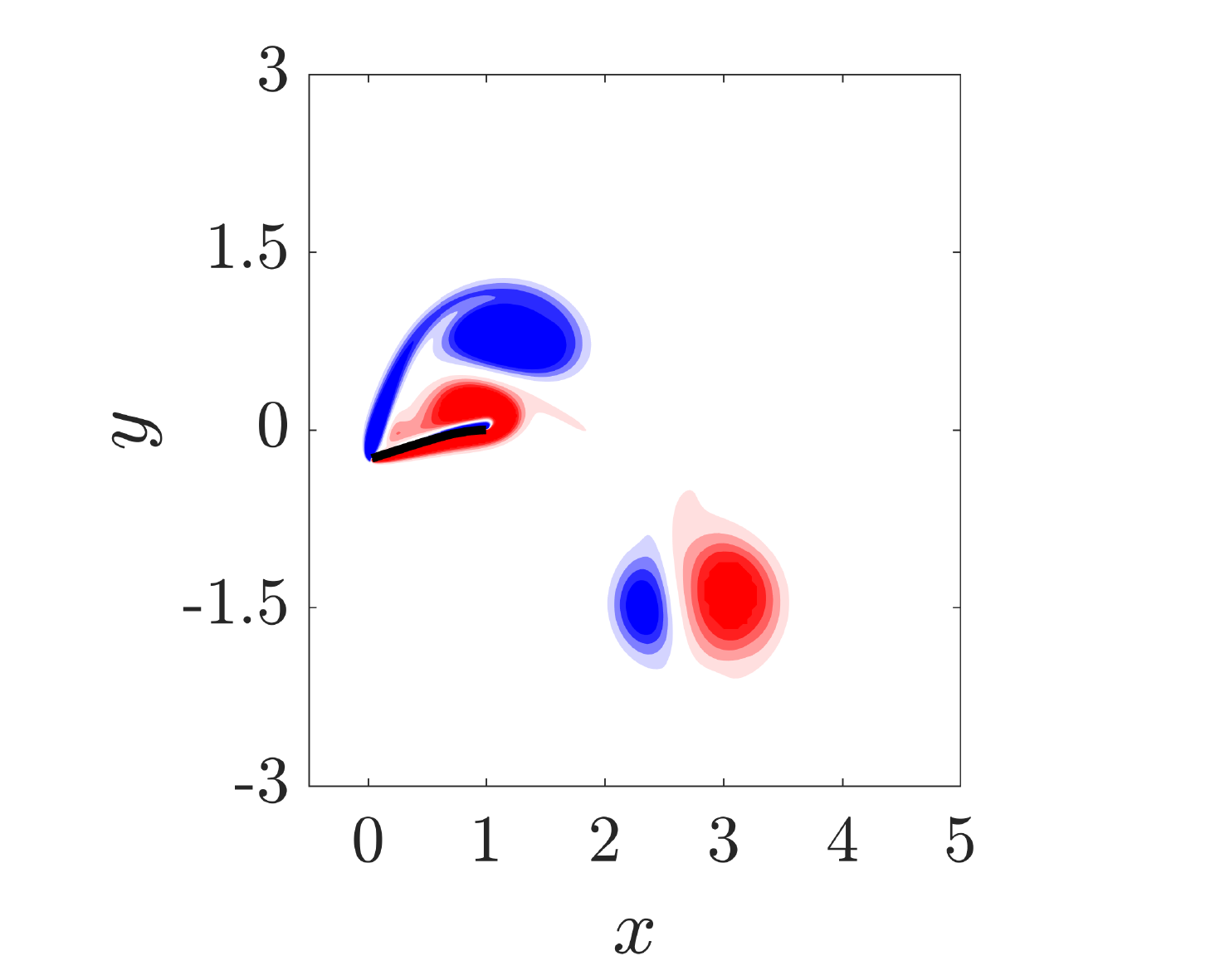}
	\end{subfigure}
	\begin{subfigure}[b]{0.245\textwidth}
        		\includegraphics[scale=0.345,trim={3.6cm 0cm 0cm 0cm},clip]{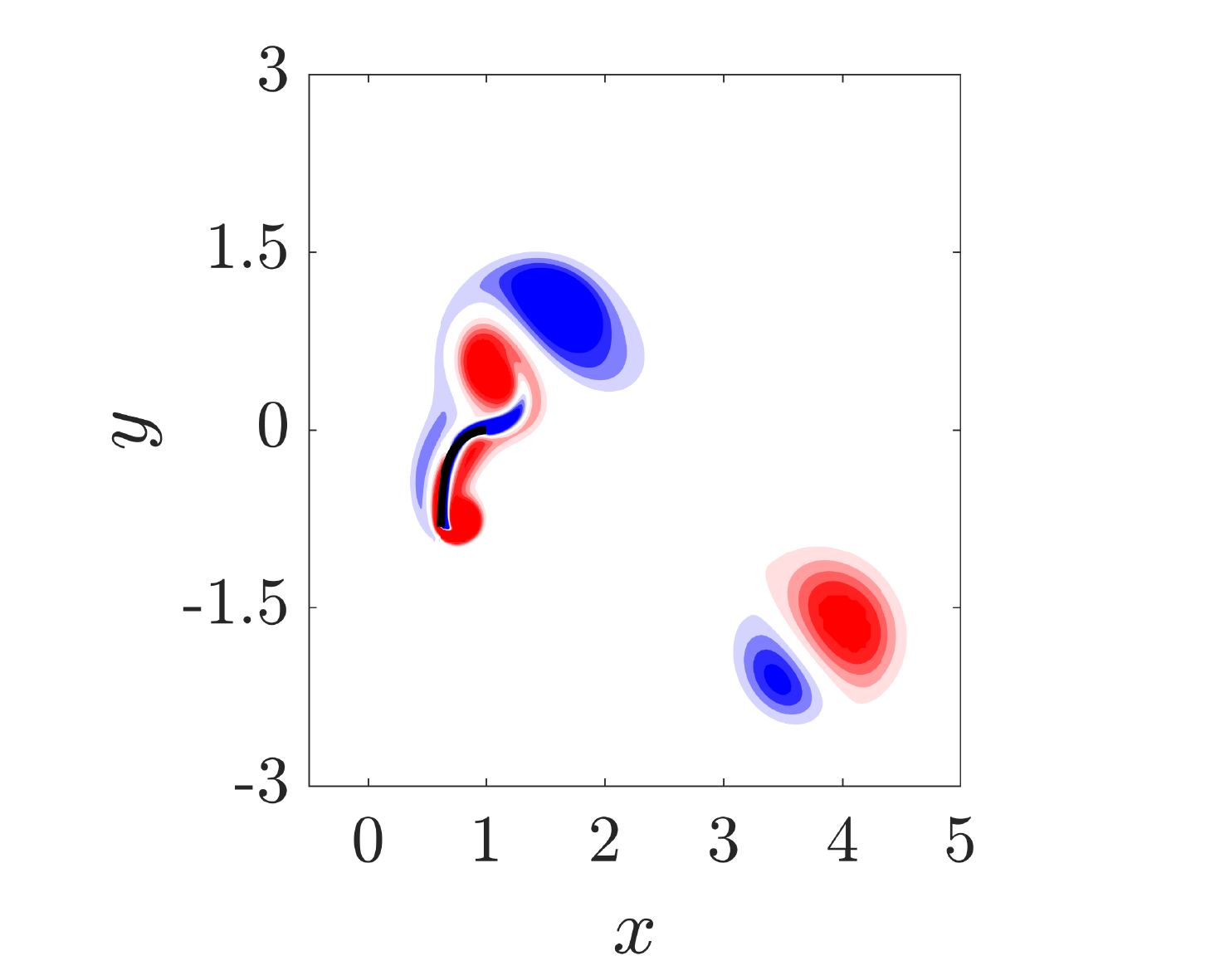}
	\end{subfigure}
	\caption{Vorticity contours at four snapshots of a flapping period of a flag in large-amplitude flapping for $M_\rho = 0.05$ (top row) and $M_\rho = 0.5$ (second row). The other parameters were $Re = 200, K_B = 0.32$. Contours are in 18 increments from -5 to 5.}
	\label{fig:large_amp_snapshots_light}
\end{figure}

For sufficiently light flags ($M_\rho \le 0.5$ in our studies), flapping was shown to synchronise with specific vortex-shedding patterns \citep{Gurugubelli2015,Shoele2016}, and \citet{Sader2016a} used experiments and a scaling analysis to argue that this regime is a VIV. To illustrate the synchronisation of vortex shedding and flapping, we show snapshots from a half-period of large-amplitude flapping for $M_\rho = 0.05, 0.5$ in figure \ref{fig:large_amp_snapshots_light}. Note that despite an order of magnitude change in mass, the vortex-shedding pattern in the top two rows of the figure is similar: when the flag reaches its peak amplitude the leading edge vortex formed during the upstroke grows (left plot); as the flag begins its downstroke the leading edge vortex is released and a trailing edge vortex forms (second from left plot); the vortices grow in size as the flag reaches its centreline position (second from right plot); while the leading and trailing edge vortices advect downstream to form a P vortex pair (see \cite{Williamson1988} for a description of this vortex characterisation), a leading edge vortex forms as the flag continues its downstroke (rightmost plot). When the flag reaches its peak position, an analogous process to the one just described occurs during the upstroke (with oppositely signed vorticity).

\begin{figure}
	\centering
	\begin{subfigure}[b]{0.495\textwidth}
		\hspace*{8mm}
		\centering
        		\includegraphics[scale=0.35,trim={0cm 0cm 0cm 0cm},clip]{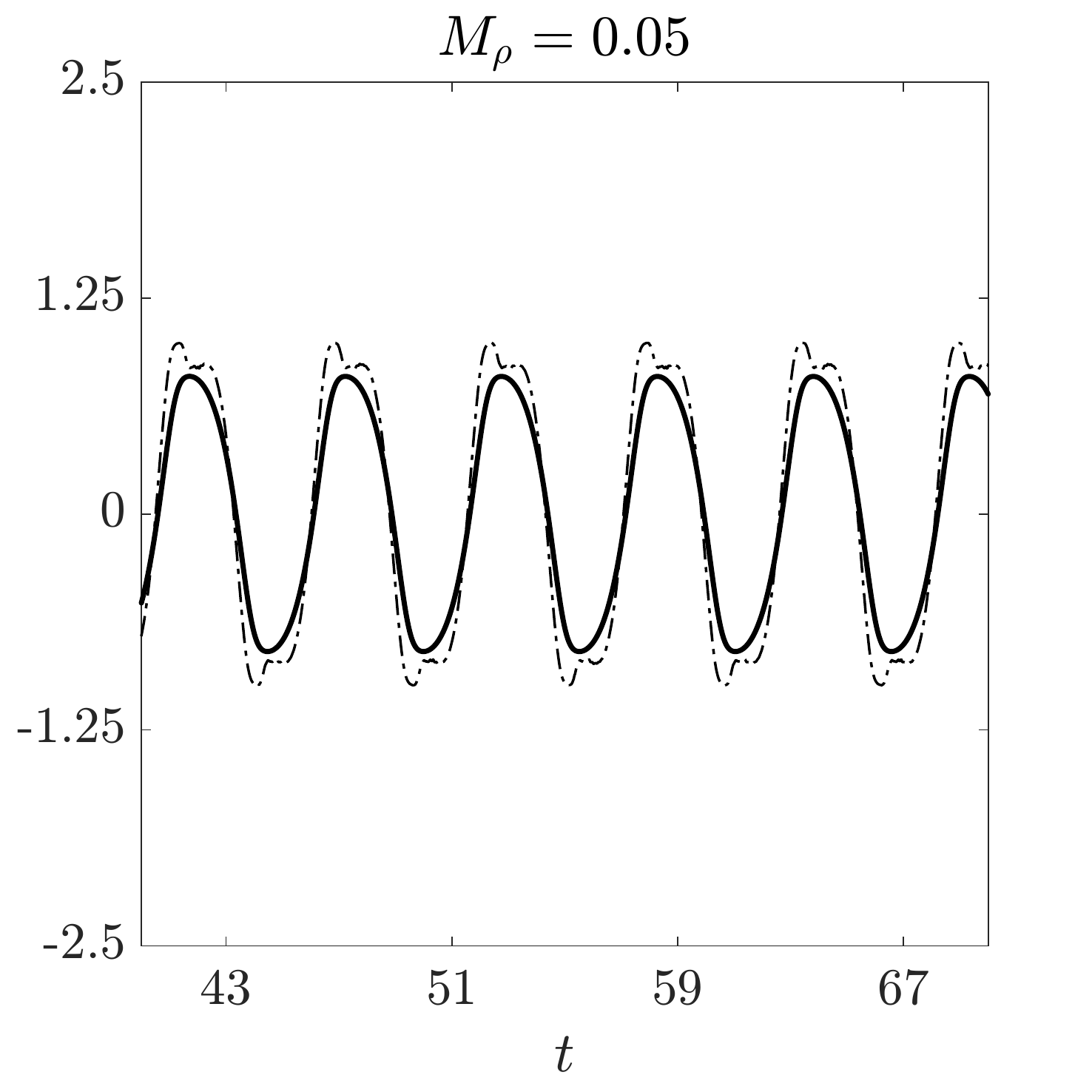}
	\end{subfigure}
	\begin{subfigure}[b]{0.495\textwidth}
		\centering
		\hspace*{-15mm}
        		\includegraphics[scale=0.35,trim={1.75cm 0cm 0cm 0cm},clip]{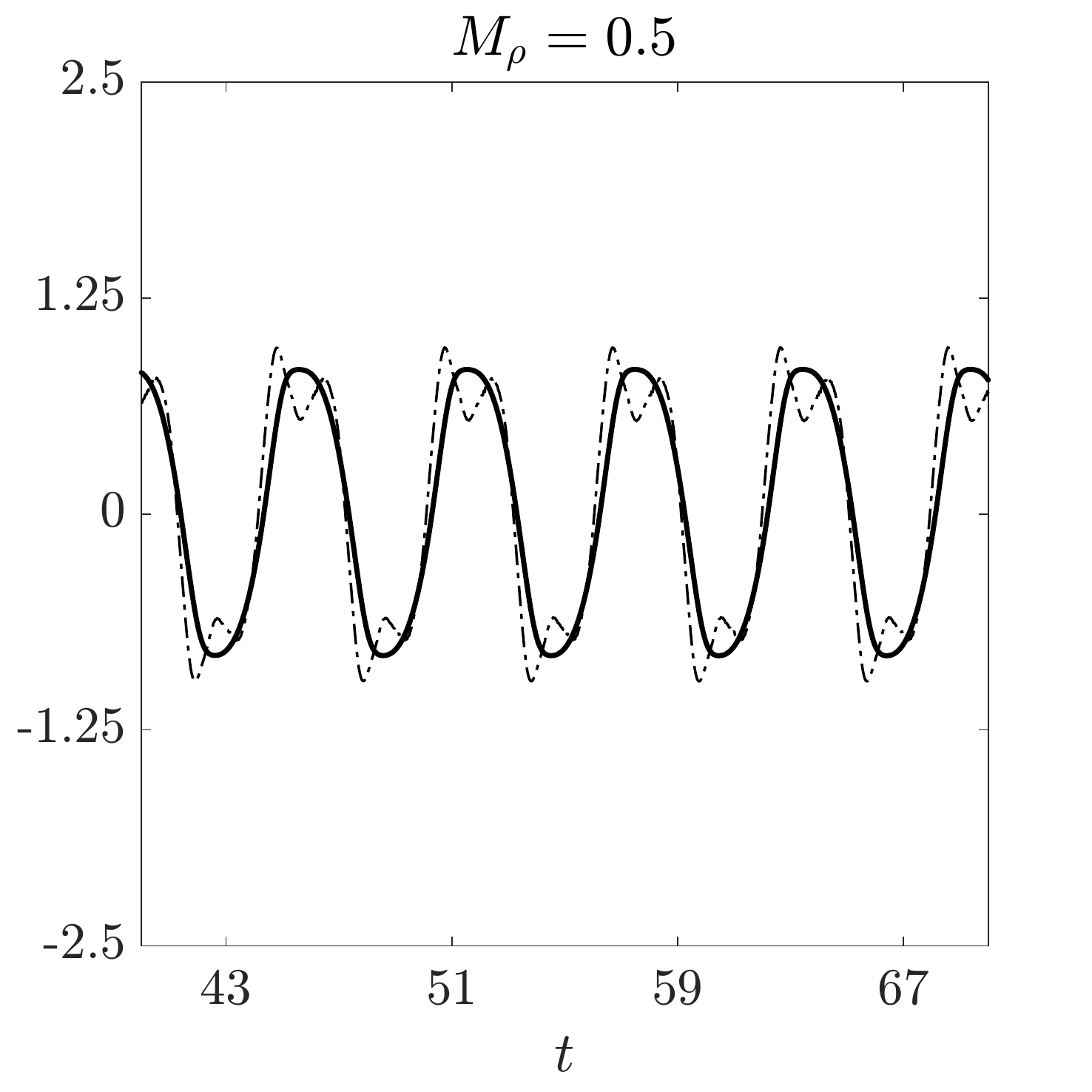}
	\end{subfigure}
	\caption{Tip displacement ($-$) and coefficient of lift ($\boldsymbol{\cdot} -$) for an inverted flag in large-amplitude flapping with $Re = 200$, $K_B = 0.32$.}
	\label{fig:largeamp_CLdisp_light}
\end{figure}

To confirm that this regime is a VIV, we show in figure \ref{fig:largeamp_CLdisp_light} time traces of the coefficient of lift and tip displacement for $M_\rho = 0.05, 0.5$. The lift and tip displacement are synchronised, and therefore satisfy the definition of VIV \citep{Khalak1999,Sarpkaya2004}.

 \subsubsection{Large-amplitude flapping of heavy flags}

\begin{figure}	
	\begin{subfigure}[b]{0.245\textwidth}
        		\includegraphics[scale=0.345,trim={0cm 2.1cm 0cm 0cm},clip]{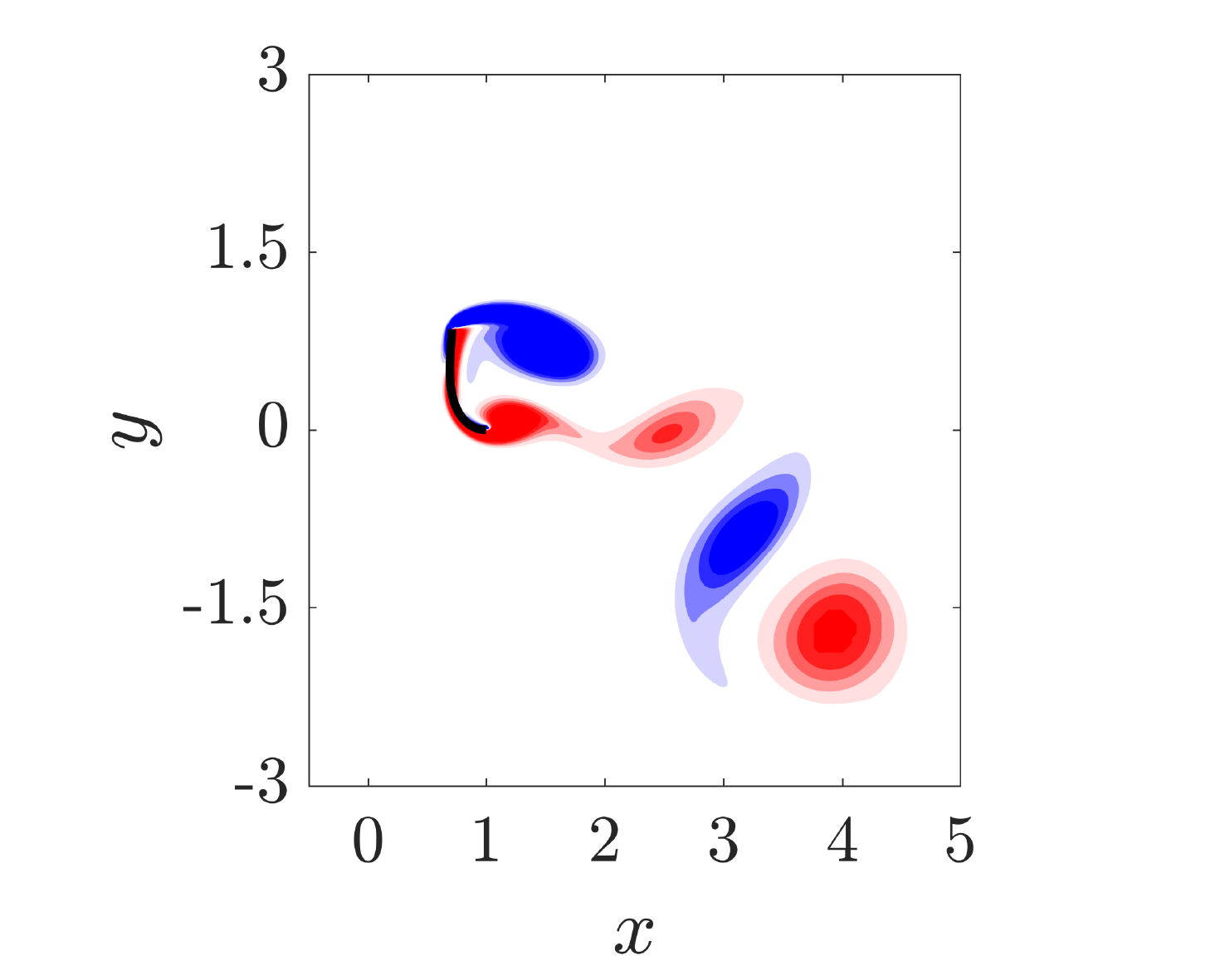}
	\end{subfigure}
	\begin{subfigure}[b]{0.245\textwidth}
		\hspace*{7.5mm}
        		\includegraphics[scale=0.345,trim={3.6cm 2.1cm 0cm 0cm},clip]{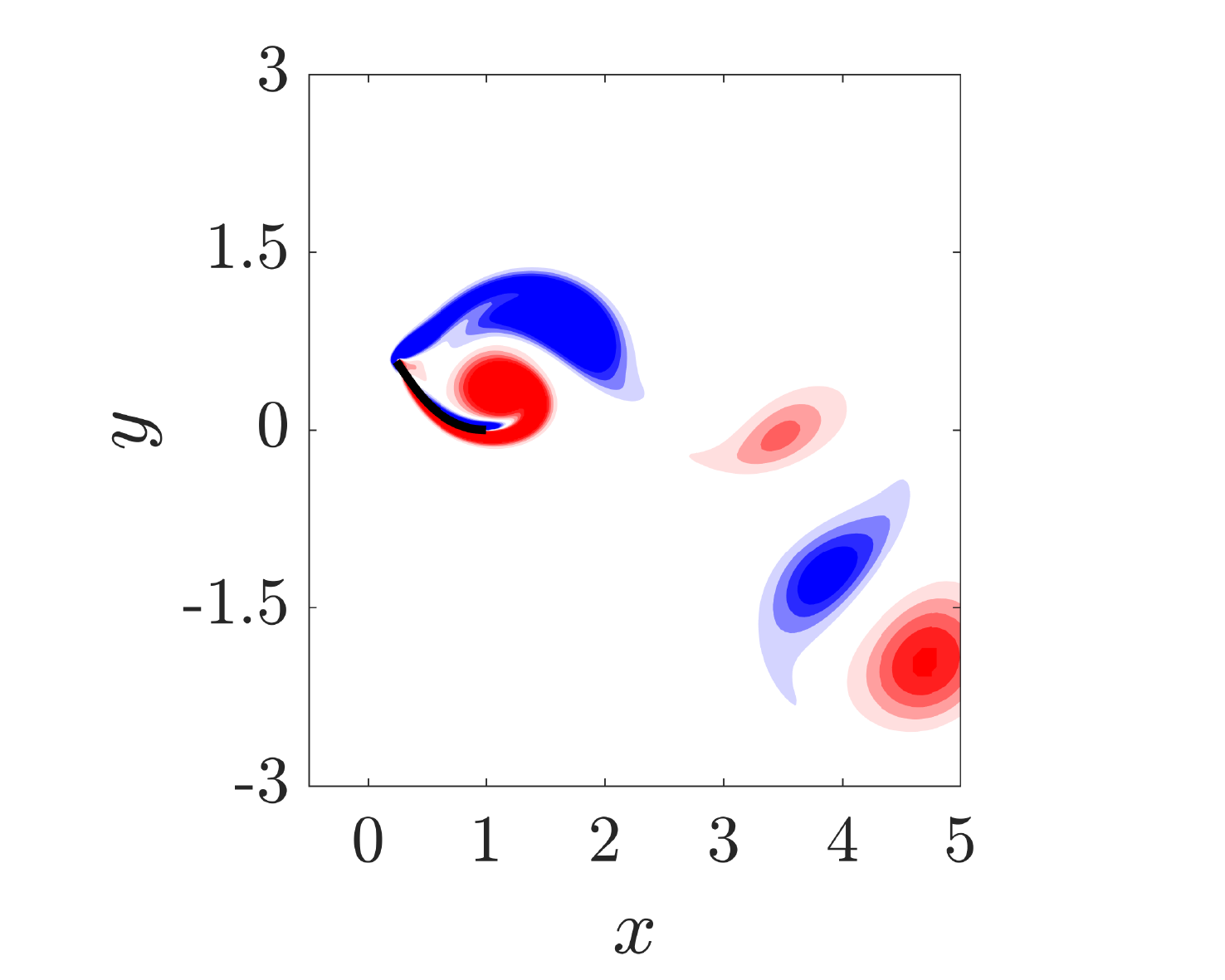}
	\end{subfigure}
    	\begin{subfigure}[b]{0.245\textwidth}
        		\hspace*{3.5mm}
        		\includegraphics[scale=0.345,trim={3.6cm 2.1cm 0cm 0cm},clip]{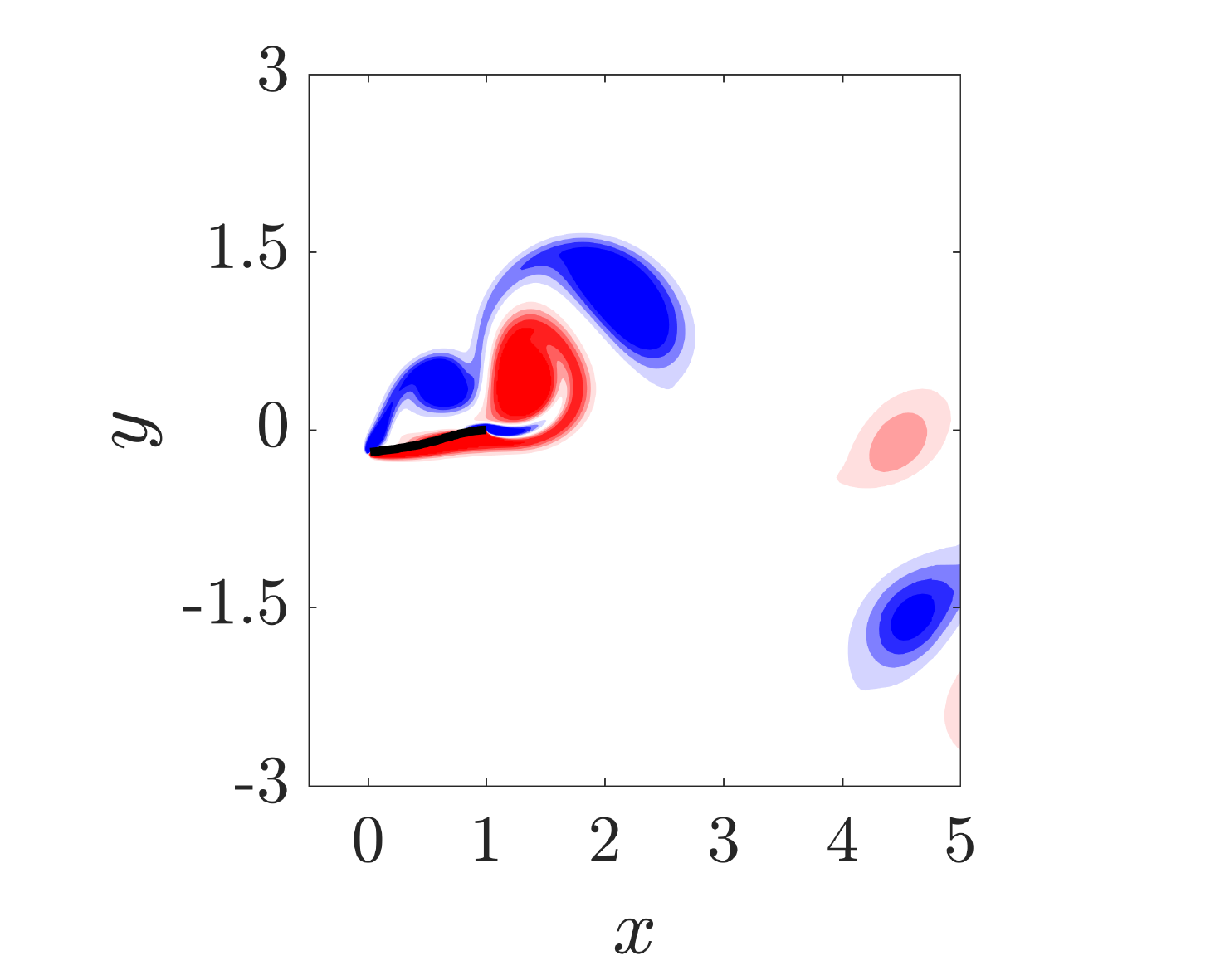}
	\end{subfigure}
	\begin{subfigure}[b]{0.245\textwidth}
        		\includegraphics[scale=0.345,trim={3.6cm 2.1cm 0cm 0cm},clip]{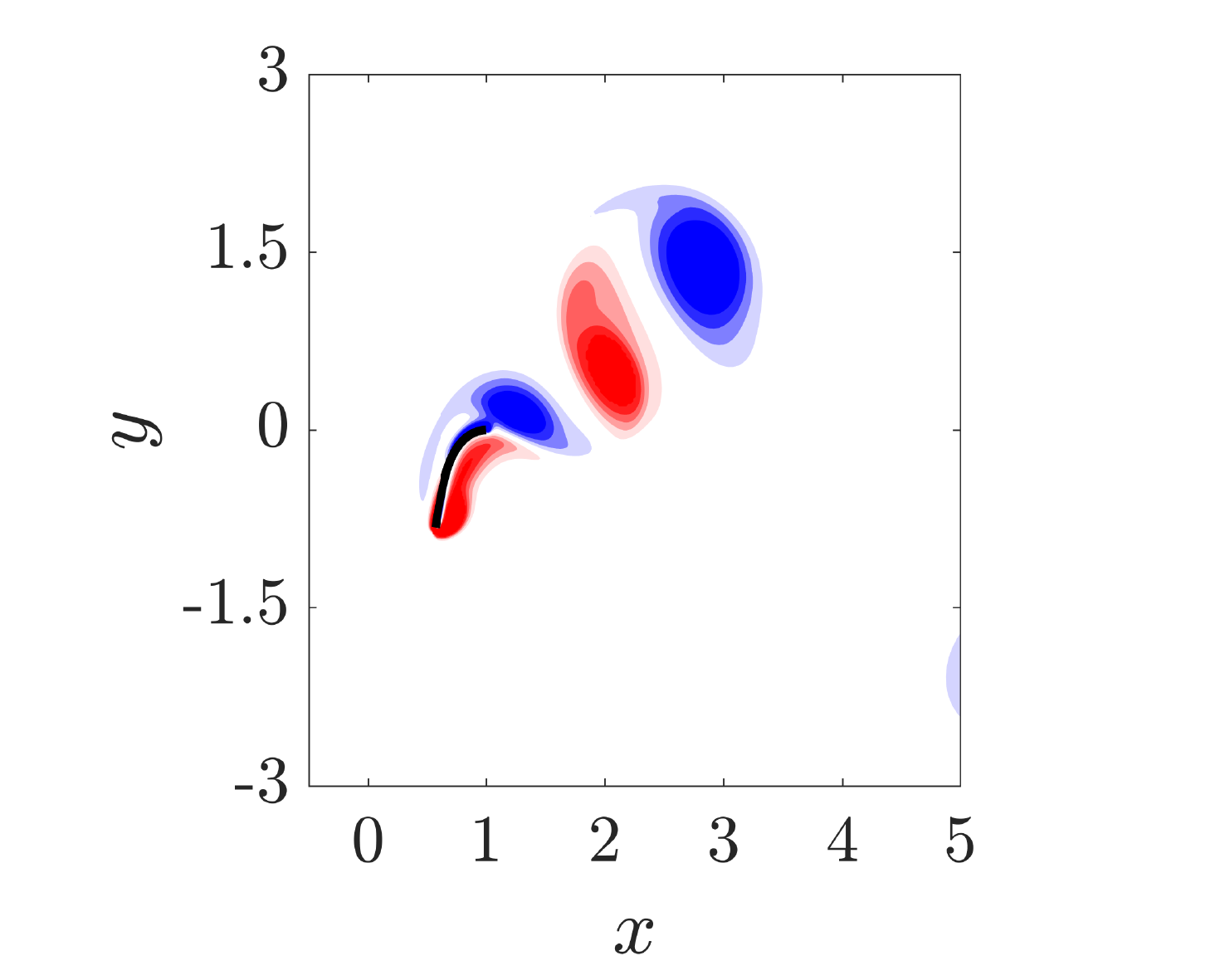}
	\end{subfigure}

	\begin{subfigure}[b]{0.245\textwidth}
        		\includegraphics[scale=0.345,trim={0cm 0cm 0cm 0cm},clip]{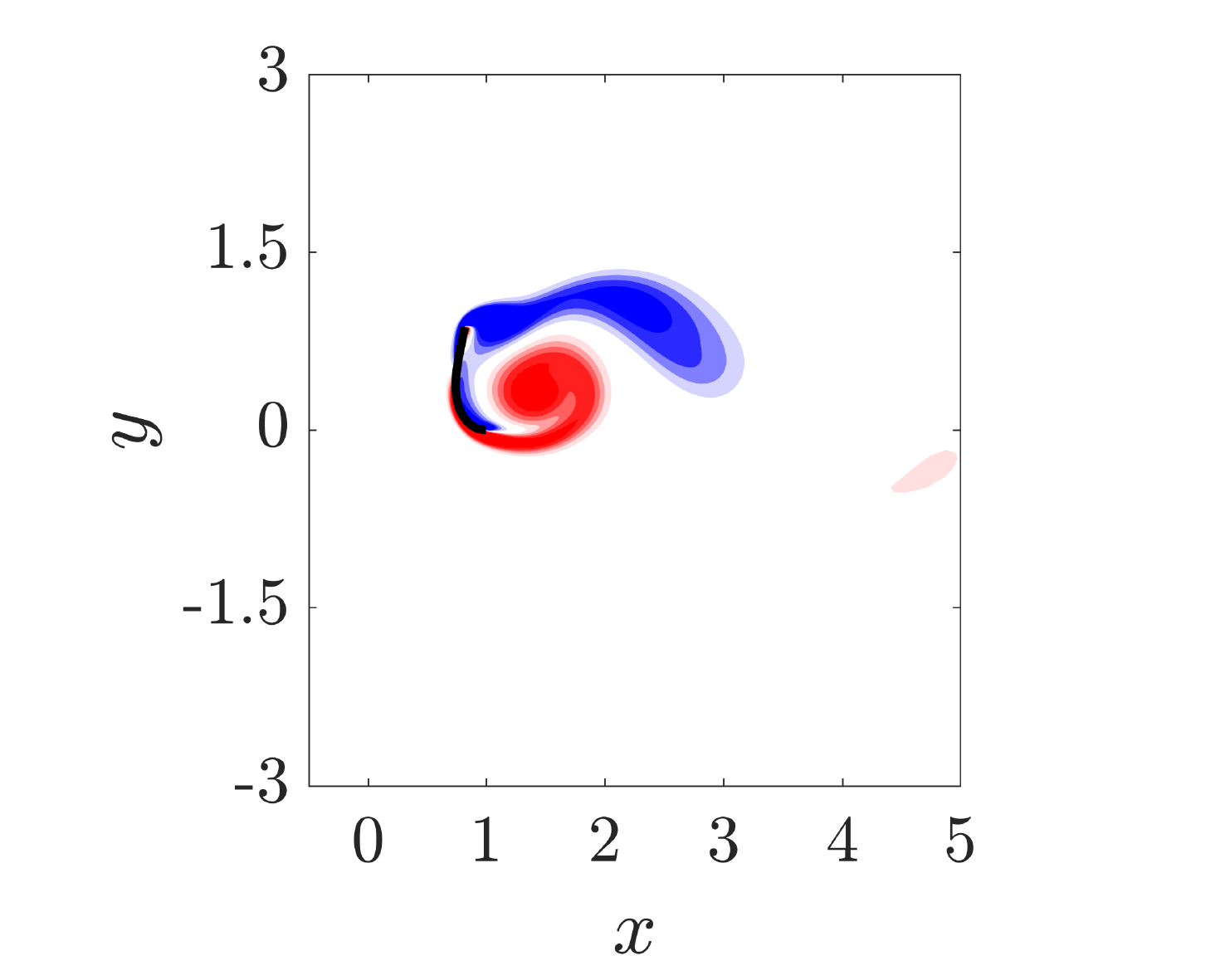}
	\end{subfigure}
	\begin{subfigure}[b]{0.245\textwidth}
		\hspace*{7.5mm}
        		\includegraphics[scale=0.345,trim={3.6cm 0cm 0cm 0cm},clip]{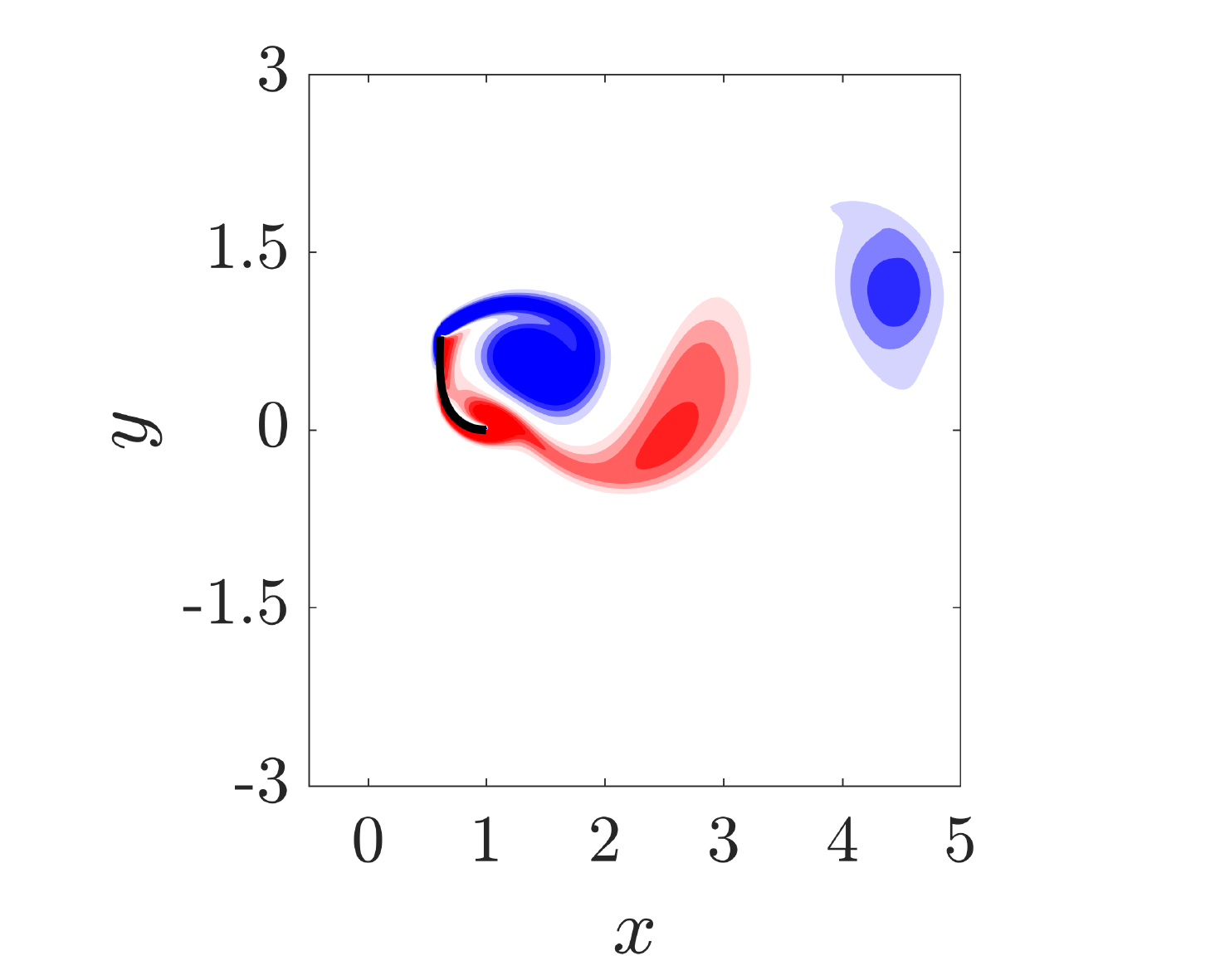}
	\end{subfigure}
    	\begin{subfigure}[b]{0.245\textwidth}
        		\hspace*{3.5mm}
        		\includegraphics[scale=0.345,trim={3.6cm 0cm 0cm 0cm},clip]{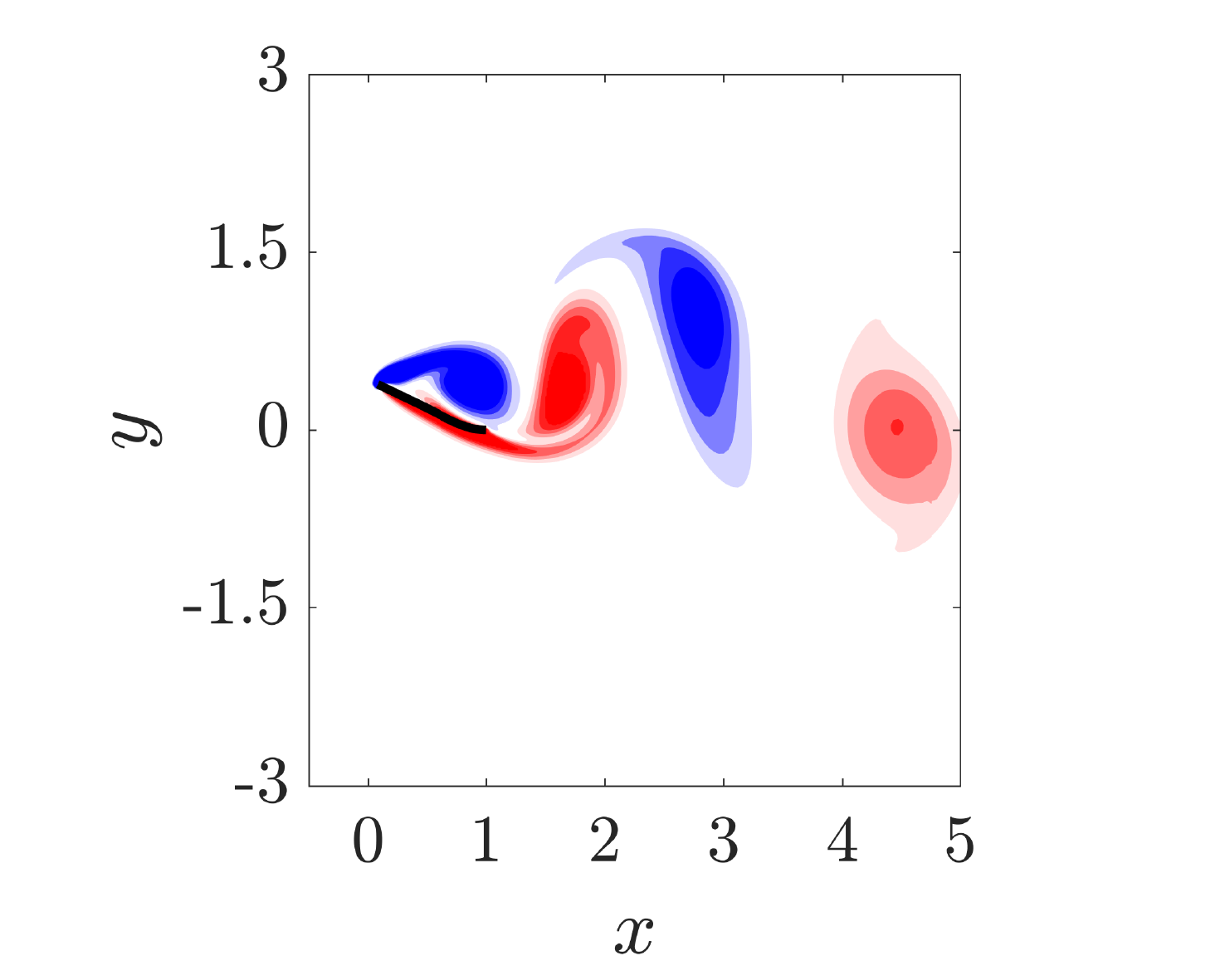}
	\end{subfigure}
	\begin{subfigure}[b]{0.245\textwidth}
        		\includegraphics[scale=0.345,trim={3.6cm 0cm 0cm 0cm},clip]{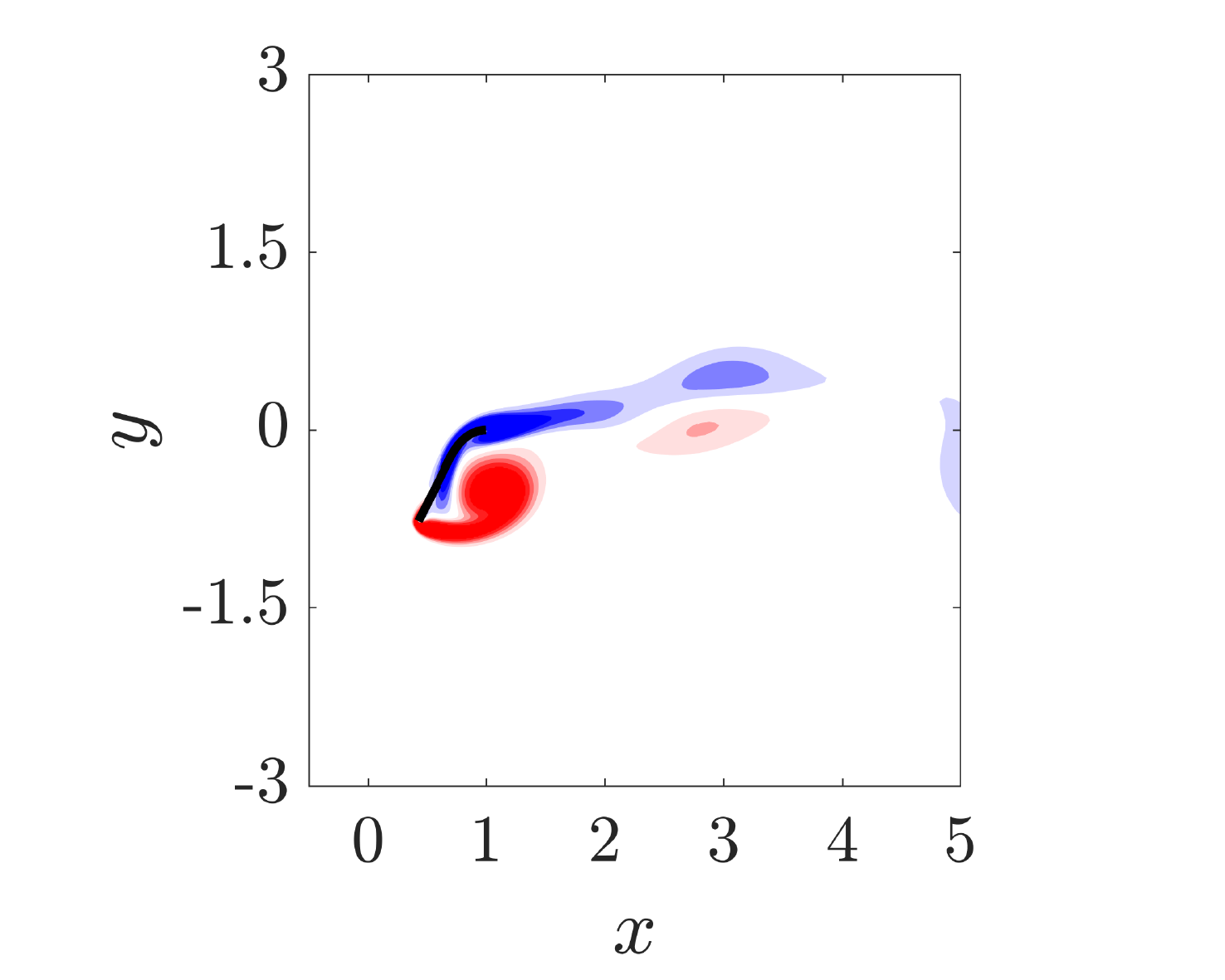}
	\end{subfigure}
	\caption{Vorticity contours at four snapshots of a flapping period of a flag in large-amplitude flapping for $M_\rho = 5$ (first row) and $M_\rho = 50$ (second row). The other parameters were $Re = 200, K_B = 0.32$. Contours are in 18 increments from -5 to 5.}
	\label{fig:large_amp_snapshots_heavy}
\end{figure}

The interaction between vortex shedding and flapping is qualitatively different as the flag mass is increased further. Figure \ref{fig:large_amp_snapshots_heavy} shows that for $M_\rho = 5, 50$ additional vortices are shed per half-period. The flapping cycle for $M_\rho = 5$ is similar to the lighter flag cases, except that during the downstroke an additional leading edge vortex forms (first row, second from rightmost plot) and advects downstream along with the original leading and trailing edge vortices (rightmost plot). The additional vortex leads to a P $+$ S wake structure. For $M_\rho = 50$, even more vortices are shed during the downstroke: the first leading and trailing edge vortices are shed when the flag is still near its peak amplitude (second row, leftmost plot); the flag begins its downstroke and another leading-trailing edge vortex pair are formed (second from leftmost plot); as the flag nears its centreline position, a third leading edge vortex forms (second from rightmost plot); this leading edge vortex combines with a newly formed trailing edge vortex during the downstroke phase to form a third pair that is advected downstream; at the end of the downstroke phase, new vortices form at the leading and trailing edge as the flag reaches its peak amplitude (rightmost plot).

\begin{figure}
    	\begin{subfigure}[b]{0.495\textwidth}
		\vspace*{2mm}
    		\hspace*{8mm}
		\centering
        		\includegraphics[scale=0.35,trim={0cm 0cm 0cm 0cm},clip]{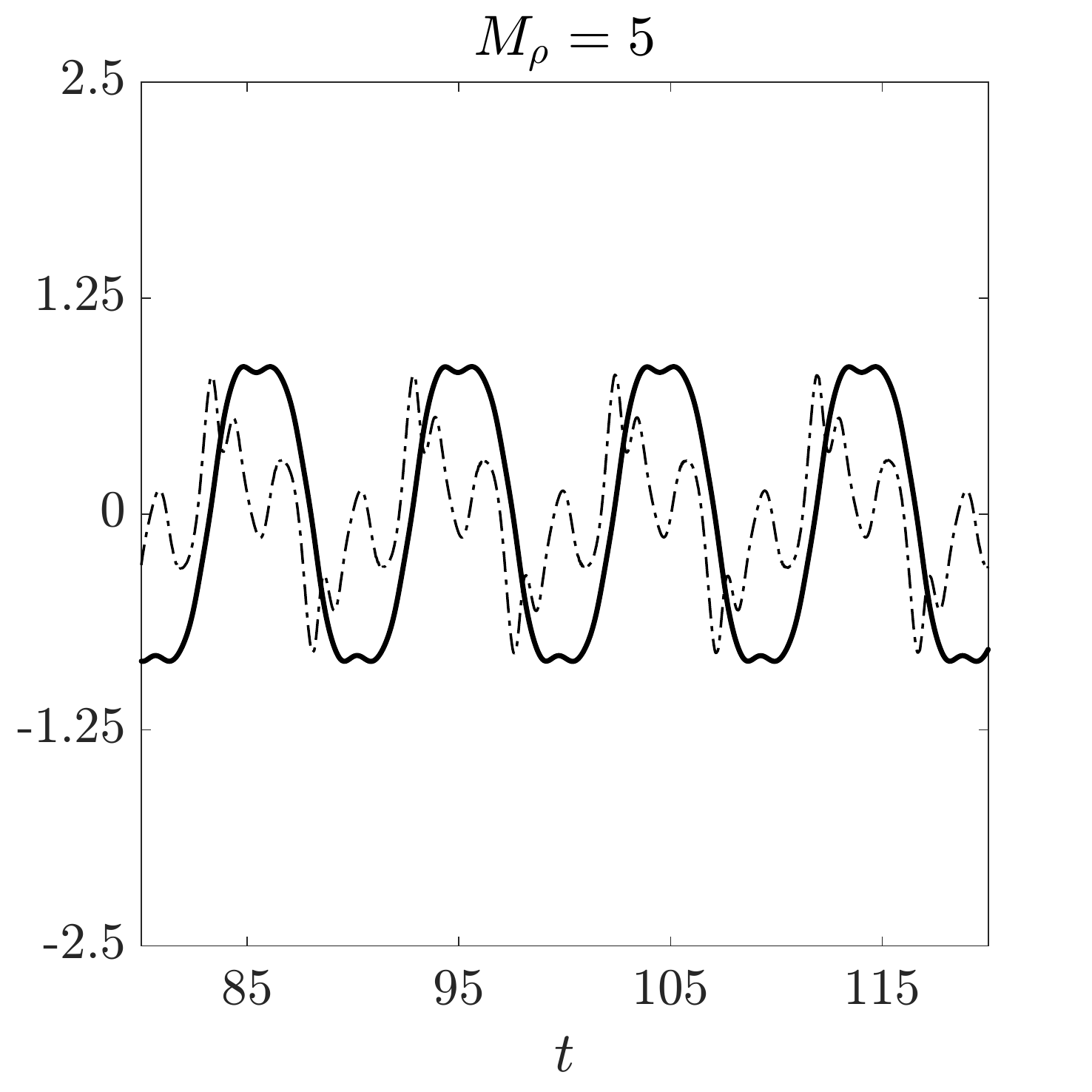}
	\end{subfigure}
	\begin{subfigure}[b]{0.495\textwidth}
		\vspace*{2mm}
		\centering
		\hspace*{-15mm}
        		\includegraphics[scale=0.35,trim={1.7cm 0cm 0cm 0cm},clip]{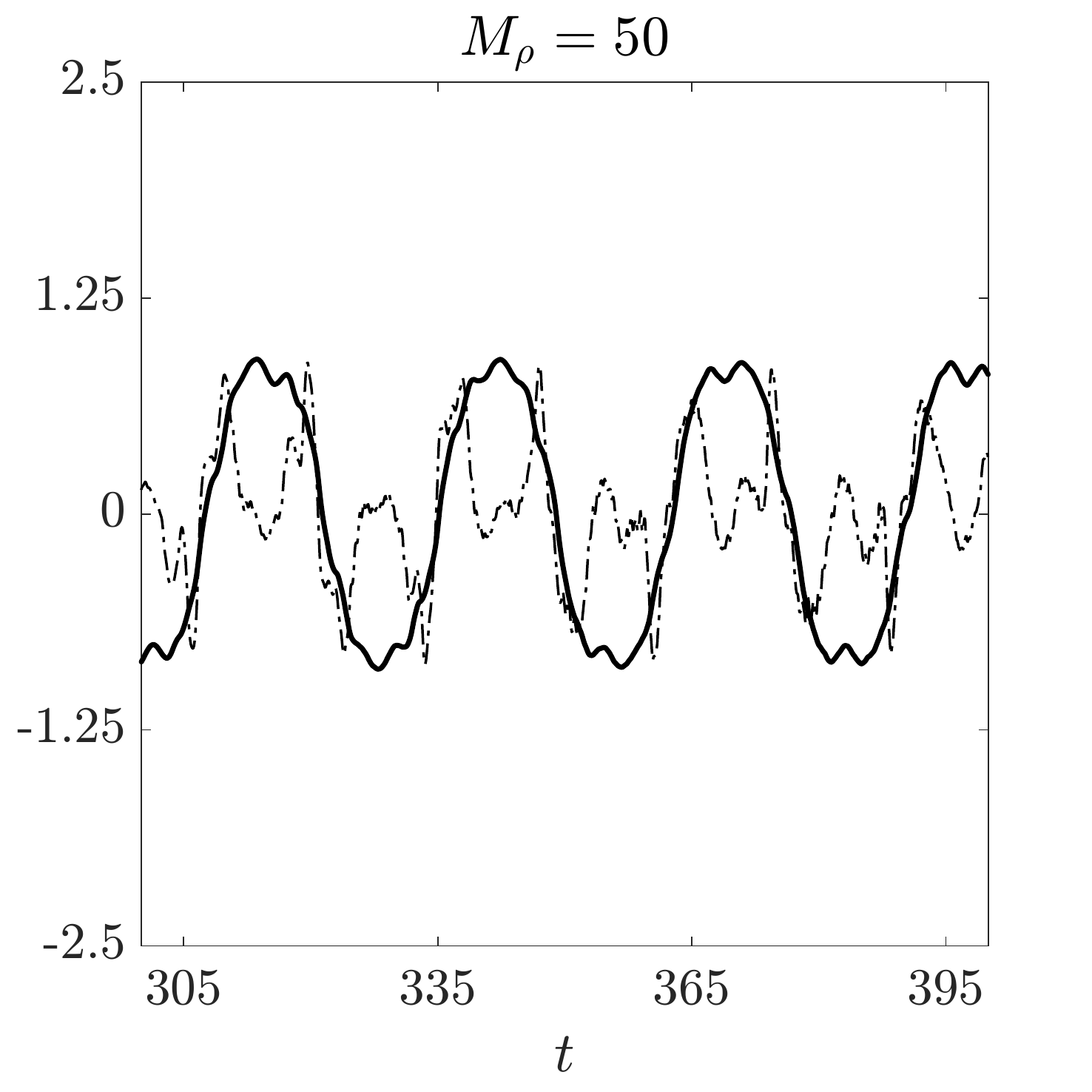}
	\end{subfigure}
	\caption{Tip displacement ($-$) and coefficient of lift ($\boldsymbol{\cdot} -$) for an inverted flag in large-amplitude flapping with $Re = 200$, $K_B = 0.32$.}
	\label{fig:largeamp_CLdisp_heavy}
\end{figure}

This desynchronisation is illustrated further through time traces of the coefficient of lift and tip displacement (see figure \ref{fig:largeamp_CLdisp_heavy}). Unlike the in-phase behaviour between lift and tip displacement seen for light flags, heavy flags contain multiple lift peaks for a given peak in tip displacement. Moreover, for $M_\rho = 50$, there is a slight departure from periodicity that can be observed in figure \ref{fig:largeamp_CLdisp_heavy} (and is reflected in the bottom right bifurcation diagram in figure \ref{fig:bif_Re200}). The break in synchronisation of vortex shedding and flapping demonstrates that sufficiently heavy flags do not undergo a VIV---a result argued by \cite{Sader2016a} through a scaling analysis.

We emphasise that large-amplitude flapping still occurs for heavy flags despite the absence of a VIV. The mechanism responsible for flapping can be explained through the deformed equilibrium of the system. In this regime, the deformed equilibrium is unstable and has a sufficiently large saturation amplitude for the flag to flap past the centreline position and into the region of attraction of the deformed equilibrium on the other side of the centreline. This newly sampled deformed equilibrium is also associated with a saturation amplitude that leads the flag to flap past the centreline, so an indefinite process ensues with flapping occurring around these two equilibria. We show in section \ref{sec:Re_20} that large-amplitude flapping also occurs for massive flags at $Re = 20$, which attests to the presence of non-VIV flapping mechanisms and to the potential for large-amplitude flapping in the absence of any vortex shedding.

%

\subsection{Deflected-mode regime}   	
\label{sec:Deflectedmode_Re200}

For low stiffnesses the system transitions to a large-deflection state about which small-amplitude flapping occurs. As seen in figure \ref{fig:bif_Re200}, this flapping is not centered about the deformed equilibrium position (\emph{i.e.}, the mean and equilibrium states are different). 

The nature of flapping in this regime is qualitatively distinct from that of small-deflection deformed flapping and large-amplitude flapping. \cite{Shoele2016} observed that when the flapping frequency is scaled by the freestream velocity and mean tip amplitude (\emph{i.e.}, the mean projected length to the flow), it agrees well with the classical 0.2 value found for vortex shedding past bluff bodies \citep{Roshko1954}. They used this finding to argue that the bluff-body wake instability is responsible for the small-amplitude flapping in this regime. 

\begin{figure}
	\centering
	\begin{subfigure}[b]{0.45\textwidth}
		\centering
        		\includegraphics[scale=0.4, trim={0cm 1.3cm 0cm 0cm},clip]{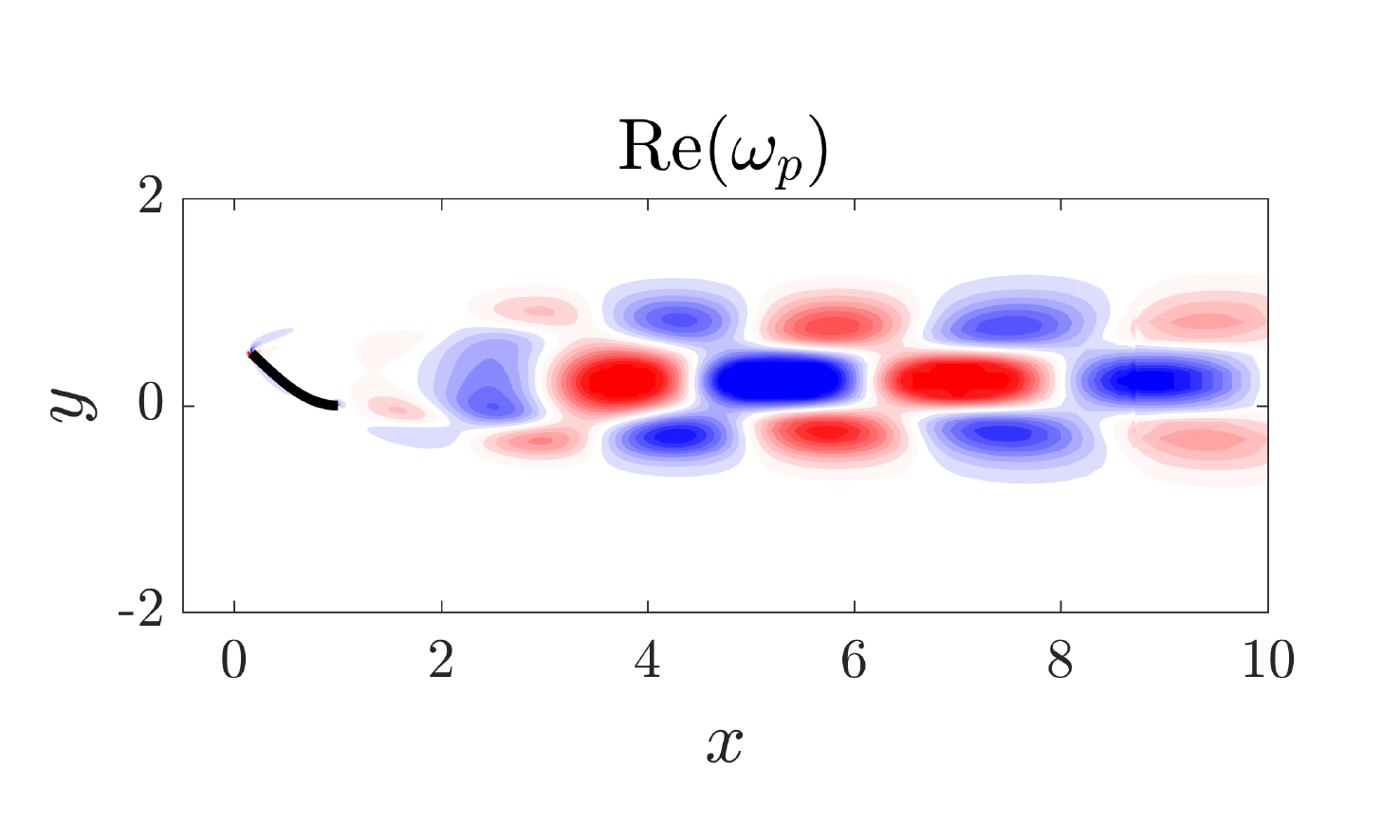}
	\end{subfigure}
	\begin{subfigure}[b]{0.45\textwidth}
		\centering
		\hspace*{3.9mm}
        		\includegraphics[scale=0.32, trim={1cm 1cm 0cm 0cm},clip]{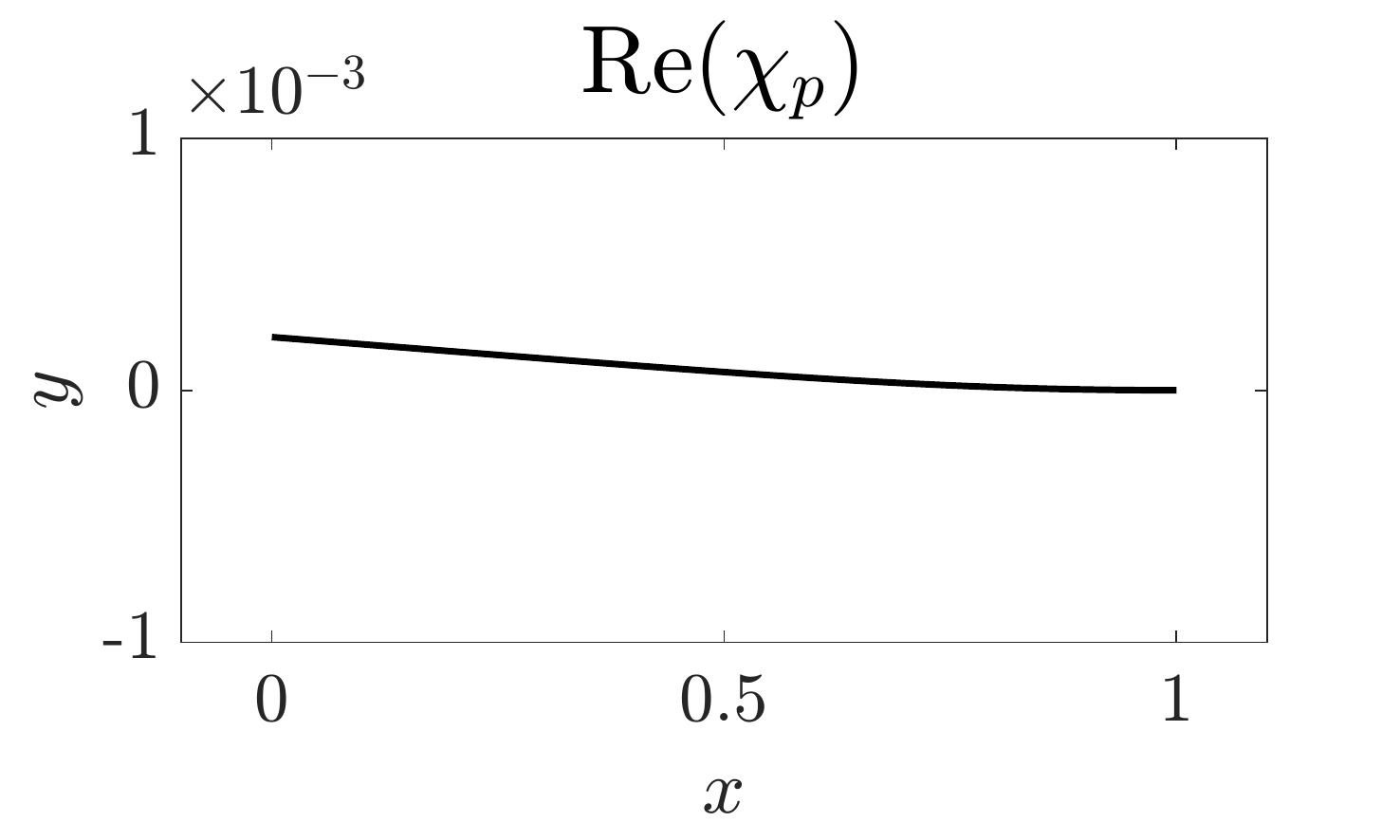}
		\vspace*{1.2mm}
	\end{subfigure}
	
    	\centering
	\begin{subfigure}[b]{0.45\textwidth}
		\centering
        		\includegraphics[scale=0.4, trim={0cm 0cm 0cm 0cm},clip]{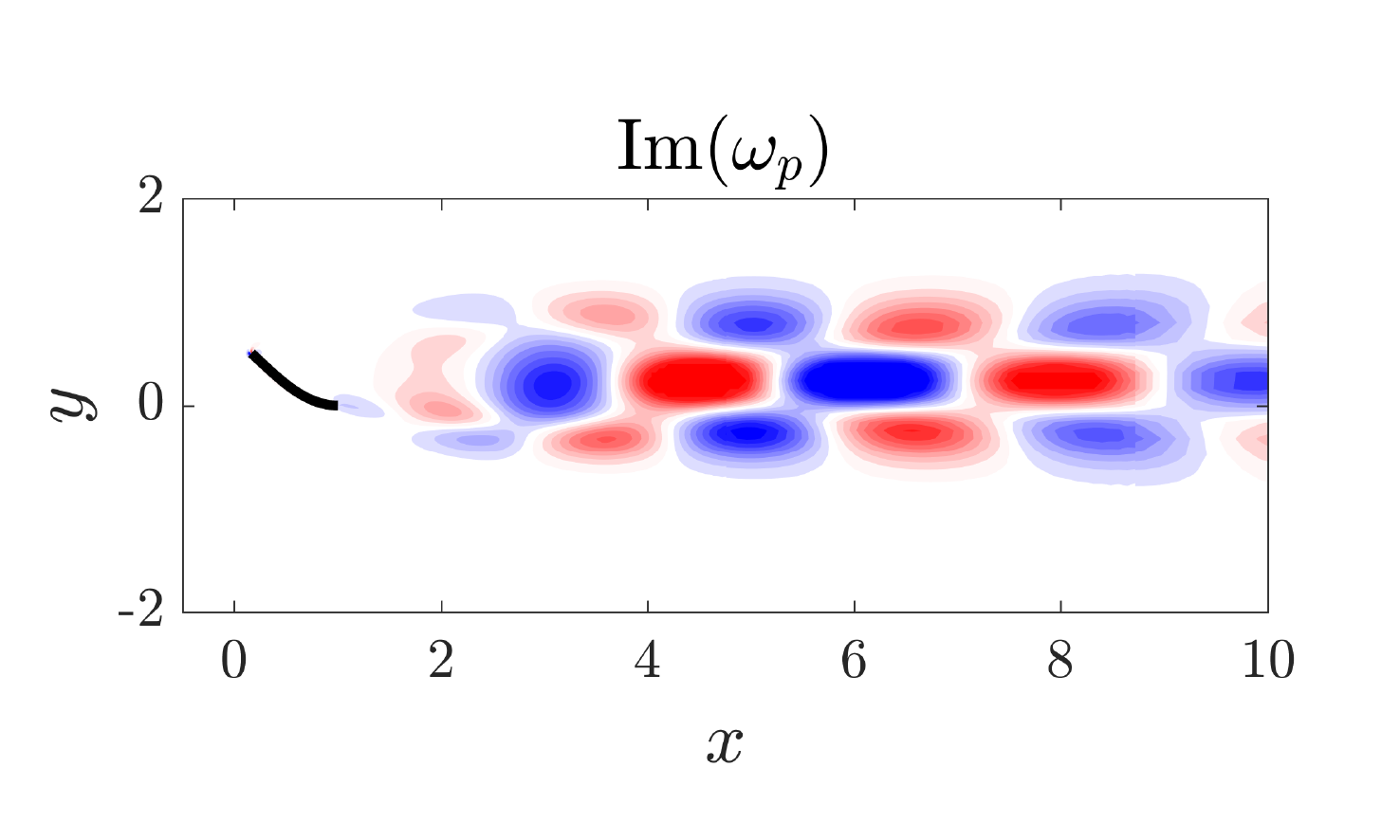}
	\end{subfigure}
	\begin{subfigure}[b]{0.45\textwidth}
		\centering
		\hspace*{3.9mm}
        		\includegraphics[scale=0.32, trim={1cm 0cm 0cm 0cm},clip]{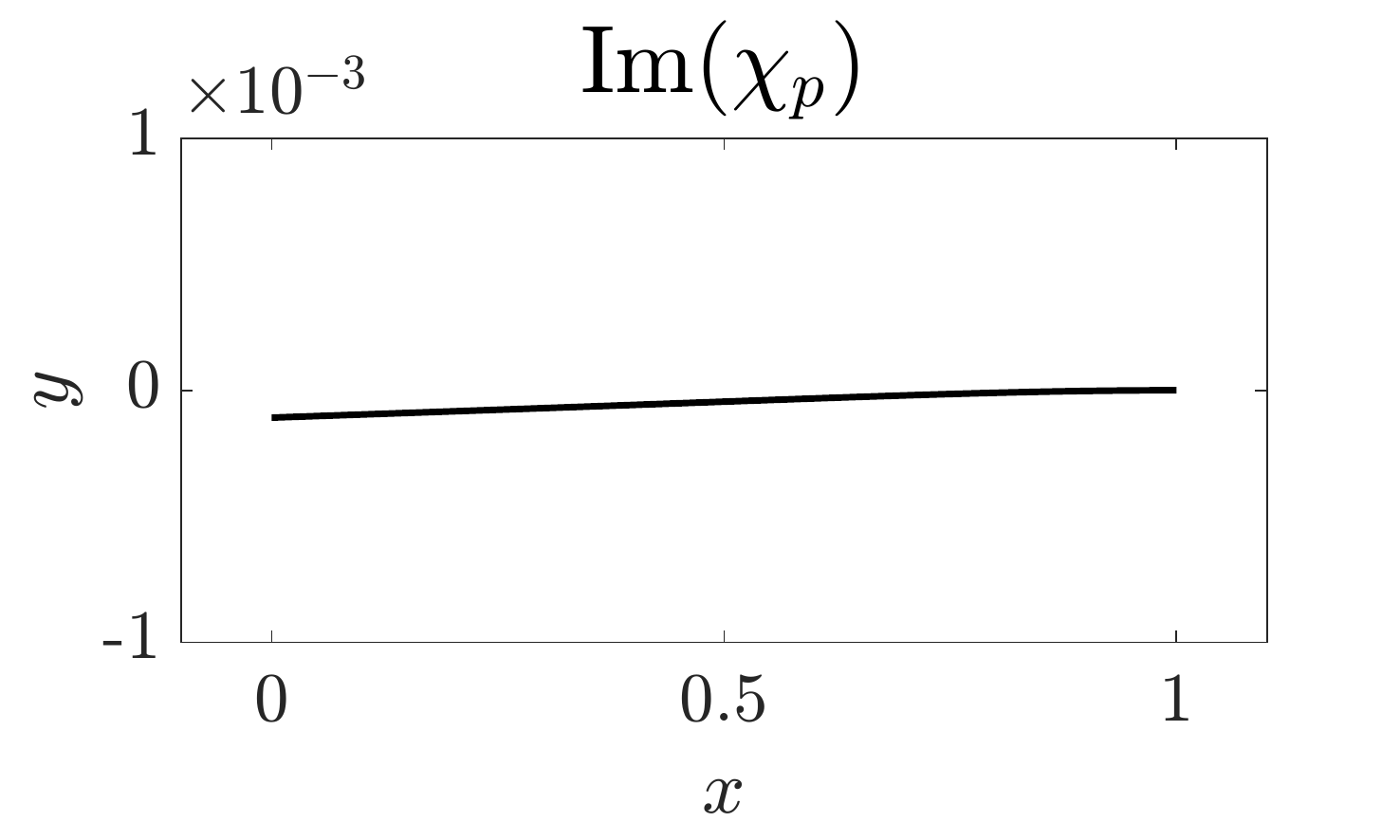}
		\vspace*{3.3mm}
	\end{subfigure}
	\caption{Real (top) and imaginary (bottom) parts of vorticity (left) and flag displacement (right) of the leading global mode of the deformed equilibrium for $M_\rho = 0.5, K_B = 0.12$ and $Re = 200$. Vorticity contours are in 20 increments from -0.2 to 0.2. The eigenvectors are identical for $M_\rho = 0.05, 0.5,$ and $50$.}
	\label{fig:large_def_mode}
\end{figure}

The global stability analysis of the deformed equilibrium confirms this previous conclusion. Figure \ref{fig:large_def_mode} shows that the least damped mode is characterised by a vortical structure in the wake of the flag similar to the leading mode of a rigid stationary cylinder \citep{Barkley2006}. This is in contrast to the leading mode found for small-deflection deformed flapping and large-amplitude flapping, which has isolated vortical structures near the flag surface\footnote{The analogous mode to the leading global mode for small-deflection deformed flapping and large-amplitude flapping is also unstable in the deflected-mode regime, but is associated with a smaller growth rate.}. Moreover, as seen in table \ref{tab:large_def_Re200}, the flapping frequency of this leading mode is independent of mass ratio ($M_\rho$). This is also distinct from the mass-dependent flapping frequency of the least damped mode for the previously discussed regimes. The similar vortical structure of the mode to other bluff-body flows and the independence of the structural parameters on the modal frequency demonstrate that the leading instability is associated with vortex shedding and is flow-driven.

%

\begin{table}
\centering
\begin{tabular}{ c c }
$M_\rho$ & {Leading mode frequency}  \\ \hline
 0.05 & 0.205 \\
 0.5 & 0.205  \\ 
 5 & 0.205 \\
 50 & 0.205
\end{tabular}
 \caption{Frequency of the leading global mode of the deformed equilibrium for $K_B = 0.12$ and four different masses. All cases correspond to the deflected-mode regime.}
\label{tab:large_def_Re200}
\end{table}

The presence of vortex shedding in this deflected-mode regime is also the cause of the difference between the mean and equilibrium flapping positions. Vortex shedding is associated with an increase in the mean forces on the flag compared with the equilibrium state (which is devoid of vortex shedding; \emph{c.f.}, figure \ref{fig:DEP_Re200}). To demonstrate this increase in fluid forces, we ran a simulation with the flag fixed in the deformed equilibrium position corresponding to $K_B = 0.1$. With the flag fixed in this position, the bluffness of the body causes the flow to enter limit-cycle vortex shedding with a mean lift and drag of 0.356 and 0.583, respectively. By contrast, when the fully-coupled system is in the deformed equilibrium and vortex shedding is absent, the lift and drag forces are 0.192 and 0.344, respectively. The increase in mean forces causes a corresponding increase in flag deflection, and thus in the nonlinear simulations flapping occurs about a mean position that is raised from the equilibrium state.

We emphasise that this vortex shedding mode is stable for small-deflection deformed flapping and large-amplitude flapping found at higher stiffnesses (and studied in the previous sections). For example, the growth rate of the vortex shedding mode for the large-amplitude flapping parameters $K_B = 0.2, M_\rho = 0.5$ is -0.392. The vortex-shedding mode is therefore not the cause of instability of the deformed equilibrium in those regimes. 


\subsection{Chaotic flapping}
\label{sec:chaos_Re200}

\begin{figure}
	\centering
	\begin{subfigure}[b]{0.45\textwidth}
		\centering
    		\includegraphics[scale=0.4,trim={0cm 0cm 0cm 0cm},clip]{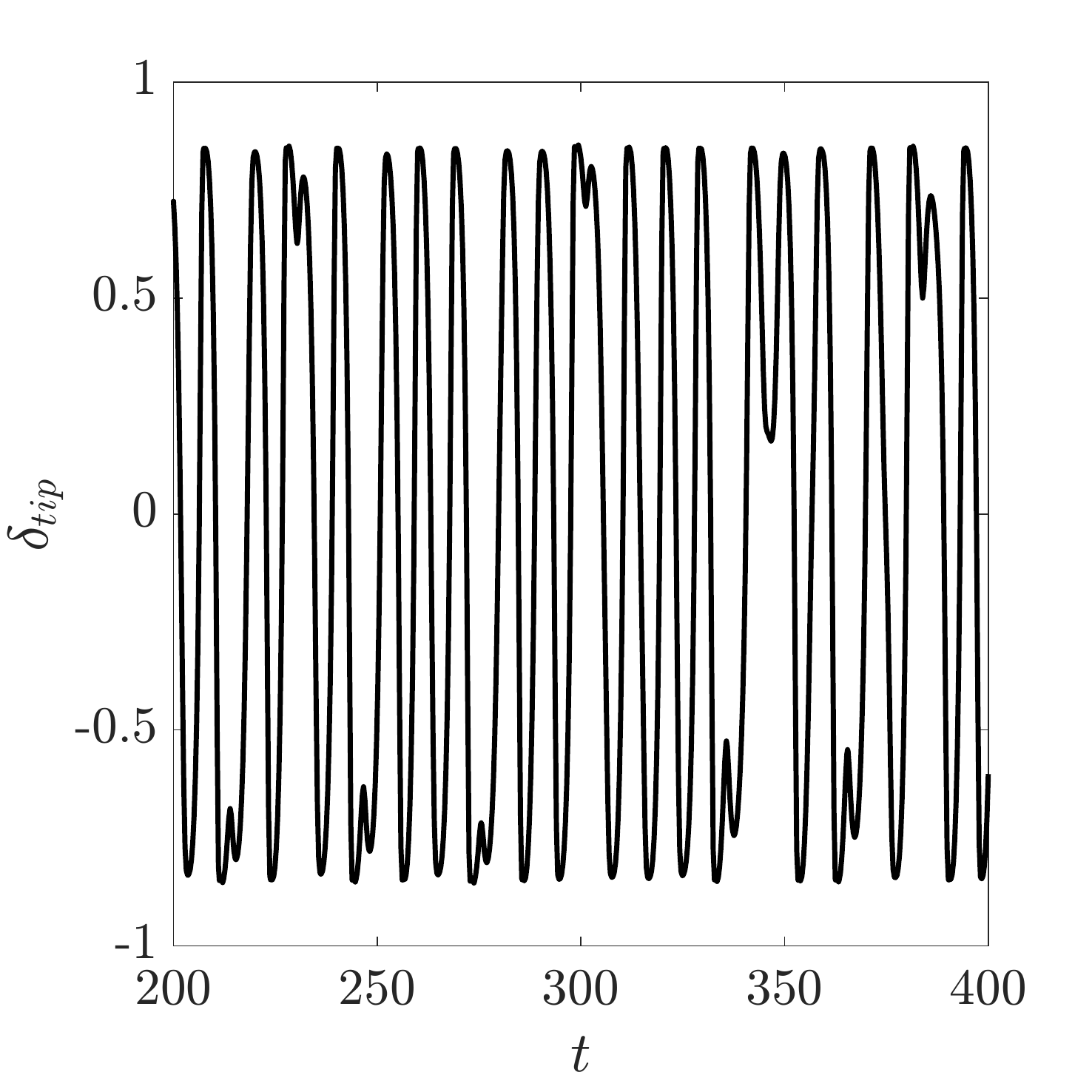}
	\end{subfigure}
	\begin{subfigure}[b]{0.45\textwidth}
		\centering
    		\includegraphics[scale=0.4,trim={0cm 0cm 0cm 0cm},clip]{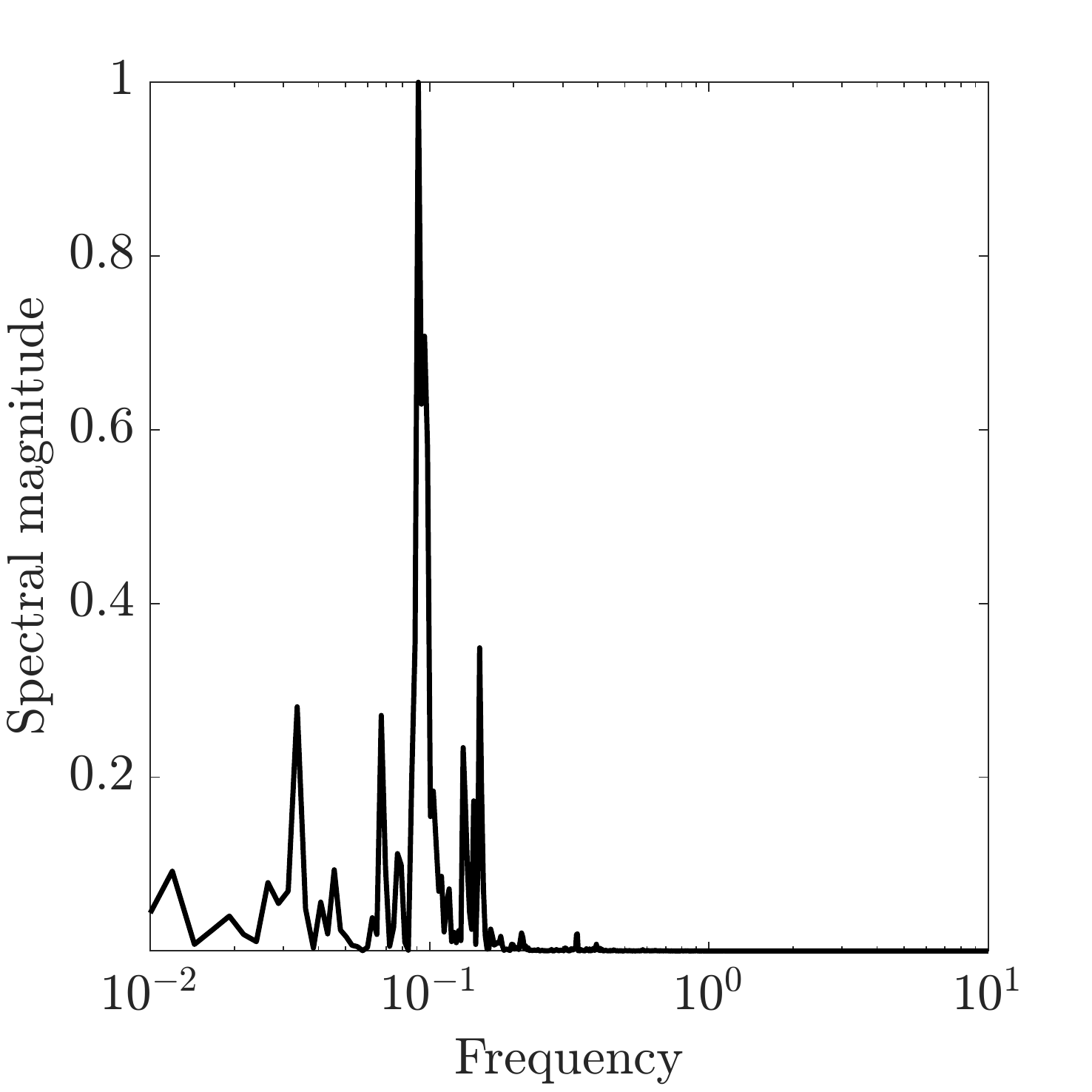}
	\end{subfigure}
	\caption{Tip displacement as a function of time (left) and spectral density of tip displacement (right) for a flag in the chaotic flapping regime, with $M_\rho = 0.05, K_B = 0.17, Re = 200$. The left plot shows a zoomed in version of the total time window used for the spectral density computation, $t\in[20,500]$.}
	\label{fig:chaotic_disp}
\end{figure}

For sufficiently light flags ($M_\rho \le 0.5$ in our studies), large-amplitude flapping (region IV in the bifurcation diagrams of figure \ref{fig:bif_Re200}) bifurcates to chaotic flapping (region V) before entering into deflected-mode flapping (region VI). Figure \ref{fig:chaotic_disp} shows that the time trace of the tip displacement is aperiodic and associated with broadband frequency content. 

To demonstrate mathematically that this behaviour is chaotic, we compute the Lyapunov exponent of the system using the time-delay method of \cite{Wolf1985}  (this approach was also used by \cite{Connell2007} to identify chaotic flapping of conventional flags). The method computes an approximation of the distance in time, $d(t)$, of two trajectories starting close to one another at an initial time $t_0$. The evolution of this distance is written as
\begin{equation}
d(t) = d(t_0) e^{\gamma (t - t_0)}
\end{equation}
where $\gamma$ is the Lyapunov exponent that represents the departure or convergence of the two trajectories. A zero value of $\gamma$ corresponds to a stationary state where the system is in limit cycle behaviour; a positive value of $\gamma$ corresponds to divergence of the two trajectories, and thus to chaotic flapping. Table \ref{tab:Lyap_exp} shows the Lyapunov exponent computed for various values of $M_\rho$ and $K_B$. For large-amplitude and deflected-mode flapping, the exponent is approximately zero, coincident with limit-cycle flapping. In the chaotic regime that occurs at stiffnesses between large-amplitude and deflected-mode flapping, the exponent is positive and larger by an order of magnitude, indicative of a transition to chaotic behaviour in this regime.

\begin{table}
    \centering
    \begin{tabular}{ c c c c }
    $M_\rho$ & $K_B$ & {Lyapunov exponent ($\gamma$)} & Flapping regime  \\ \hline
     0.05 & 0.35 & -0.0012 & Large-amplitude flapping (region IV) \\
     0.05 & 0.17 & 0.068 & Chaotic flapping (region V) \\ 
     0.05 & 0.08 & -0.0023 & Deflected-mode (region VI) \\
     0.5 & 0.35 & -0.0015 & Large-amplitude flapping (region IV) \\
     0.5 & 0.17 & 0.059 & Chaotic flapping (region V) \\ 
     0.5 & 0.08 & 0.0009 & Deflected-mode (region VI)
    \end{tabular}
     \caption{Lyapunov exponents for different flapping regimes. For each regime, the corresponding region from the bifurcation diagram of figure \ref{fig:bif_Re200} is indicated in parentheses.}
\label{tab:Lyap_exp}
\end{table}

The bifurcation diagrams in figure \ref{fig:bif_Re200} demonstrate that increasing mass reduces the chaotic behaviour. In moving from $M_\rho = 0.05$ to $M_\rho = 0.5$, there were certain stiffnesses within the chaotic flapping regime that exhibited periodic flapping instead of chaotic flapping (see figure \ref{fig:bif_Re200}). We believe that this is an artifact of only running the simulations for finite time, but the absence of chaotic flapping over a minimum of 55 flapping periods for certain stiffnesses at $M_\rho = 0.5$ speaks to the effect of increasing inertia on reducing the chaotic behaviour. For the heavier flag cases of $M_\rho = 5, 50$, chaotic flapping disappears altogether. Thus, chaotic flapping is only associated with mass ratios ($M_\rho$) for which VIV flapping occurs. 

\begin{figure}
	\centering
	\begin{subfigure}[b]{0.32\textwidth}
		\centering
    		\includegraphics[scale=0.35,trim={0cm 0cm 0cm 0cm},clip]{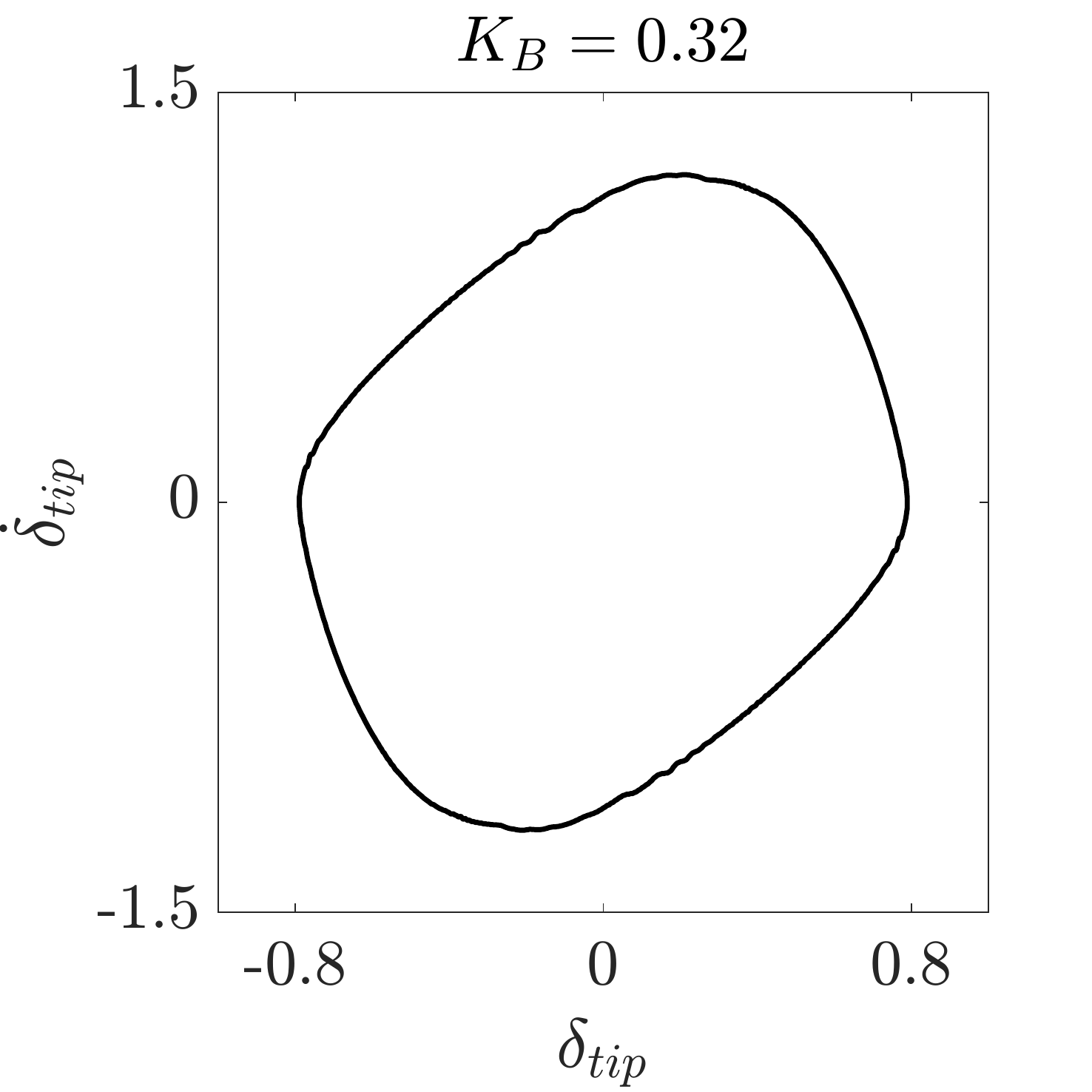}
	\end{subfigure}
	\begin{subfigure}[b]{0.32\textwidth}
		\hspace*{4.7mm}
		\centering
    		\includegraphics[scale=0.35,trim={2.9cm 0cm 0cm 0cm},clip]{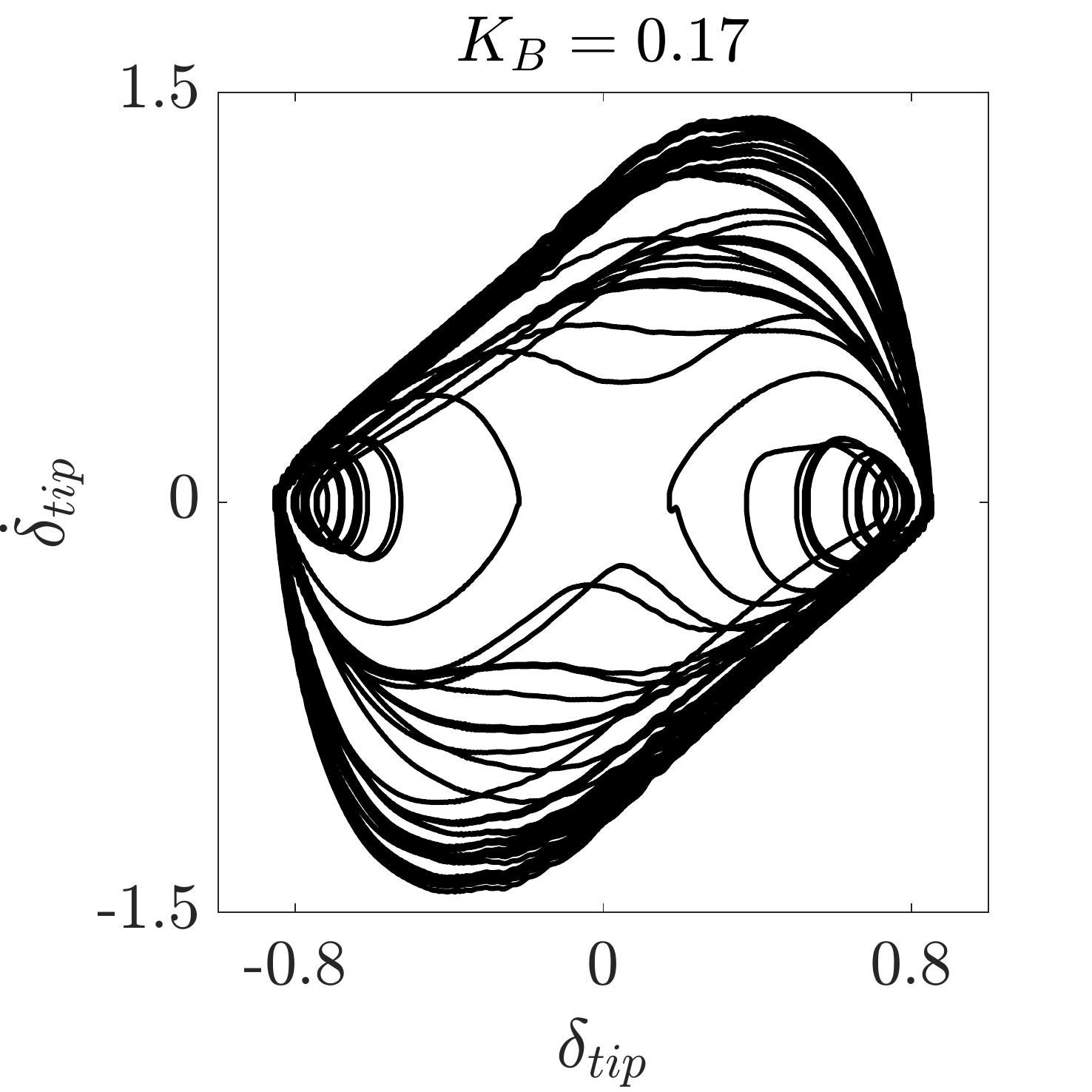}
	\end{subfigure}
	\begin{subfigure}[b]{0.32\textwidth}
		\centering
    		\includegraphics[scale=0.35,trim={2.8cm 0cm 0cm 0cm},clip]{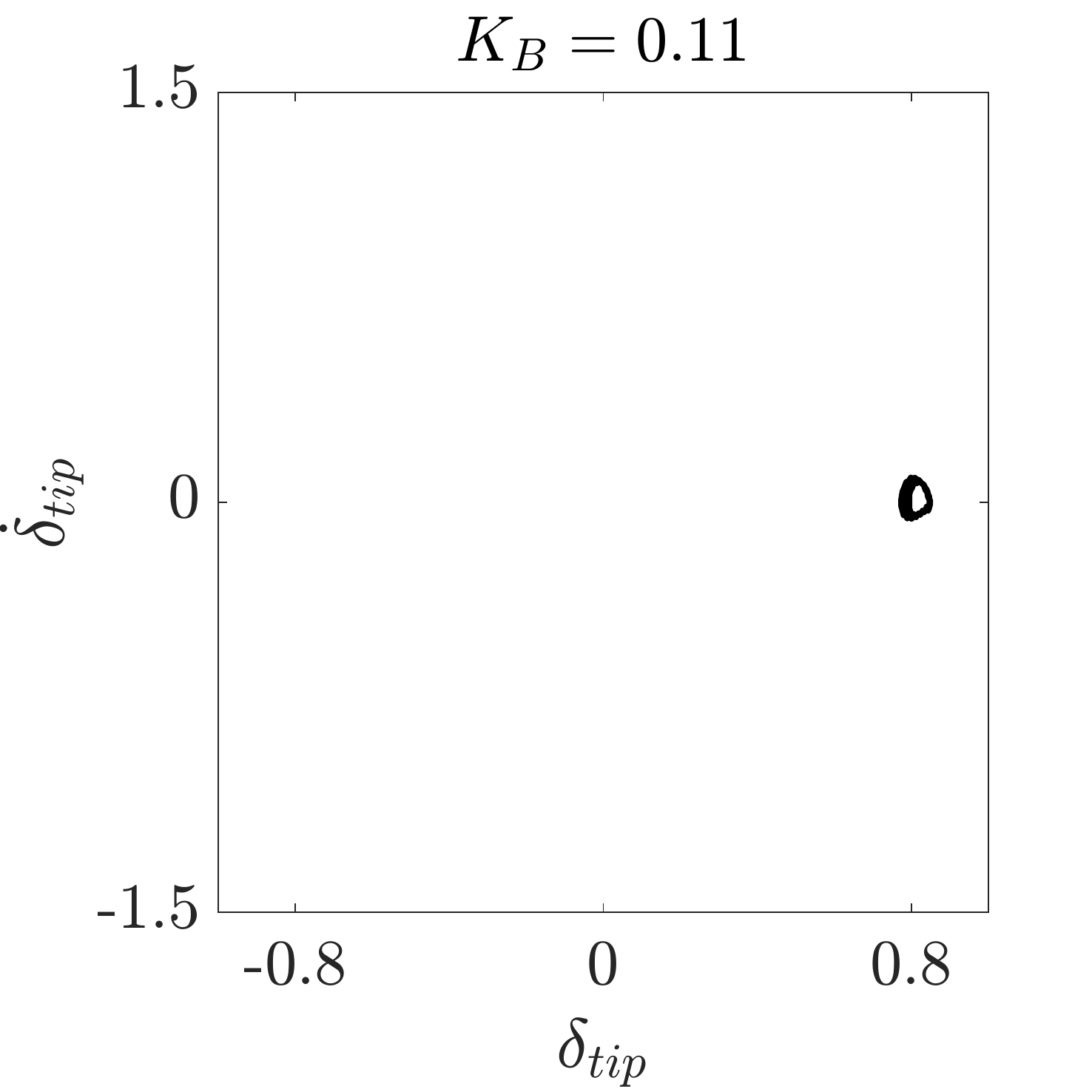}
	\end{subfigure}
	\caption{Tip velocity-tip displacement phase portraits for $M_\rho = 0.05$ and $K_B = 0.32$ (left), $K_B = 0.17$ (middle), and $K_B = 0.11$ (right).}
	\label{fig:chaotic_phase}
\end{figure}

To elucidate the nature of chaotic flapping, we show in figure \ref{fig:chaotic_phase} phase portraits of tip velocity versus tip displacement for inverted flags in the large-amplitude flapping, chaotic flapping, and deflected-mode regimes. The figures demonstrate that the chaotic flapping phase portrait contains both the large periodic orbit of large-amplitude flapping and the small-amplitude large-deflection periodic orbit of deflected-mode flapping. Thus, chaotic flapping is a regime in which large-amplitude flapping and the deflected mode hybridise to form a new strange attractor involving both states. The chaotic nature of the regime is associated with the apparent randomness in switching between these two orbits. 

\section{Dynamics for $Re = 20$}
\label{sec:Re_20}

Previous numerical simulations of \citet{Ryu2015} demonstrated the absence of flapping for flags with $M_\rho \le O(1)$ and $Re < 50$. We now consider $Re =20$ to investigate the stability and dynamics of the inverted-flag system below this previously identified critical Reynolds number. In agreement with \citet{Ryu2015}, we find that light flags with $M_\rho = 0.05, 0.5$ do not flap. Heavy flags with $M_\rho = 5, 50$ are shown to undergo both small-deflection deformed flapping and large-amplitude flapping. Such behaviour has yet to be reported, and we demonstrate that for this heavy flag case neither flapping regime is a VIV. As was observed for $Re = 200$, small-deflection flapping is caused by a supercritical Hopf bifurcation of the deformed equilibrium associated with the transition to instability of the least damped mode of the deformed equilibrium. Large-amplitude flapping is characterised by an increase in saturation amplitude of small-deflection flapping until eventually the flag swings past the centreline and begins a process where it samples both deflected equilibria. Finally, we show that at this low Reynolds number the deflected-mode state is not associated with flapping, and is instead a formal stable equilibrium of the fully-coupled system.

\subsection{Bifurcation diagrams and general observations}

\begin{figure}
    \begin{subfigure}[b]{0.5\textwidth}
        \includegraphics[scale=0.5,trim={0cm 1.7cm 0cm 0cm},clip]{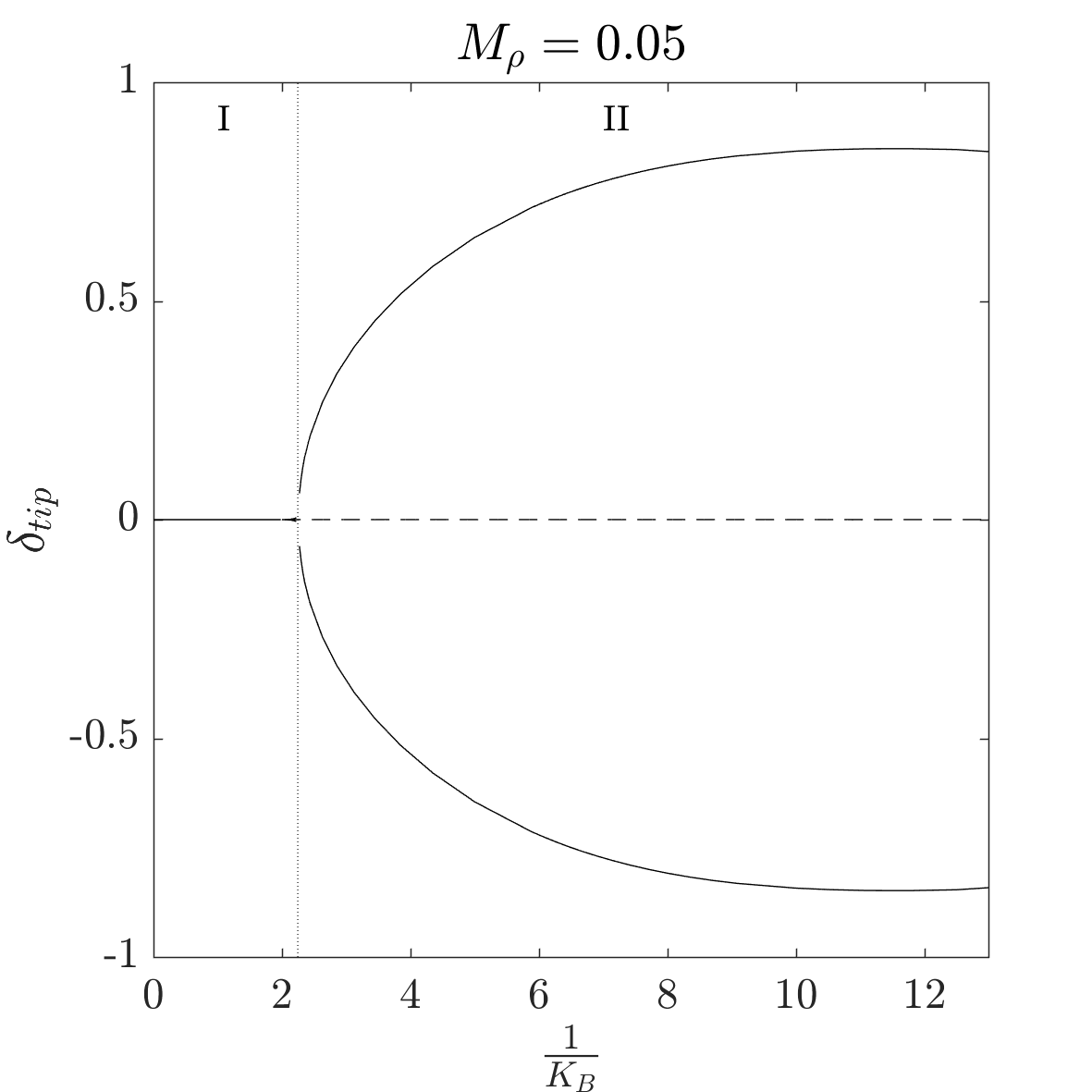}
    \end{subfigure}
    \hspace*{3mm}
    \begin{subfigure}[b]{0.5\textwidth}
        \includegraphics[scale=0.5,trim={2cm 1.7cm 0cm 0cm},clip]{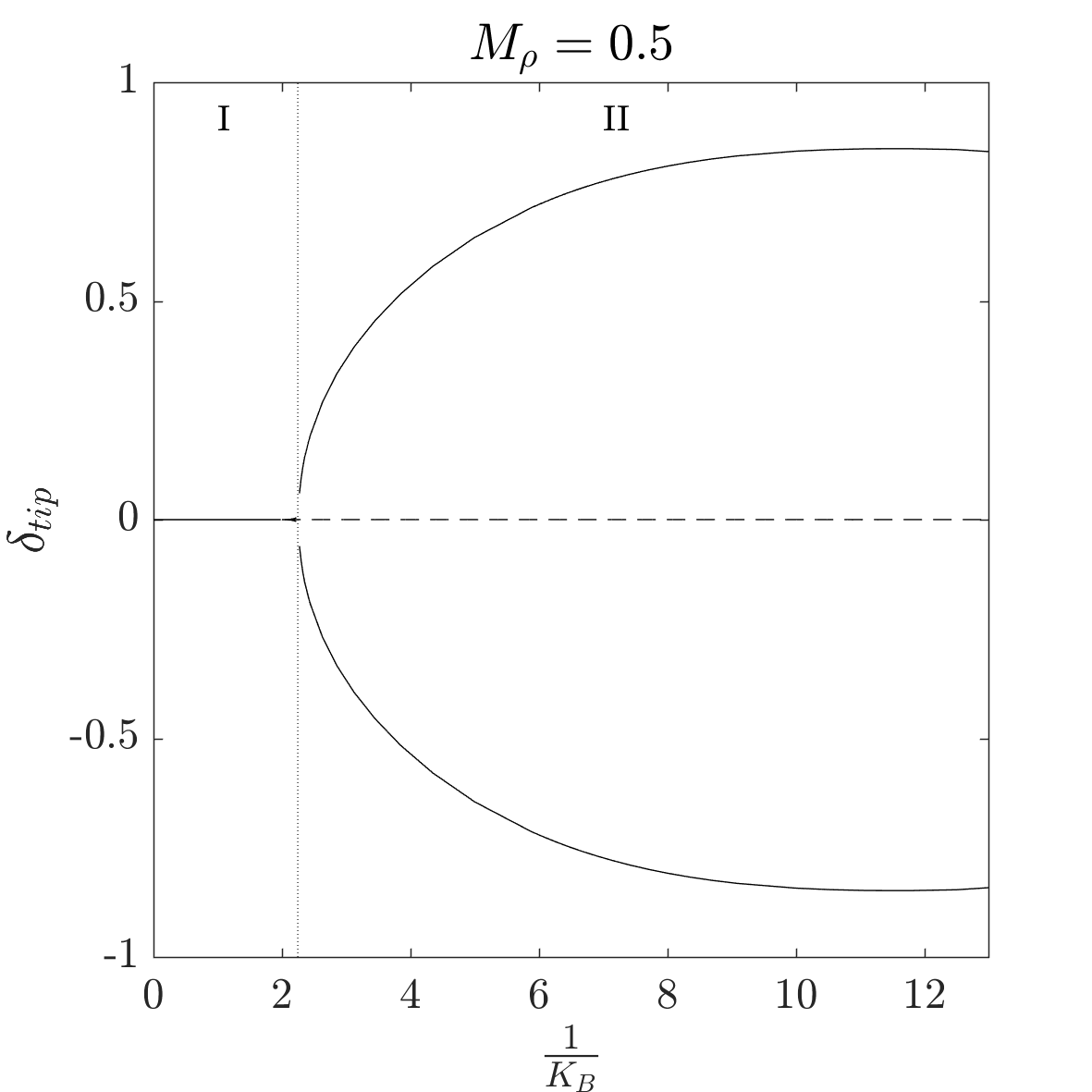}
    \end{subfigure}

	\begin{subfigure}[b]{0.5\textwidth}
	\vspace*{2mm}
        \includegraphics[scale=0.5,trim={0cm 0cm 0cm 0cm},clip]{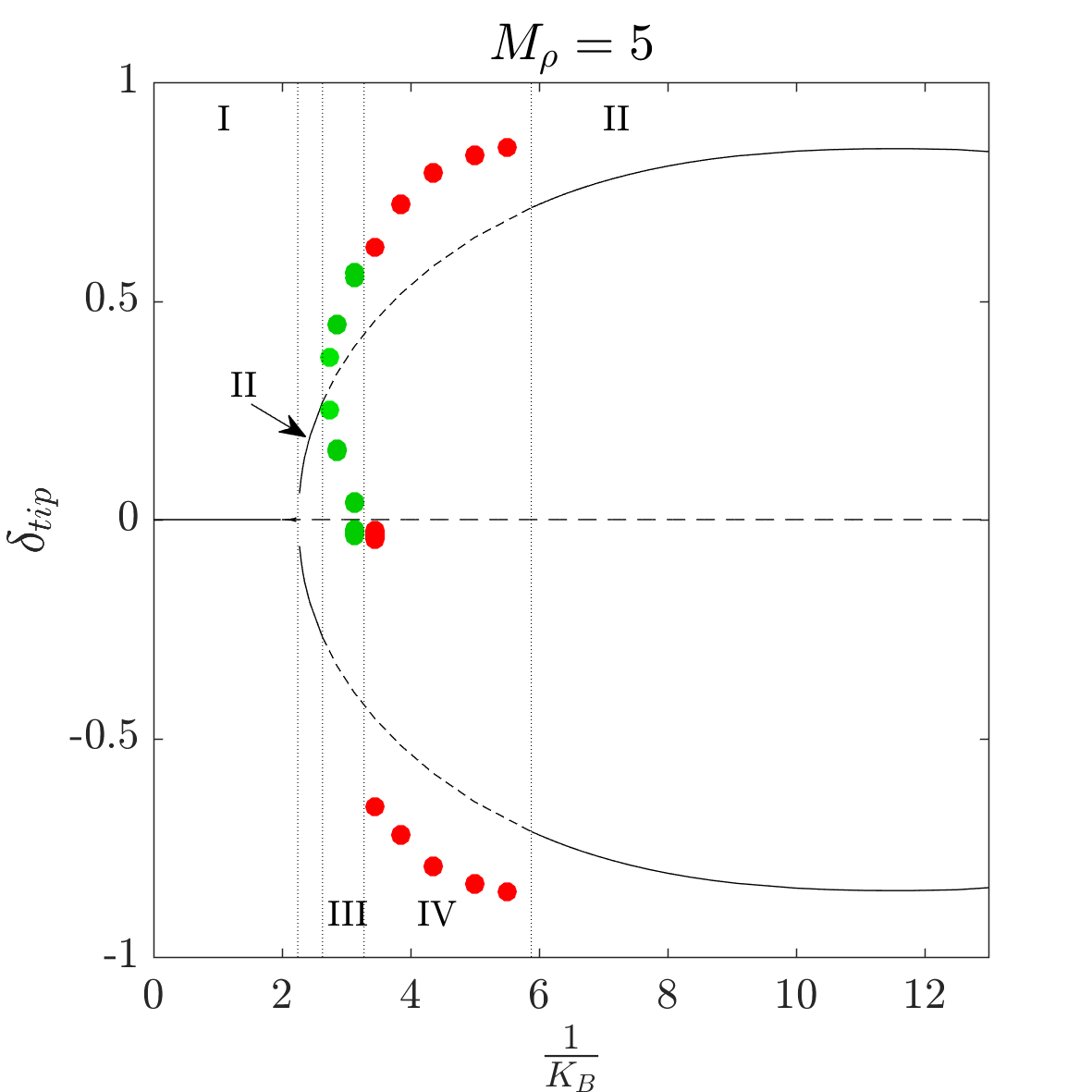}
    \end{subfigure}
        \hspace*{3mm}
	\begin{subfigure}[b]{0.5\textwidth}
       \vspace*{2mm}
	 \includegraphics[scale=0.5,trim={2cm 0cm 0cm 0cm},clip]{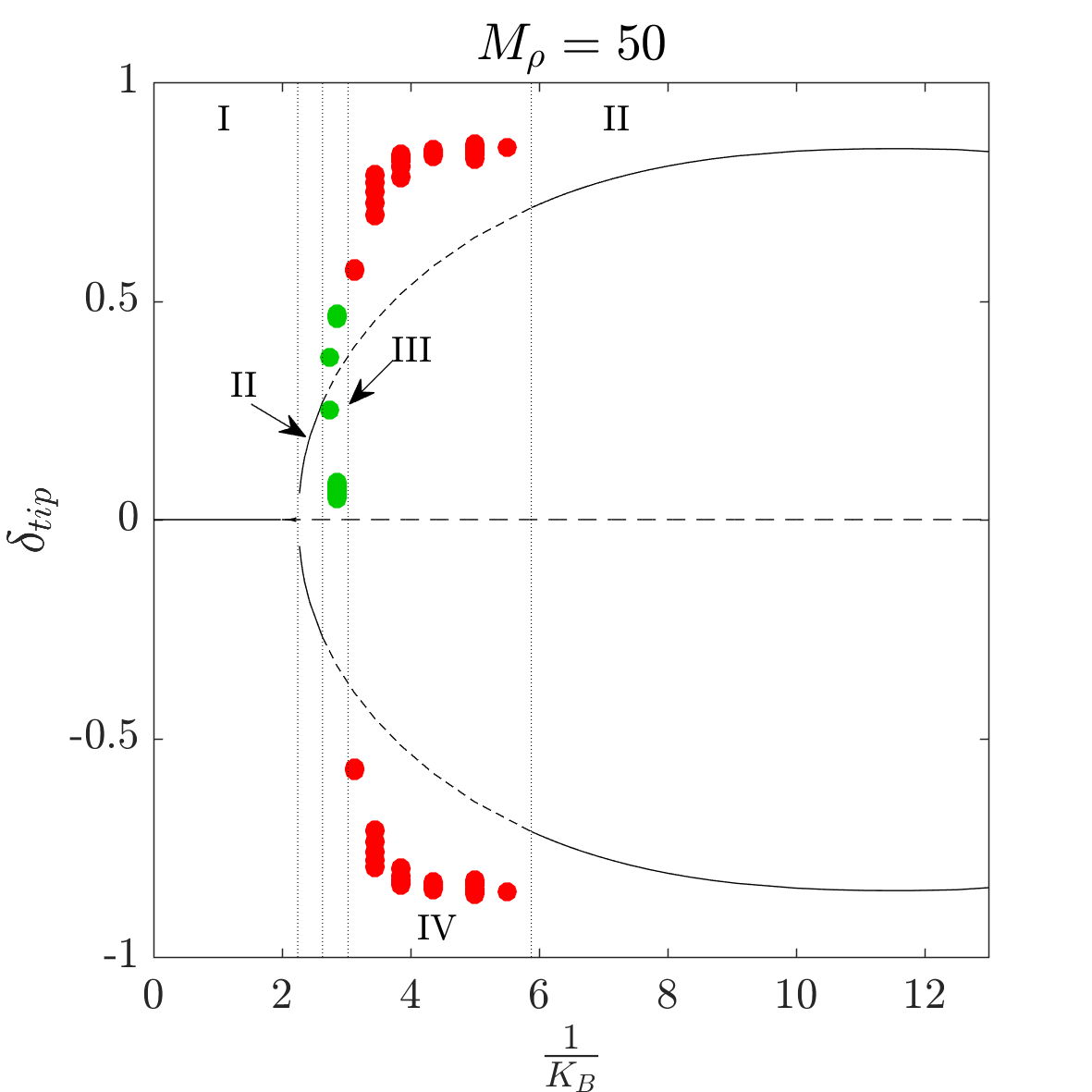}
    \end{subfigure}
       \caption{Bifurcation diagrams of inverted-flag dynamics at $Re = 20$ that show leading edge transverse displacement (tip deflection, $\delta_{tip}$) versus inverse stiffness ($1/K_B$). I: undeformed equilibrium, II: deformed equilibrium, III: small-deflection deformed flapping, IV: large-amplitude flapping. See the main text for a description of the various lines and markers and details on how the diagrams were constructed.}
    \label{fig:bif_Re20}
\end{figure}

Figure \ref{fig:bif_Re20} gives bifurcation diagrams of the inverted-flag system at four different masses. These figures were plotted as described in section \ref{sec:Re_200}. The bifurcation diagrams reveal four distinct regimes: a stable undeformed equilibrium (I), a stable deformed equilibrium (II), small-deflection deformed flapping (III), and large-amplitude flapping (IV). 

While many of the same bifurcations found at $Re = 200$ remain for $Re = 20$, there are also distinctions between them that are visible through the bifurcation diagrams. First, flapping does not occur for all masses considered at this lower Reynolds number, which demonstrates the stabilising effect of fluid diffusion for inverted-flag dynamics. Second, the deflected-mode state no longer corresponds to flapping at this lower Reynolds number, and is instead a formal stable equilibrium of the fully-coupled fluid-structure equations of motion. Finally, chaotic flapping does not occur for any of the considered values of $M_\rho$ at this lower Reynolds number.

As was seen for $Re = 200$, the divergence instability of the undeformed equilibrium (caused by decreasing $K_B$) leads to a stable deformed equilibrium that is independent of the mass ratio, $M_\rho$. This follows from the fact that the deformed equilibria are steady state solutions and therefore do not depend on flag inertia.

\begin{figure}
	\begin{subfigure}[b]{0.245\textwidth}
        		\includegraphics[scale=0.28]{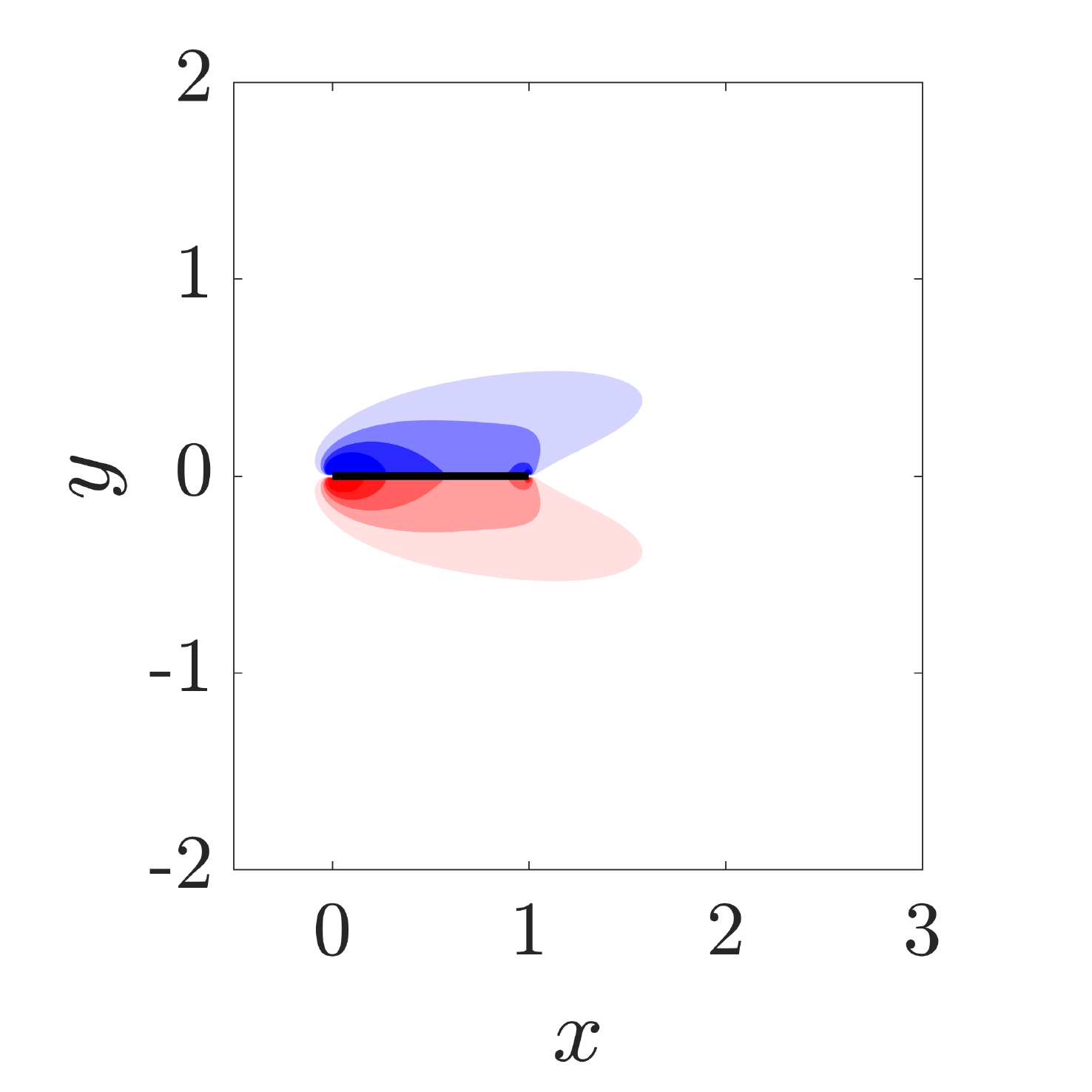}
	\end{subfigure}
	\begin{subfigure}[b]{0.245\textwidth}
		\hspace*{4.7mm}
        		\includegraphics[scale=0.28,trim={2.95cm 0cm 0cm 0cm},clip]{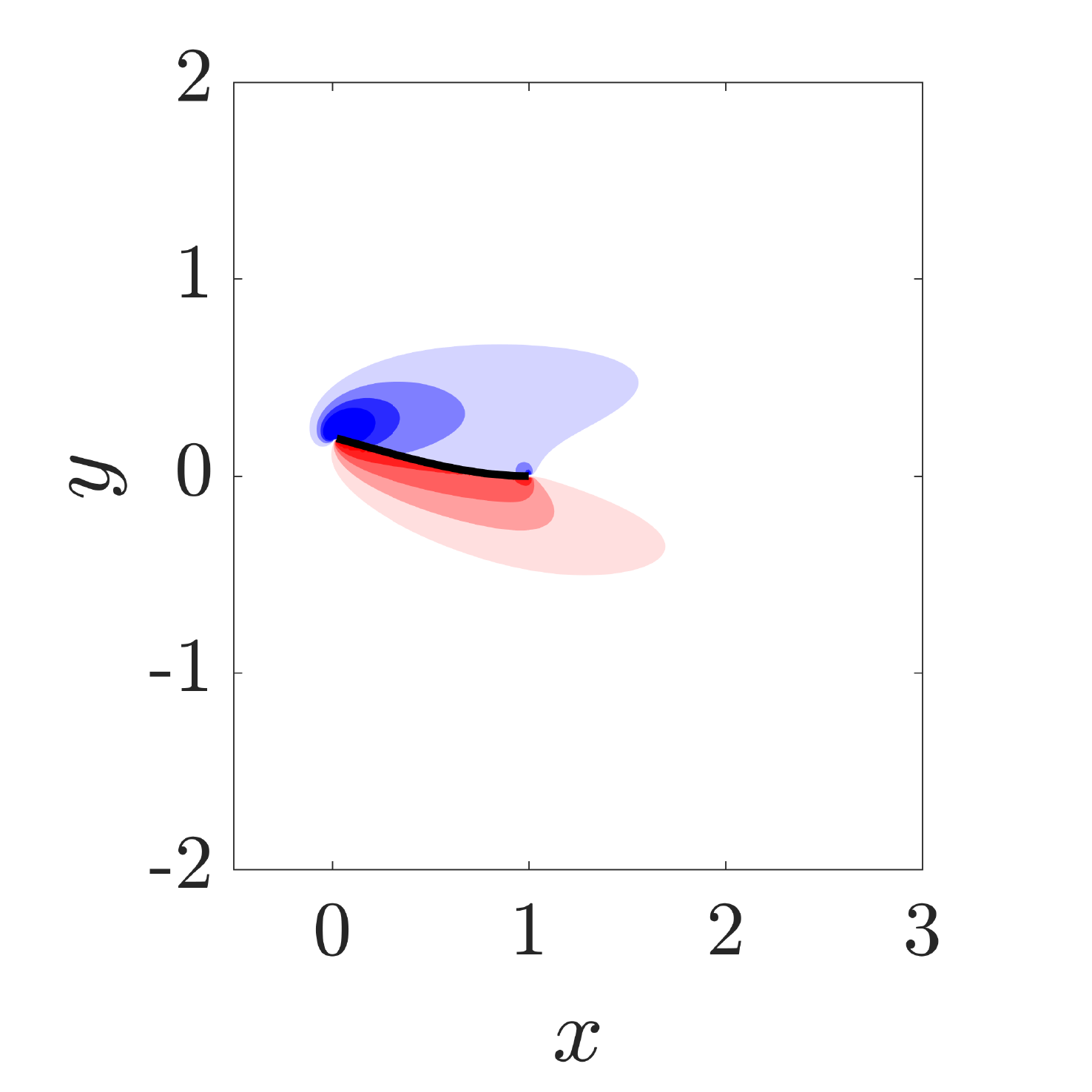}
	\end{subfigure}
    	\begin{subfigure}[b]{0.245\textwidth}
        		\hspace*{2mm}
        		\includegraphics[scale=0.28,trim={2.95cm 0cm 0cm 0cm},clip]{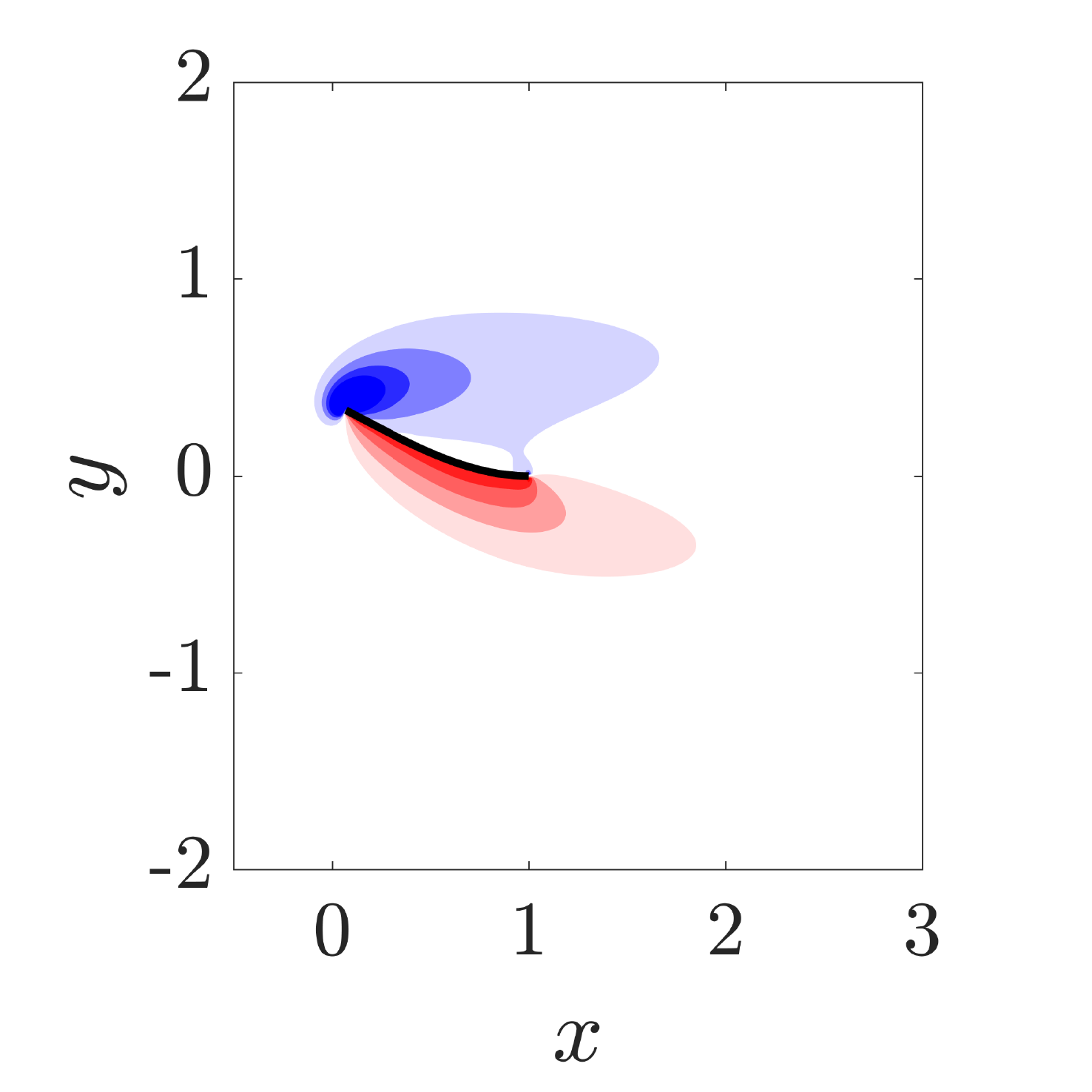}
	\end{subfigure}
	\begin{subfigure}[b]{0.245\textwidth}
        		\includegraphics[scale=0.28,trim={2.95cm 0cm 0cm 0cm},clip]{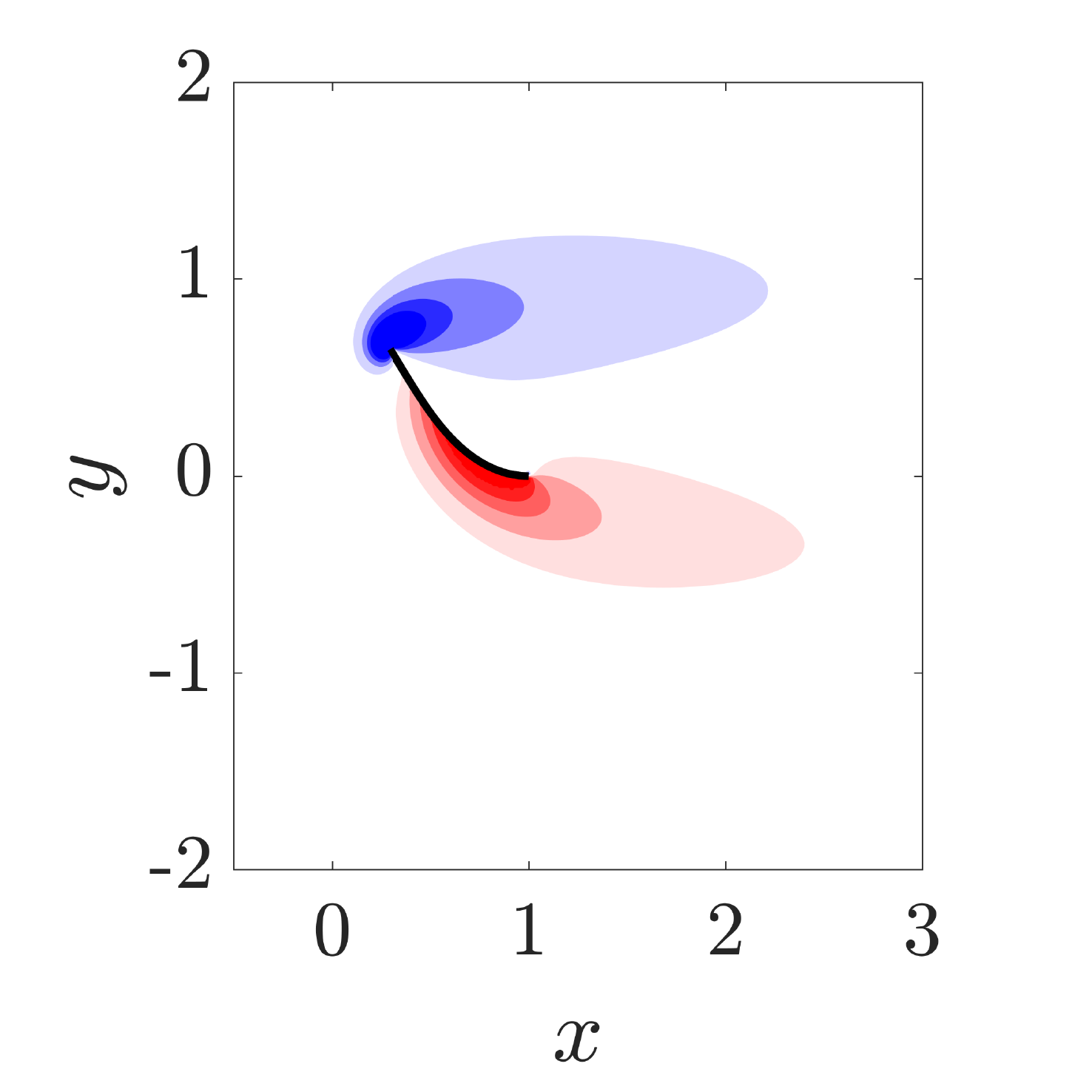}
	\end{subfigure}
    \caption{Vorticity contours for equilibrium states of the inverted-flag system at $Re = 20$. From left to right: $K_B = 0.5, 0.41, 0.35, 0.2$.  The two rightmost equilibria are unstable for $M_\rho = 5, 50$. Contours are in 18 increments from -5 to 5.}
	\label{fig:DEP}
\end{figure}

\begin{figure}
    	\begin{subfigure}[b]{0.5\textwidth}
	\vspace*{2mm}
        \includegraphics[scale=0.5,trim={0cm 0cm 0cm 0cm},clip]{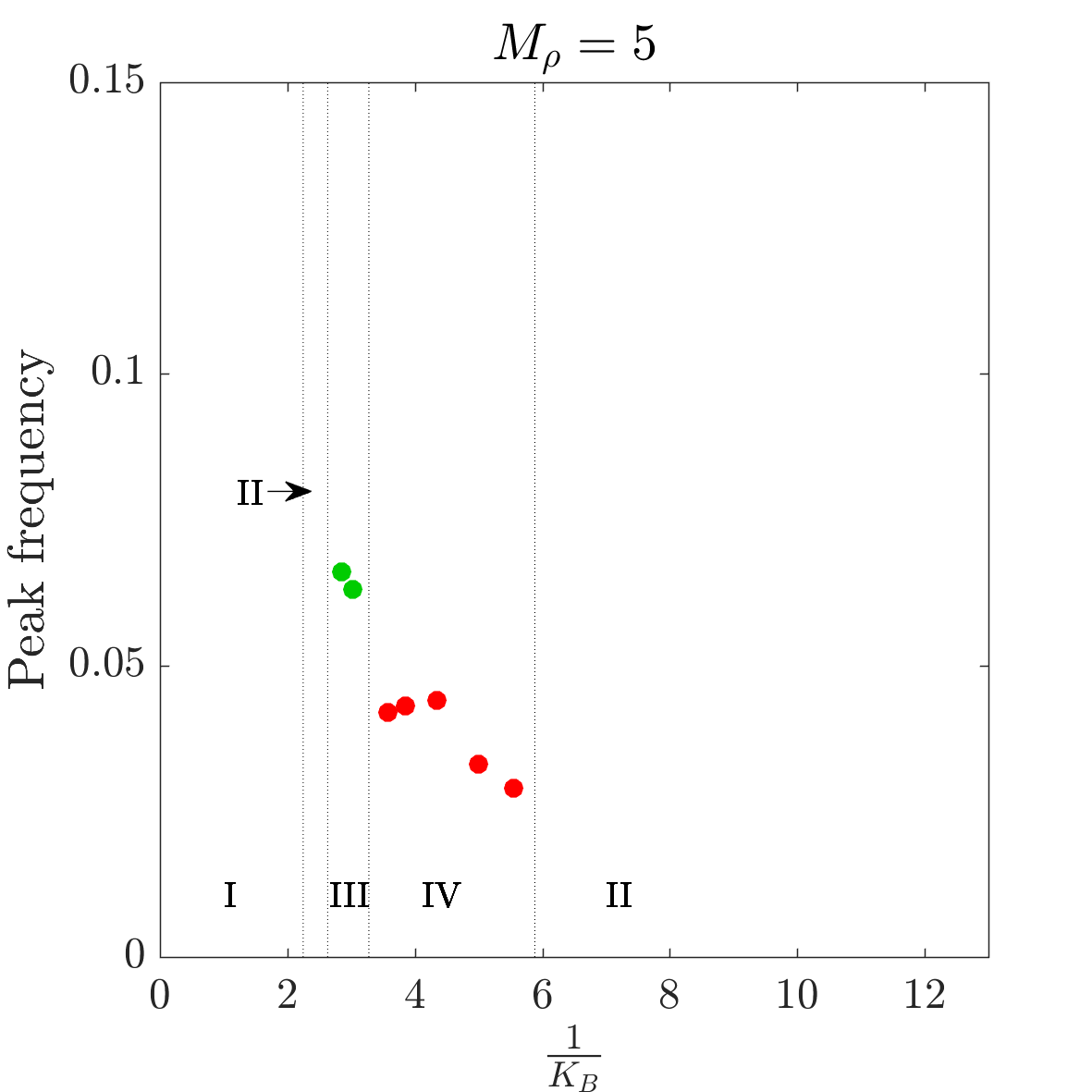}
    \end{subfigure}
        \hspace*{3mm}
	\begin{subfigure}[b]{0.5\textwidth}
       \vspace*{2mm}
	 \includegraphics[scale=0.5,trim={2.1cm 0cm 0cm 0cm},clip]{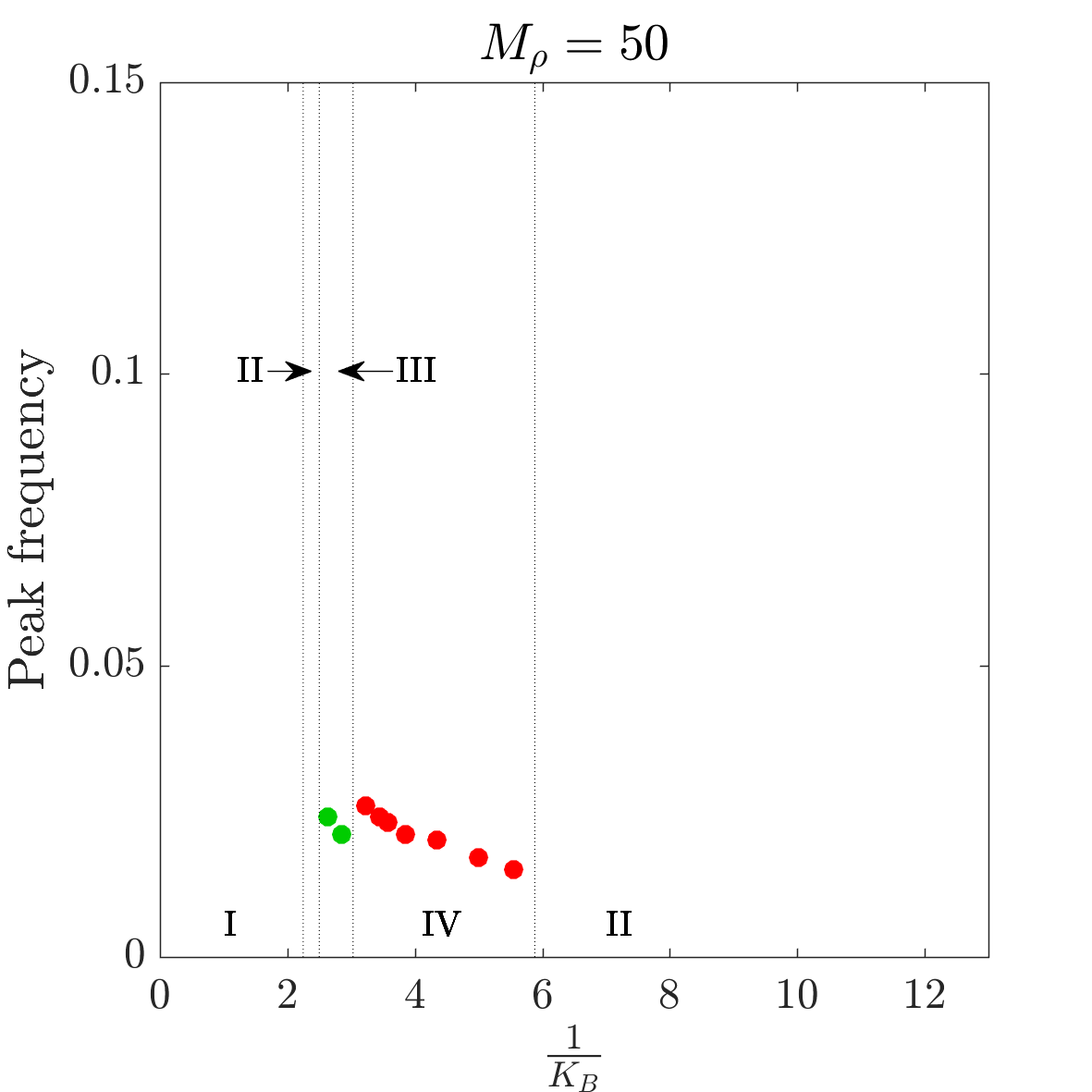}
    \end{subfigure}
       \caption{Markers: peak flapping frequency at $Re = 20$ for the parameters corresponding to the bifurcation diagrams shown in figure \ref{fig:bif_Re20}. }
    \label{fig:freq_Re20}
\end{figure}

As stiffness is decreased, light flags remain in this deformed equilibrium regime---no flapping occurs at this Reynolds number for $M_\rho =0.05, 0.5$. Moreover, since the equilibrium states do not depend on flag inertia, their bifurcation diagrams are identical. By contrast, with decreasing stiffness heavy flags transitioned from the deformed equilibrium to (respectively) small-deflection flapping and large-amplitude flapping before returning at even lower stiffnesses to a stable deformed equilibrium. To demonstrate the non-VIV nature of flapping at this low Reynolds number, we show in figure \ref{fig:freq_Re20} the peak flapping frequency for the parameters corresponding to the bifurcation diagrams in figure \ref{fig:bif_Re20}. For all cases, the flapping frequency from the nonlinear simulations (denoted by the markers) is substantially different from the bluff-body vortex-shedding frequency. In the remainder of this section we explore the physical mechanisms behind the various regimes and the transitions between them.

\subsection{Small-deflection deformed flapping}

We show in table \ref{tab:Hopf_Re20} that the transition from the deformed equilibrium to small-deflection deformed flapping is associated with the least damped global mode of the deformed equilibrium becoming unstable. Thus, as was seen for $Re = 200$, small-deflection deformed flapping is a supercritical Hopf bifurcation of the deformed equilibrium state.  Table \ref{tab:Hopf_Re20} also shows that the corresponding eigenvalue accurately predicts the flapping frequency of the nonlinear simulations near the stability boundary. 

\begin{table}
\centering
\begin{tabular}{ c c c c c }
\multirow{2}{*}{$M_\rho$} & \multirow{2}{*}{$K_B$} & \multicolumn{2}{ c }{Leading mode} & {Peak frequency of } \\ 
                        &                  & \scriptsize{Growth rate} & \scriptsize{Frequency}                                &    nonlinear simulation              \\ \hline
 5 & 0.374 & -0.0061 & 0.083 & N/A (stable equilibrium) \\
 5 & 0.371 & 0.0039 & 0.082 & 0.080 \\ 
 50 & 0.40 & -0.003 & 0.031 & N/A (stable equilibrium) \\
 50 & 0.397 & 0.001 & 0.030 & 0.030
\end{tabular}
 \caption{Growth rate and frequency of the leading global mode of the deformed equilibrium compared with nonlinear behaviour for parameters near the onset of small-deflection deformed flapping.}
\label{tab:Hopf_Re20}
\end{table}

\begin{figure}
	\centering
	\begin{subfigure}[b]{0.45\textwidth}
		\centering
        		\includegraphics[scale=0.3, trim={0cm 1cm 0cm 0cm},clip]{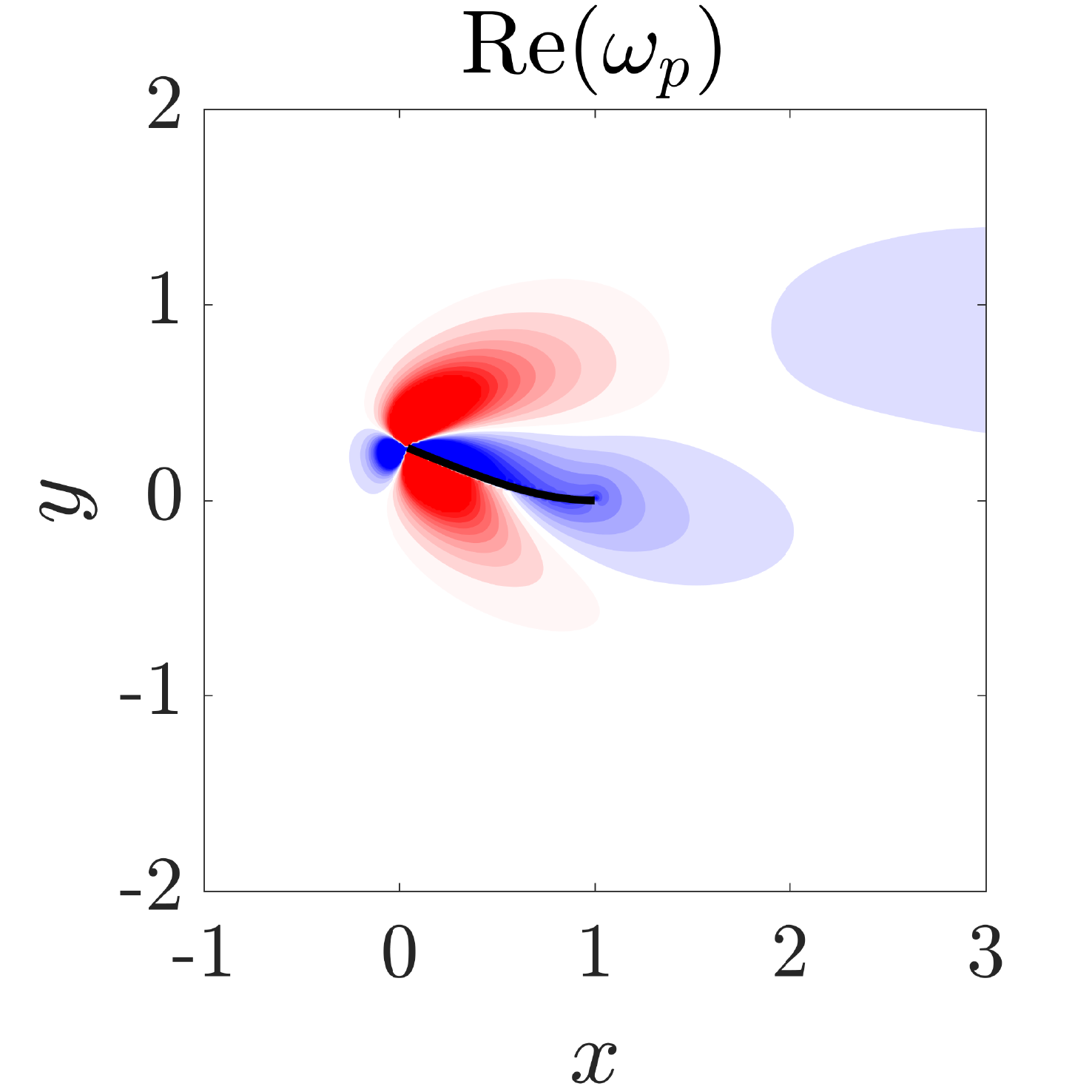}
	\end{subfigure}
	\begin{subfigure}[b]{0.45\textwidth}
		\centering
        		\includegraphics[scale=0.305, trim={0cm 1cm 0cm 0cm},clip]{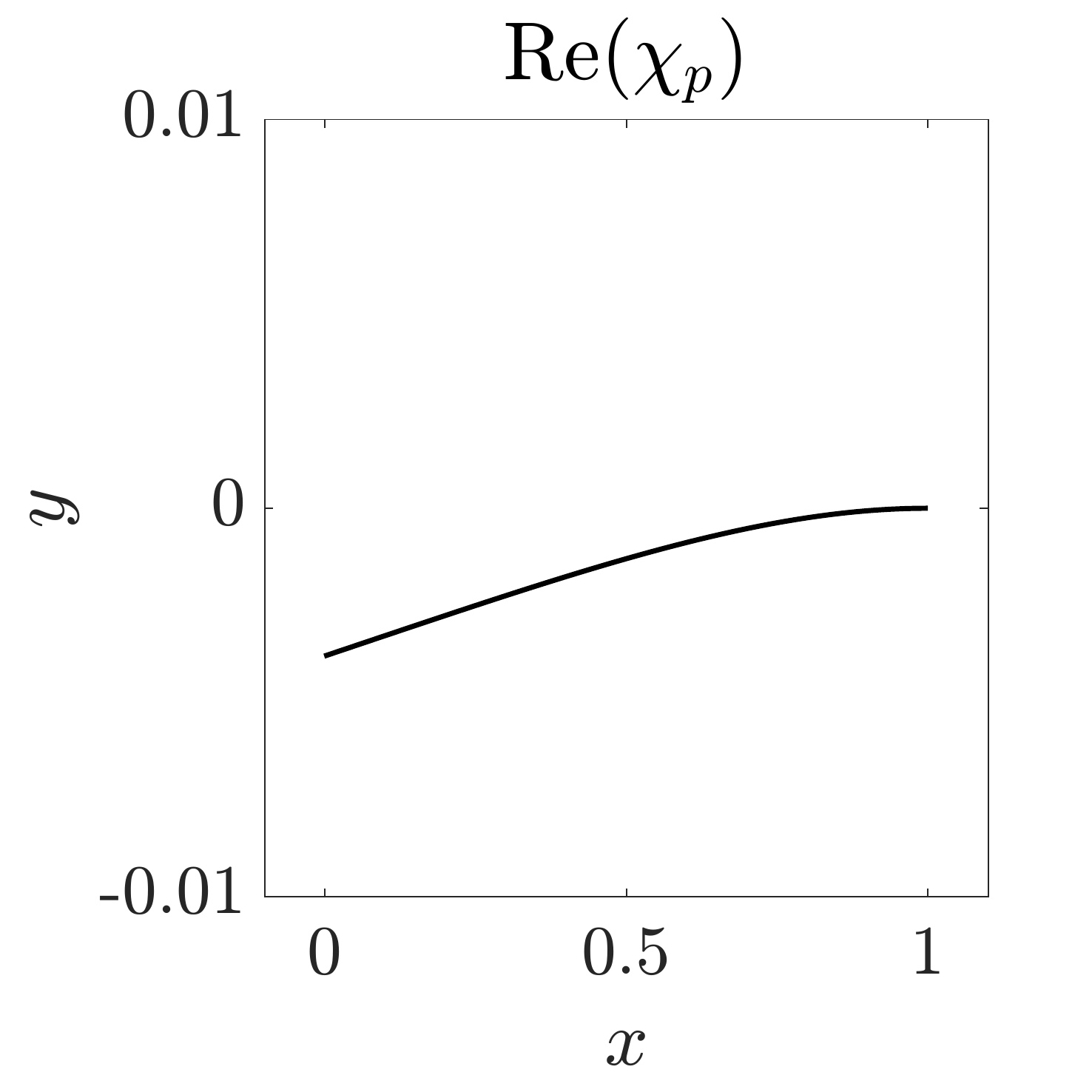}
		\vspace*{-0.3mm}
	\end{subfigure}
	
    	\begin{subfigure}[b]{0.45\textwidth}
		\vspace*{3mm}
		\centering
        		\includegraphics[scale=0.3]{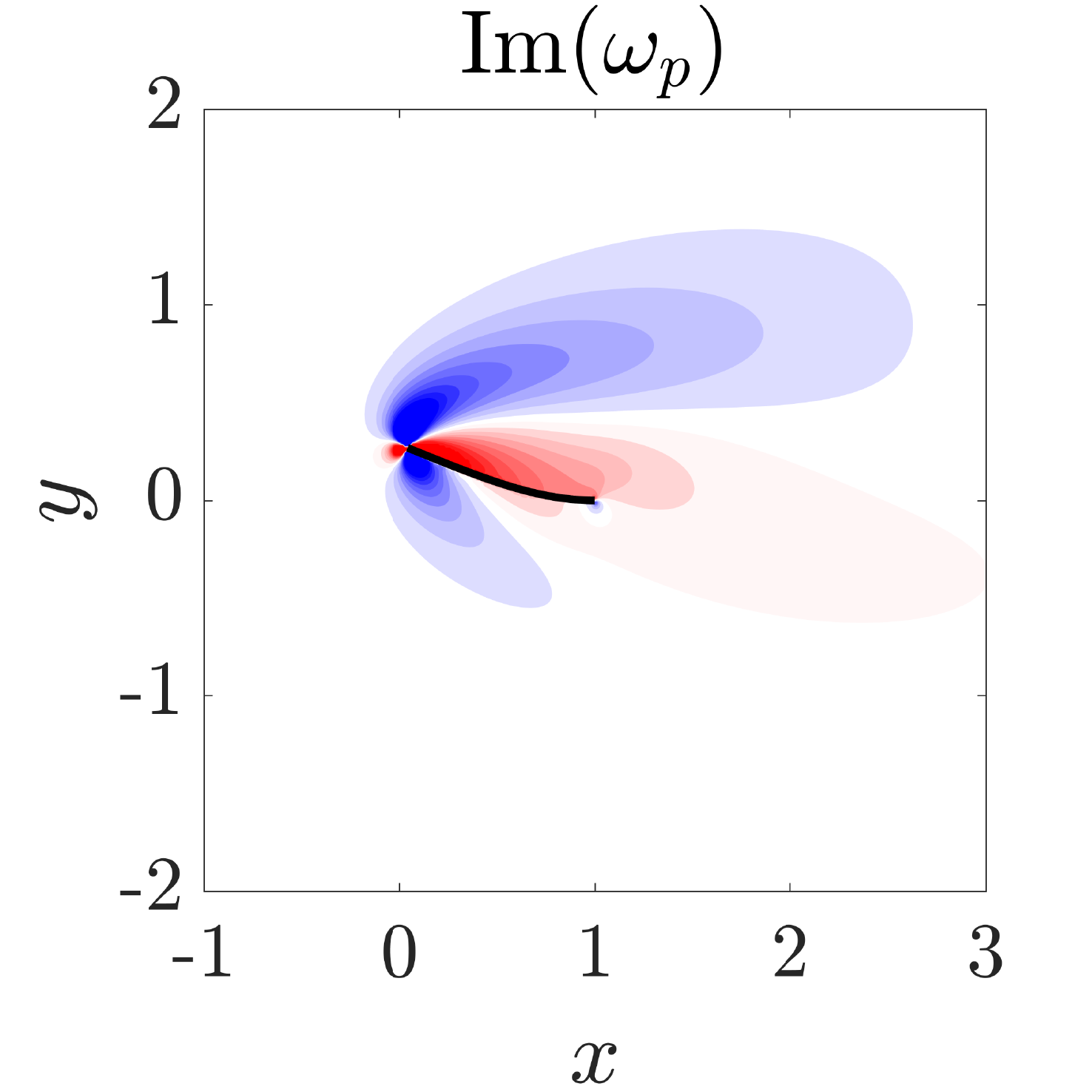}
	\end{subfigure}
	\begin{subfigure}[b]{0.45\textwidth}
		\vspace*{3mm}
		\centering
        		\includegraphics[scale=0.305, trim={0cm 0cm 0cm 0cm},clip]{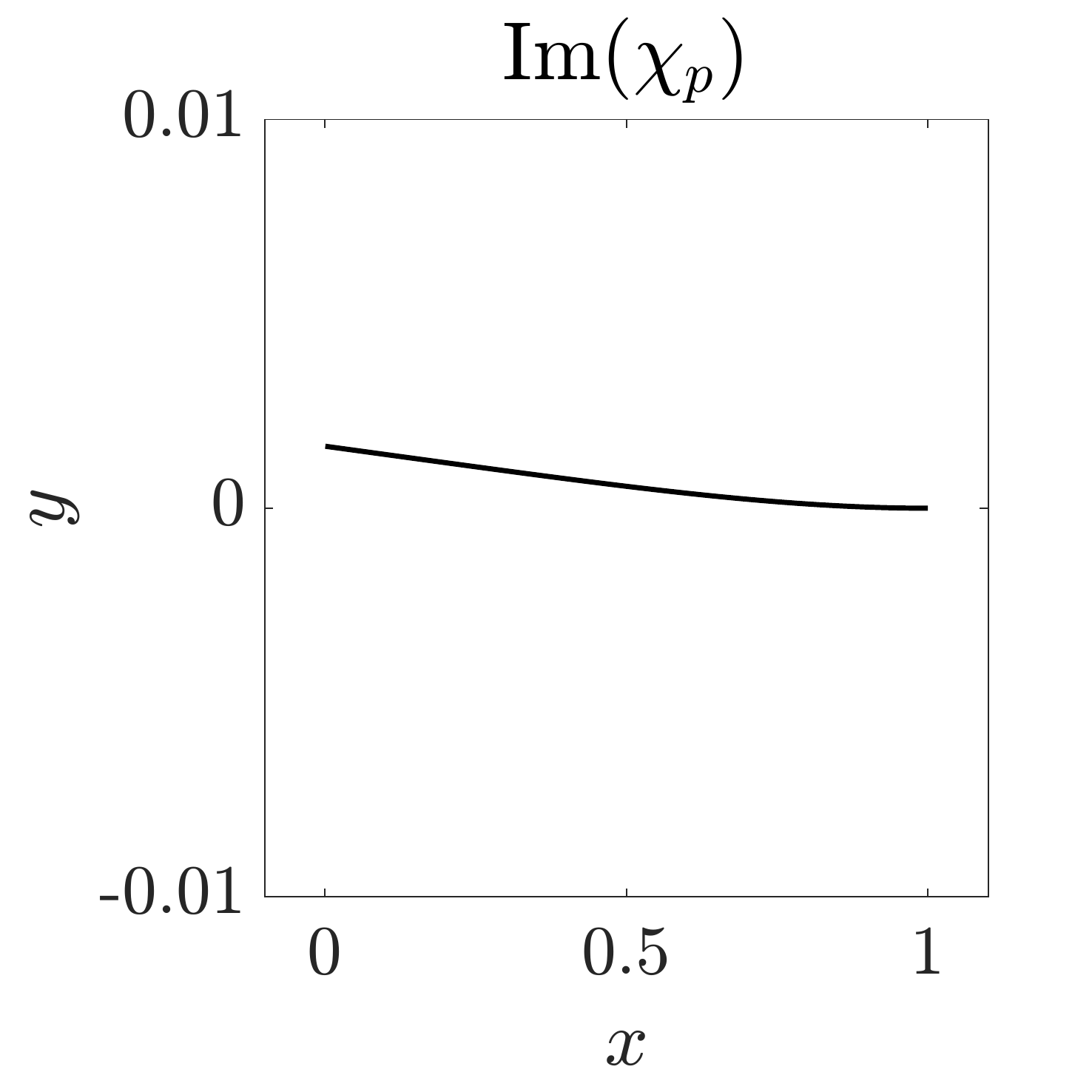}
		\vspace*{-0.3mm}
	\end{subfigure}
   	\caption{Real (top) and imaginary (bottom) parts of vorticity (left) and flag displacement (right) of the leading global mode of the deformed equilibrium for $M_\rho = 5, K_B = 0.37$ and $Re = 20$. Vorticity contours are in 20 increments from -0.05 to 0.05.}
\label{fig:mode_Re20_def}
\end{figure}

To illustrate the vortical structures and flag shapes associated with the instability mechanism at this lower Reynolds number, we plot the real and imaginary parts of the leading global mode near the critical stiffness for $M_\rho = 5$ in figure \ref{fig:mode_Re20_def} (the plot is similar for $M_\rho = 50$). Flag motion is associated with four vortical structures isolated near the flag surface. 

We emphasise that a linear stability analysis of the undeformed equilibrium state is associated with a zero-frequency (non-flapping) unstable mode, and therefore does not capture the flapping behaviour observed in the nonlinear simulations. This demonstrates that the divergence instability derived by \citet{Sader2016a} for inviscid fluids persists at lower Reynolds numbers. 

\begin{figure}
	\centering
	\begin{subfigure}[b]{0.45\textwidth}
		\centering
        		\includegraphics[scale=0.3, trim={0cm 0cm 0cm 0cm},clip]{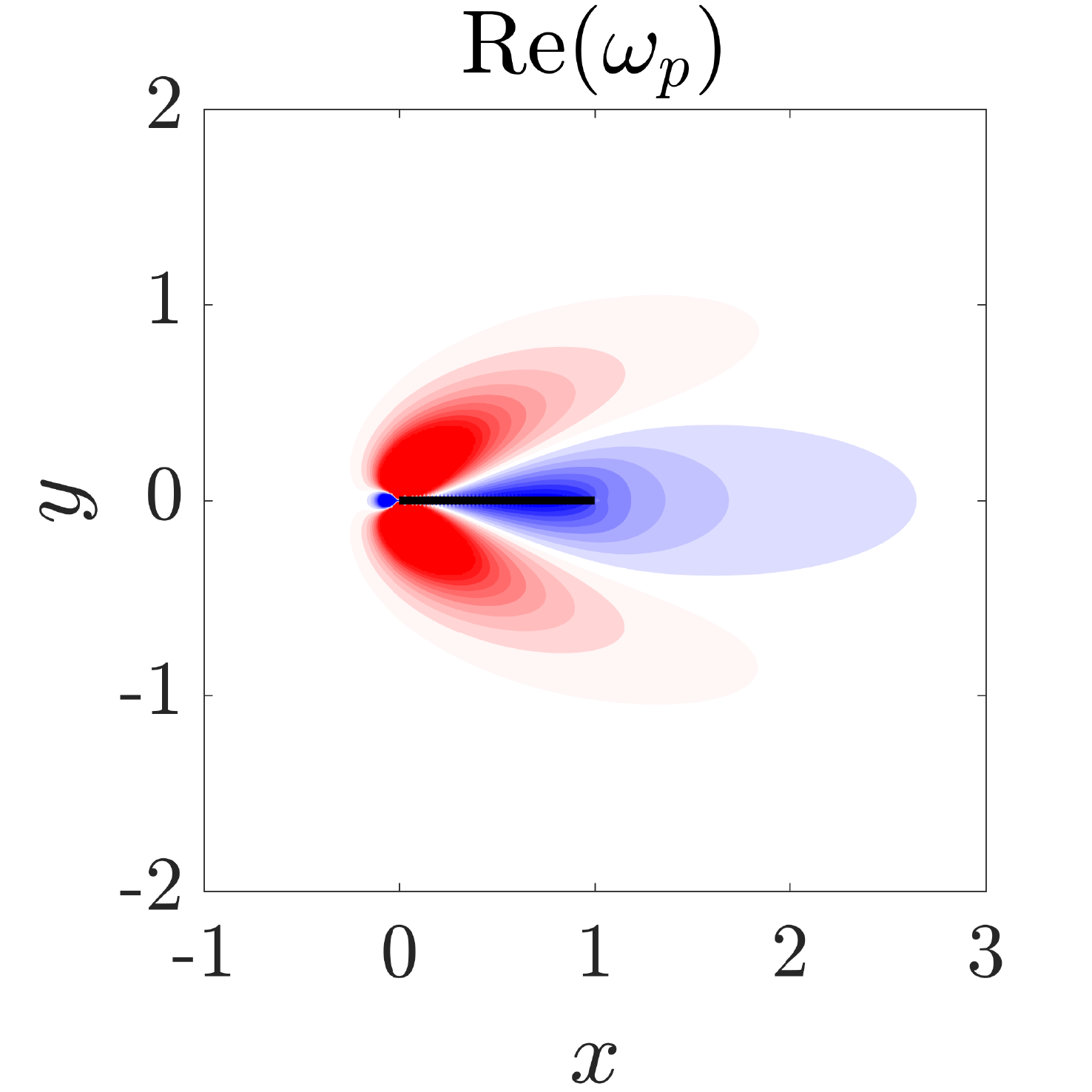}
	\end{subfigure}
	\begin{subfigure}[b]{0.45\textwidth}
		\centering
        		\includegraphics[scale=0.305, trim={0cm 0cm 0cm 0cm},clip]{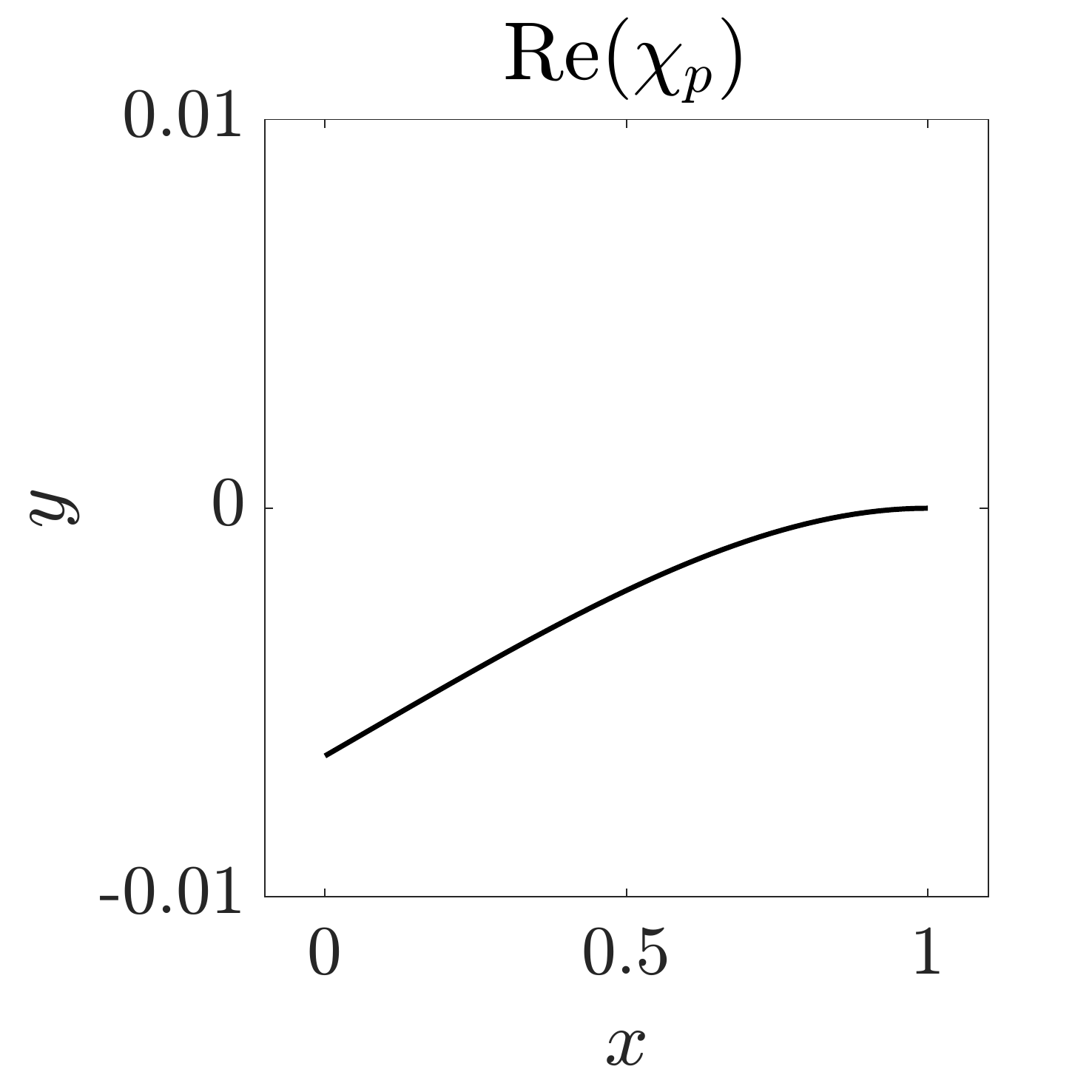}
		\vspace*{-0.3mm}
	\end{subfigure}
\caption{Real part of vorticity (left) and flag displacement (right) of the leading global mode of the undeformed equilibrium for $M_\rho = 5, K_B = 0.37$ and $Re = 20$. Vorticity contours are in 20 increments from -0.05 to 0.05.}
\label{fig:mode_Re20_undef}
\end{figure}

Figure \ref{fig:mode_Re20_undef} shows the leading global mode of the undeformed equilibrium. The mode has a similar flag shape and set of vortical structures to the real part of the leading mode of the deformed equilibrium. A noticeable distinction between the two, however, is that the vortical structures of the undeformed equilibrium mode are symmetric about the equilibrium flag position while those of the deformed equilibrium mode are not. The presence of asymmetry associated with flapping is indicative of the interplay between fluid forces, flag inertia, and internal flag stresses necessary to sustain flapping. To explore this interplay, consider a perturbation of the deformed equilibrium that sets the flag into motion in the direction of increasing deflection. This causes an increase in internal flag stresses that act to restore the flag to its deformed equilibrium. These stresses are opposed by the flag inertia and by forces from the oncoming fluid, which tend to destabilise the system further away from its deformed equilibrium. By contrast, if the flag is set into motion the other direction (towards the undeformed state), the fluid forces act to restore the flag to its deformed equilibrium and the flag inertia and internal flag stresses act as destabilising forces. The exchange of internal flag stresses and fluid forces as destabilising quantities is a unique feature of flapping about the deflected state---an analogous perturbation to a flag in the undeformed equilibrium results in flag stresses that are always restoring and fluid forces that are always destabilising.

\subsection{Large-amplitude flapping}

We now consider the transition from small-deflection deformed flapping to large-amplitude flapping. Within the small-deflection deformed flapping regime, the bifurcation diagrams show that decreasing stiffness causes an increase in flapping amplitude. This is associated with an increase in growth rate of the leading mode (see table \ref{tab:grow_Re20}). The mechanism through which the increasingly unstable leading mode develops into large-amplitude flapping is similar to what was discussed for heavy flags at $Re = 200$. Eventually, the growth in saturation amplitude leads the flag to deform past the centreline position and into the region of attraction of the deformed equilibrium on the other side of the centreline. This newly sampled deformed equilibrium is also associated with a saturation amplitude that leads the flag to flap past the centreline, and indefinite flapping occurs around these two equilibria. We emphasise that at this low Reynolds number, vortex shedding does not occur, and the flapping frequency from figure \ref{fig:freq_Re20} demonstrates that flapping is not a VIV in this regime. The nonlinear behaviour characterised by flapping around the two unstable deformed equilibria provides the necessary non-VIV flapping mechanism.

\begin{table}
\centering
\begin{tabular}{ c c }
{$K_B$} & {Leading mode growth rate}  \\ \hline
 0.37 & 0.008 \\
 0.35 & 0.039  \\ 
 0.32 & 0.120 \\
 0.29 & 0.342
\end{tabular}
 \caption{Growth rate of the leading global mode of the deformed equilibrium for $M_\rho = 5$ for stiffnesses in the small-deflection deformed flapping and large-amplitude flapping regimes.}
\label{tab:grow_Re20}
\end{table}

We note that this non-VIV flapping mechanism is distinct from what is observed in large-amplitude oscillations of elastically mounted cylinders at subcritical Reynolds numbers. In the elastically mounted cylinder case, \citet{Mittal2005} showed through nonlinear simulations and a global stability analysis that VIV persists at subcritical Reynolds numbers for certain parameters, and that large-amplitude vibrations are a result of this VIV. They moreover demonstrated that for these parameters the vibration frequency matched the bluff-body shedding frequency. By contrast, in the case of large-amplitude inverted-flag flapping, the flapping frequency is substantially smaller than the bluff-body vortex-shedding frequency.  

\subsection{Large-deflection equilibrium (deflected-mode regime)}

A continued decrease in stiffness leads to a bifurcation from large-amplitude flapping back to a stable deformed equilibrium with large deflection. This transition corresponds to the re-stabilisation of the leading global mode (\emph{e.g.}, for $M_\rho = 5$, $K_B = 0.17$ the growth rate of the leading mode is -0.032). Note that this deflected-mode state is distinct from that found at higher Reynolds numbers, where the flag undergoes small-amplitude oscillations driven by vortex shedding \citep{Sader2016a,Shoele2016}. Since vortex shedding is absent at $Re = 20$, the deflected-mode regime is a formal equilibrium of the fully-coupled equations of motion.

\section{Conclusions}

We used 2D high-fidelity nonlinear simulations and a global linear stability analysis of inverted-flag flapping to (i) investigate the physical mechanisms responsible for the onset of flapping, (ii) study the role of vortex shedding in large-amplitude flapping, and (iii) further characterise various regime bifurcations that were previously identified and explored \citep{Kim2013,Gurugubelli2015,Ryu2015,Shoele2016,Sader2016a}. We performed studies at $Re = 20$ and 200 for a wide range of $K_B$ and over a four-order-of-magnitude range of $M_\rho$. For $Re = 20$ and $M_\rho \le O(1)$, no flapping occurs and the flag transitions with decreasing stiffness from an undeformed equilibrium to a deformed equilibrium. For all other combinations of $Re$ and $M_\rho$ considered, with decreasing flag stiffness the system transitions from a stable undeformed equilibrium to a stable deflected equilibrium via a divergence instability, to an unstable deformed equilibrium through a supercritical Hopf bifurcation that exhibits small-amplitude flapping, to large-amplitude flapping, and finally to a deflected-mode state. Below we summarise the key features of each of these regimes.

\textit{Stable deflected equilibrium:} we demonstrated that for all parameters considered the stationary deflected state identified by \citet{Gurugubelli2015, Ryu2015} is a formal equilibrium of the fully-coupled equations, and that even when flapping occurs this equilibrium persists as an unstable steady-state. A similar deformed equilibrium was found at $Re = O(30,000)$ by \citet{Sader2016a} through the addition of damping; establishing similarities between these findings is an area for future work.

\textit{Small-deflection deformed flapping:} the deformed equilibrium becomes unstable and transitions to small-deflection deformed flapping with decreasing stiffness ($K_B$). This occurred at $Re =200$ for all mass ratios considered and at $Re=20$ for heavy flags ($M_\rho > O(1)$). This transition was shown to be initiated by a supercritical Hopf bifurcation of the deformed equilibrium state (i.e. a complex-conjugate set of eigenvectors becomes unstable). For all parameters that exhibited this small-deflection flapping regime, the leading mode and ensuing nonlinear behaviour are both devoid of vortex shedding and have a flapping frequency that is not commensurate with a VIV.

\textit{Large-amplitude flapping:} light flags ($M_\rho <O(1)$) at $Re = 200$ exhibit VIV behaviour in which the fluid forces on the flag are synchronised to the flag motion. This coincides with the arguments of \citet{Sader2016a} based on experimental measurements and a scaling analysis. By contrast, heavy flags ($M_\rho > O(1) $) did not exhibit VIV behaviour---also consistent with the scaling analysis of \citet{Sader2016a}. Yet, we found that large-amplitude flapping persists at these large values of $M_\rho$ despite the absence of VIV, and demonstrated that this flapping is due to the instability of the two deformed equilibria on either side of the centreline. To further highlight the potential for large-amplitude flapping without VIV mechanisms, we also showed that large-amplitude flapping occurs for heavy flags at $Re = 20$. No flapping was observed for flags with $M_\rho < O(1)$ at $Re = 20$, which is in agreement with the simulations of \citet{Ryu2015}.

\textit{Deflected-mode:} for $Re = 200$ we used a global stability analysis to confirm the argument of \cite{Shoele2016} that this regime is driven by the canonical bluff-body wake instability. For all masses considered, the leading mode has vortical structures similar to the leading global mode found in canonical bluff-bodies \citep{Barkley2006} and a flapping frequency commensurate with the $St \sim 0.2$ bluff-body scaling \citep{Roshko1954}. We then showed that the deflected mode does not exhibit any flapping at any mass ratio for $Re =20$, and the system is instead in a large-deflection equilibrium state. 

\textit{Chaotic flapping:} we identified chaotic flapping for light flags at $Re = 200$ and characterised this regime by switching between large-amplitude flapping and the deflected-mode regime. No chaotic flapping was observed at $Re = 200$ for $M_\rho > O(1)$ or at $Re = 20$ for any of the mass ratios considered.

These findings demonstrate a wide range of physical mechanisms that drive the various dynamical regimes of the inverted flag system. Moreover, they highlight that the system dynamics depend on both the Reynolds number and mass ratio. At the same time, these results motivate future work that compares our low-to-moderate Reynolds number computational findings with results at higher Reynolds numbers and in three dimensions.

\section{Acknowledgments}

AJG and TC gratefully acknowledge funding through the Bosch BERN program and through the AFOSR (grant number FA9550-14-1-0328). JES thanks the ARC Centre of Excellence in Exciton Science and the Australian Research Council Grants Scheme.

\bibliographystyle{jfm}
\bibliography{inv_bib}{}

\end{document}